\documentclass[aip,pop,reprint,groupaddress,noshowpacs,noshowkeys]{revtex4-2}

\usepackage{amsmath,amssymb,mathrsfs}

\usepackage{graphicx}% Include figure files
\usepackage{dcolumn}% Align table columns on decimal point
\usepackage{bm}% bold math
\usepackage{xcolor}
\usepackage[pdftex,
  hidelinks,
  colorlinks=true,
  linkcolor=blue,
  anchorcolor=black,
  citecolor=blue,
  urlcolor=blue,
  pdfauthor= {Michael Glinsky},
  pdfsubject = {Coordinate-free plasma theory},
  pdfdisplaydoctitle,
  bookmarks=true,
  bookmarksopen=true]{hyperref}

\begin{document}
\title{A coordinate-free expression of plasma theory}

\author{Michael E. Glinsky}
\affiliation{BNZ Energy Inc., Santa Fe, NM, USA}

\begin{abstract}
The theory of plasmas, that is collectives of charged particles, is developed using the coordinate-free and geometric methods of exterior calculus.  This dramatically simplifies the algebra and gives a geometric physical interpretation.  The fundamental foundation on which the theory is built is the conservation of phase space volume expressed by the Generalized Liouville Equation in terms of the Lie derivative.  The theory is expanded both in the order of the correlation, $n$, and in the weakness of the correlation, $\Gamma \ll 1$.  This gives a Generalized BBGKY (Bogoliubov-Born-Green-Kirkwood-Yvon) Hierarchy.  The derivation continues to give a new generalized formula for the Variational Theory of Reaction Rates (VTRR).  Pullbacks of the generalized formulas to generic canonical coordinates, $(p,q)$, and Poisson brackets are done.  Where appropriate, the canonical coordinates are assumed to be ``action-angle'' coordinates, $(J,\psi)$ or $(P,Q)$, that are generated by the solution to the Hamilton-Jacobi equation, the action $S(P,q;E(P),\tau)$.  Finally, generalized forms of all the common kinetic equations are derived:  the Vlasov Equation, the Boltzmann Equation, the Master Equation, the Fokker-Planck Equation, the Vlasov-Fokker-Planck (VFP) Equation, the Fluid Equations, and the MagnetoHydroDynamic (MHD) Equations.  Specific examples are given of these equations.  The application of the VTRR to three-body recombination in a strong magnetic field is shown.
\end{abstract}

\maketitle

\section{Introduction}
\label{sec:intro}
The traditional way that the theory of a collective of charged particles, a plasma, has been developed is to use the mathematical machinery of vector calculus in a Cartesian coordinate system, subject to an ElectroMagnetic (EM) force.  A continuity-like equation for the $N$-particle distribution function of coordinates and velocities, called the Liouville Equation, is justified based on an analysis of orbits as is done by Nicholson,\citep{nicholson.83}  
\begin{equation}
    \frac{\partial f_N}{\partial t} + \sum_{i=1}^N{\mathbf{v}_i \cdot \nabla_{\mathbf{x}_i} f_N} + \sum_{i=1}^N{\dot{\mathbf{V}}_i \cdot \nabla_{\mathbf{v}_i} f_N} = 0,
\end{equation}
where $f_N(\mathbf{x}_1,\dots,\mathbf{x}_N;\mathbf{v}_1,\dots,\mathbf{v}_N;t)$ is the $N$-particle distribution function and
\begin{equation}
    m_i \, \dot{\mathbf{V}}_i = q_i \left[ \mathbf{E}(\mathbf{x}_i,t) + \frac{\mathbf{v}_i}{c} \times \mathbf{B}(\mathbf{x}_i,t) \right]
\end{equation}
is the EM force on the $i^\text{th}$ particle.  A Mayer Cluster Expansion \citep{uhlenbeck.63} is then made in the order of the correlation, $n$, and the weakness of the correlation, $\Gamma \ll 1$.  Continuing to use vector calculus in a Cartesian coordinate system, subject to a Coulomb model (self generated electric field), the BBGKY (Bogoliubov-Born-Green-Kirkwood-Yvon) Hierarchy of equations are developed,
\begin{equation}
\begin{split}
    \frac{\partial f_k}{\partial t} &+ \sum_{i=1}^k{\mathbf{v}_i \cdot \nabla_{\mathbf{x}_i} f_k} + \sum_{i=1}^k{\sum_{j=1}^k{\mathbf{a}_{ij} \cdot \nabla_{\mathbf{v}_i} f_k}} \\
    &+ \frac{N-k}{V} \sum_{i=1}^k{\int{d\mathbf{x}_{k+1} \, d\mathbf{v}_{k+1} \, \mathbf{a}_{i,k+1} \cdot \nabla_{\mathbf{v}_i} f_{k+1}}},
\end{split}
\end{equation}
where 
\begin{multline}
    f_k(\mathbf{x}_1,\dots,\mathbf{x}_k;\mathbf{v}_1,\dots,\mathbf{v}_k;t) \\
    \equiv V^k \int{d\mathbf{x}_{k+1} \, d\mathbf{v}_{k+1} \dots d\mathbf{x}_N \, d\mathbf{v}_N \, f_N}
\end{multline}
is the $k$-particle distribution function and
\begin{equation}
    m_i \, \mathbf{a}_{ij} = \frac{q_i \, q_j}{|\mathbf{x}_i - \mathbf{x}_j|^3} \, (\mathbf{x}_i - \mathbf{x}_j)
\end{equation}
is the Coulomb force of the $j^\text{th}$ particle on the $i^\text{th}$ particle.

It should be noted that there are three significant issues with the BBGKY expansion.  The first is the super-convergence of the Mayer Cluster Expansion is reduced to the only asymptotically convergent BBGKY expansion.  The second is that, many times (in fact the most interesting), the plasma is not weakly correlated, $\Gamma \gtrsim 1$.  This leads to the third problem that draconian closure assumptions must be made.  In response to this, a new theory of collective behavior has been developed.\citep{glinsky.24a}  It is based on expressing the Mayer Cluster Expansion as the functional Taylor expansion, that is Heisenberg's S-matrix,\citep{heisenberg.43} of the generating functional, $S(p)[f(x)]$ or action, of the dynamics of the collective field, $f(x)$.  It does not suffer from any of the shortcomings of the BBGKY expansion.  

Despite this new development, there is still significant benefit to continuing to remedy many of the other limitations of traditional plasma theory.  These are the lack of a geometric interpretation of plasma theory.  We will find that the geometric interpretation leads to an understanding of the Variational Theory of Reaction Rates (VTRR) that allows it to be generalized.  The coordinate-free geometric interpretation will also allow the application to other coordinate systems and the inclusion of atomic structure and Hamiltonians other than a free particle subject to a self generated Coulomb force.  It will also allow collisions to be included in a systematic way without resorting to \textit{ad hoc} arguments.

This paper develops a coordinate-free expression of plasma and kinetic theory that uses the methods of exterior calculus as presented by Frankel.\citep{frankel2011geometry}  We explicitly use the symplectic, that is canonical, structure of the physical dynamics.  The starting point is the statement of conservation of probability or phase space volume for an ensemble of $N$ conservatively interacting particles with Hamiltonian, $H^{(N)}$, as shown in Eq.~\eqref{eqn:n_hamiltonian}.  This is expressed compactly as the Generalized Liouville Equation, in terms of the Lie derivative in Eq.~\eqref{eqn:gle}.  It continues with the coordinate-free development of the Generalized BBGKY Hierarchy, by expanding both in terms of the order of correlation $n$ and the weakness of correlation $\Gamma \ll 1$, as shown in Eq.~\eqref{eqn:bbgky}.  The practical pullback of the first two equations of the Generalized BBGKY Hierarchy, in terms of generic canonical coordinates $x=(p,q)$ and Poisson brackets, are shown in Eqns.~\eqref{eqn:1d_bbgky_1} and \eqref{eqn:1d_bbgky_2}.  The derivations are shown in Sec.~\ref{sec:bbgky}.

A new Generalized Variational Theory of Reaction Rates (VTRR) is developed in Sec.~\ref{sec:vrates}, starting with the Generalized BBGKY Hierarchy.  The coordinate-free expression for the Generalized VTRR is shown in Eq.~\eqref{eqn:vtrr.bound}, and the pullback to generic canonical coordinates is shown in Eq.~\eqref{eqn:vtrr.pullback}.

The Generalized BBGKY Hierarchy is then reduced, in Sec.~\ref{sec:kinetic}, to the various kinetic equations:  the Vlasov Equation, the Boltzmann Equation, the Master Equation, the Fokker-Planck Equation, the Vlasov-Fokker-Plank (VFP) Equation, the Fluid Equations, and the MagnetoHydroDynamic (MHD) Equations.  These equations are expressed in terms of the pullbacks to generic canonical coordinates and Poisson brackets, and, where appropriate, to ``action-angle'' or cyclic coordinates, $(P,Q)$ or $(J,\psi)$, that are generated by the solution to the Hamilton-Jacobi Equation
\begin{equation}
\label{eqn:HJ}
    \frac{\partial S(q,\tau)}{\partial \tau} + H \left(\frac{\partial S}{\partial q},q \right) = 0
\end{equation}
where $S(P,q;E(P),\tau)=S_P(q) - \tau \, E(P)$ is the action on extended phase space, and $p=\partial S/\partial q$ is the canonical momentum.  Note that the action, $S=\int{\lambda}=\int{p \, dq - H \, d\tau}$, where $\lambda$ is the Poincaré one-form on extended phase space or Lagrangian.  The equation of motion for the ``action-angle'' coordinates, $(P,Q)$, are
\begin{equation}
    \frac{dP}{d\tau} = 0
\end{equation}
and
\begin{equation}
    \frac{dQ}{d\tau} = \frac{\partial E(P)}{\partial P} \equiv \omega_Q(P).
\end{equation}
The action, $P$ or $J$, is a constant and the angle, $Q$ or $\psi$, evolves at a constant angular frequency, $\omega_Q$, and cycles between $0$ and $2\pi$.

Specific examples, such as a free particle with an electrostatic field, a free particle in an electromagnetic field, and an ion and guiding center electron in a strong magnetic field, are given in Sec.~\ref{sec:examples}.  Three different closures for the Fluid Equations will be shown:  a cold plasma, an isotropic velocity distribution, and a drifting Maxwellian (both isothermal and adiabatic).

Section~\ref{sec:app_vtrr} applies the new Generalized VTRR theory of Sec.~\ref{sec:vrates} to estimate the three-body recombination for an ion and guiding center electron in a strong magnetic field.  It is first done for the case of an infinitely massive ion, with terms for both the axial and perpendicular drift motion, where the strong magnetic field is in the axial direction.  This is followed by a derivation of a rough scaling for the case of a finite ion and electron mass, and the development of a two bottleneck model.  The summary and conclusions are given in Sec.~\ref{sec:conclusions}.

\begin{figure}
\noindent\includegraphics[width=\columnwidth]{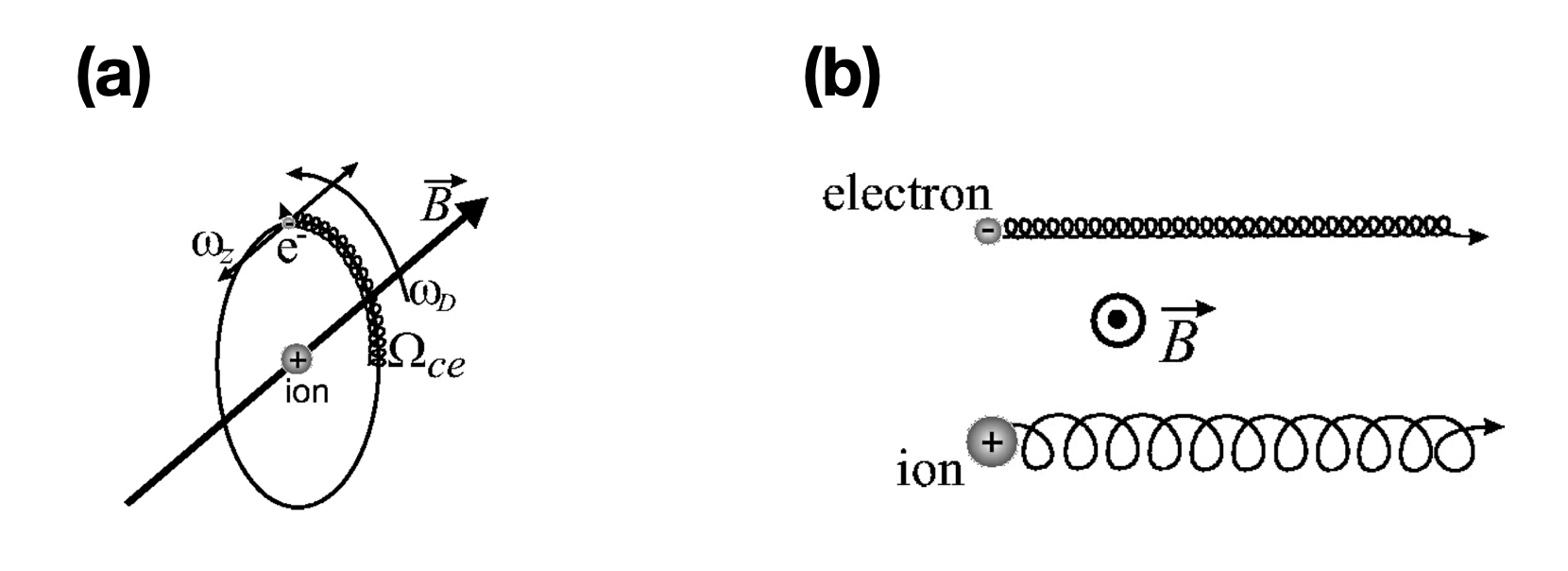}
\caption{\label{fig:gca.dp} Bound motion of: (a) the Guiding Center Atom (GCA), and (b) the Drifting Pair (DP).}
\end{figure}

This is a physical system of an electron and an ion in a constant magnetic field.  It is assumed that the magnetic field is strong enough that the electron undergoes guiding center motion, that is the electron cyclotron motion is an adiabatic invariant.  It is also assumed that this magnetic field is strong enough that the motion of the electron along the magnetic field is also adiabatic.  This gives two types of bound motion.  The first is called Guiding Center Atoms (GCAs),\citep{glinsky1991guiding} where the ion is assumed to be infinitely massive.  The second is called Drifting Pairs (DPs),\citep{kuzmin2004guiding} where the electron is assumed to have no mass.  These two bound states are shown in Fig.~\ref{fig:gca.dp}.
%===============================%
%===============================%

\begin{figure}
\noindent\includegraphics[width=17pc]{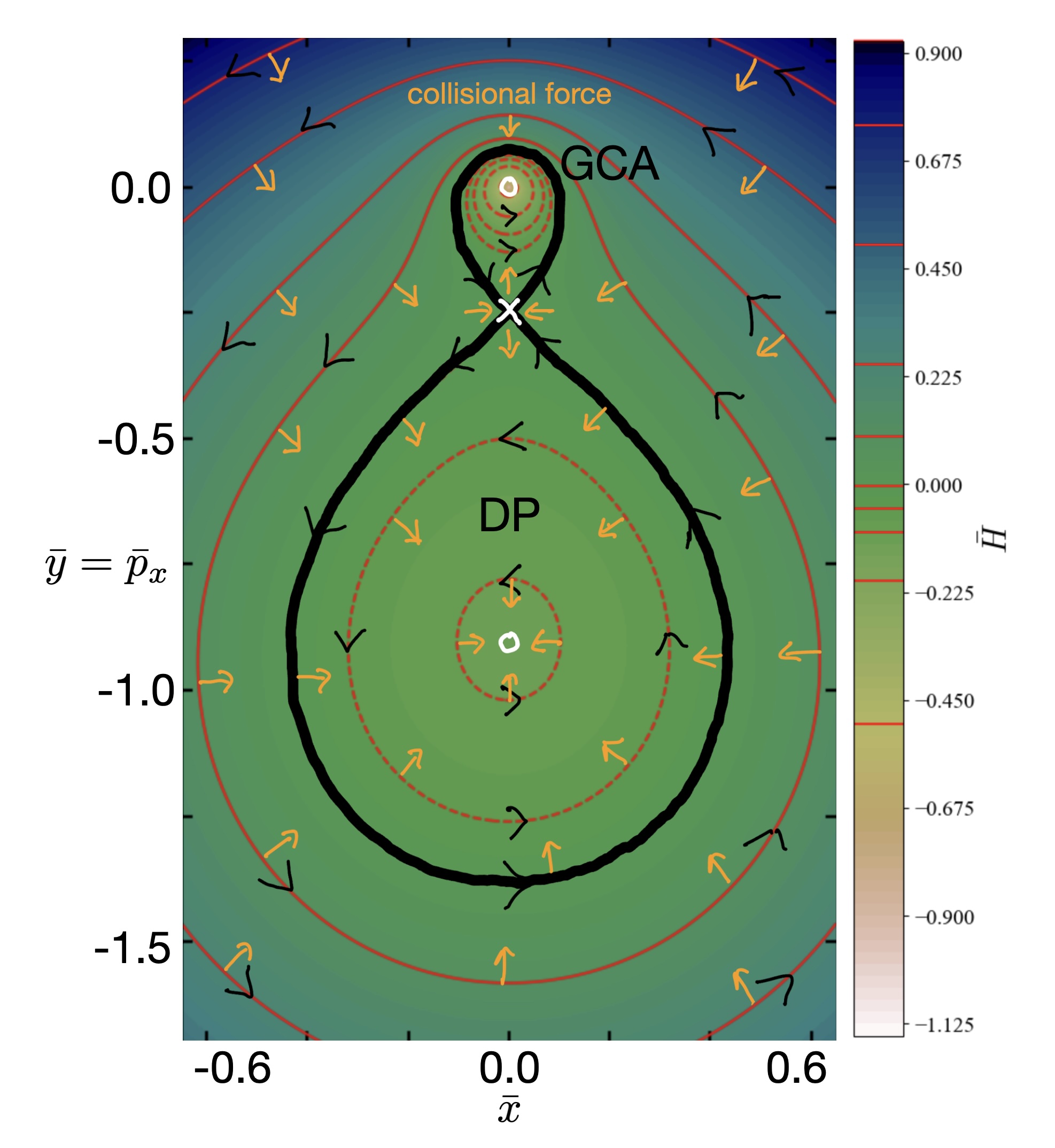}
\caption{\label{fig:gca.dp.plot} Plot of phase space trajectories shown by red lines with black arrows of an electron motion about an ion in a plane perpendicular to a strong magnetic field (solid for $\bar{H}>0$ and dashed for $\bar{H}<0$).  Note the boundary (thick black line) between two basins with the basin centers indicated by the white o-points, and the mountain heights with a mountain pass indicated by the white x-point.  The motion will circulate around the red lines and slowly relax in the directions shown by the orange arrows due to collisional forces.  The motion will relax from the mountain heights and eventually end up at the mountain pass (x-point), but this point is an unstable equilibrium and the motion (without control) will relax across the boundary into one of the two basins depending on how it approaches the mountain pass.  The motion will then continue to relax to the basin center at the o-point.  These are stable equilibriums.  The evolution of the one-particle distribution function is governed by the Collisional Vlasov Equation, $\partial f_1/\partial \tau + \left\{ \bar{H}, f_1 \right\} = \mathcal{C}[\bar{f}_1]$, where the LHS is advection by the motion of phase space trajectories indicated by the red lines with black arrows, and the RHS is the collisional motion indicated by the yellow arrows.}
\end{figure}

The phase space for this system is shown in Fig.~\ref{fig:gca.dp.plot}.  This system has two stable equilibriums, shown as the o-points, and two unstable equilibriums, shown as the x-points.  One set is for the electron at $0$ and $-0.3$, and another set is for the ion at $-0.9$ and $\infty$.  There are basins for the GCA (at the top) and the DP bound motions separated from the free motion by the black boundary called the separatrix.  The o-points are stable local minimums in the energy, and the x-points are unstable saddle points that are local maximums in the energy in the vertical direction.  It will take an infinite amount of time to approach the saddle points so that they will be metastable.  An external thermal force coming from a heat bath will exert a force perpendicular to the motion towards lower energy.  This will cause the motion to descend from the mountain top, spiraling around the mountain until the mountain pass, that is the saddle point, is reached.  It will take a long time to reach the mountain pass, and an equally long time to move away from the mountain pass.  The motion will then fall into one of the two basins of bound motion and spiral down to the valley center, that is the stable equilibrium.  The motion will eventually wander around the valley center in thermal equilibrium.  

The motion is generated, infinitesimally, by the Hamiltonian, $\bar{H}$, given in Eq.~\eqref{eqn:ei.full.hamiltonian}.  In general, this can be viewed as $\bar{H}=H_1+\left< H_{12} \right>$, where $H_1$ is the one-particle Hamiltonian, and $\left< H_{12} \right>$ is the two-particle interaction Hamiltonian averaged over the motion of the second particle.  The evolution of the one-particle distribution function, $f_1(P,Q)$, will be shown to be governed by the Collisional Vlasov Equation
\begin{equation}
    \frac{\partial f_1}{\partial \tau} + \left\{ \bar{H}, f_1 \right\} = \mathcal{C}[\bar{f}_1],
\end{equation}
where $\bar{f}_1(P)=\int{f_1 \, dQ}$ is averaged over the trajectory generated by $\bar{H}$, and $\mathcal{C}[\bar{f}_1]$ is the collision operator with a third body given by either the Boltzmann, Master, or Fokker-Planck operator.
%===============================%
%===============================%

\section{Derivation of the BBGKY Hierarchy}
\label{sec:bbgky}
We consider the Hamiltonian flow in the phase space of $N$ identical particles moving on the same base manifold, $M$.  This is to say that there exits an N particle Hamiltonian function
\begin{displaymath}
H^{(N)}:T^*M^N  \to \mathbb{R},
\end{displaymath}
where $T^*M^N$ is the product space of $N$ symplectic cotangent bundles, $\left( T^*M, \omega \right)$, and $\omega$ is a non-degenerate Poincaré two-form or sympletic metric on $T^*M$ that is exact, $\omega = d \lambda$, where $\lambda$ is the Poincaré one-form.  In addition, the flow on $T^*M^N$ is given by the vector field $U$ defined by
\begin{equation}
\label{eqn:ham_vector_def}
\text{i}_U\omega^{(N)} = -d H^{(N)},
\end{equation}
where
\begin{equation}
\omega^{(N)} \equiv \sum_{i=1}^N \omega_i
\end{equation}
and $\omega_i$ acts on $T^*M$ of the $i^\text{th}$ particle.  We restrict $H^{(N)}$ to cause only binary interactions, that is to say, we can write
\begin{equation}
\label{eqn:n_hamiltonian}
H^{(N)} =  \sum_{i=1}^N H_1(x_i) +   \sum_{i<j}^N H_2(x_i,x_j),
\end{equation}
where $x_i$ is a point in $T^*M$ of the $i^\text{th}$ particle and $H_1$ and $H_2$ are arbitrary functions such that
\begin{displaymath}
H_1:T^*M  \to \mathbb{R},
\end{displaymath}
\begin{displaymath}
H_2:T^*M^2  \to \mathbb{R}.
\end{displaymath}

We wish to describe the evolution of an ``ensemble'' of such $N$ particle systems.  A particular ensemble is specified by a $2N$-form on $T^*M^N$ which is symmetric with respect to particle interchange.  We call this form the \underline{$N$-particle distribution form}, $\rho^{(N)}$.  One can write the form in the following way
\begin{equation}
\rho^{(N)} = f_N (x_1,\dots,x_N) \; \text{vol}^{(N)},
\end{equation}
where
\begin{equation}
\text{vol}^{(N)} = \prod_{i=1}^N \wedge \omega_i
\end{equation}
and $f_N$ is symmetric with respect to $x_i \leftrightarrow x_j$ interchange.  The function $f_N$ is what is commonly called the $N$-particle distribution function.  It is physically the probability of finding the system of $N$ in a particular state specified by a point in $T^*M^N$.  We adopt the standard normalization of $\rho^{(N)}$ used by most plasma physicists
\begin{equation}
N = \frac{1}{V^N} \int_{T^*M^N} \rho^{(N)},
\end{equation}
where $V$ is a real number usually identified as the volume of $M$.

Assuming that there is no physical mechanism that can create or destroy particles (\textit{i.e.,} $dN/dt=0$), $\rho^{(N)}$ must be invariant under the flow generated by the vector field $U$ given in Eq.~\eqref{eqn:ham_vector_def},
\begin{equation}
\label{eqn:gle}
\boxed{\frac{\partial \rho^{(N)}}{\partial \tau} + \mathscr{L}_U \rho^{(N)} = 0,}
\end{equation}
where $\mathscr{L}_U$ is the Lie derivative.  It can be identified as a \underline{Generalized Liouville Equation}.  This equation, in principle, tells one exactly how the ensemble evolves.  The practical conundrum is that this equation gives too much information.  If one would write Eq.~\eqref{eqn:gle} in terms of coordinates, one would have a PDE for a function, $f_N$ of $2N$ variables.  Since $N$ can easily be of order $10^{23}$ or greater, a reduced description is obviously necessary to have a computationally manageable problem.  Therefore, we introduce the $n$-particle distribution form
\begin{equation}
\label{eqn:def_rhon}
\rho^{(n)} \equiv \frac{1}{V^{N-n}} \int_{T^*M^{N-n}} \rho^{(N)}.
\end{equation}

A few words are necessary to explain what is meant by the integration in Eq.~\eqref{eqn:def_rhon}.  Obviously the dimension of $T^*M^{N-n}$ does not match the order of the form $\rho^{(N)}$;  but because of the product structure of $T^*M^N$ one can always expand $T^*M^{N-n}$ to $R \times T^*M^{N-n} \subset T^*M^N$, where $R$ is an arbitrary region in $T^*M^n$.  The meaning of Eq.~\eqref{eqn:def_rhon} can now be stated as
\begin{displaymath}
\int_{R} \rho^{(n)} = \frac{1}{V^{N-n}} \int_{R \times T^*M^{N-n}} \rho^{(N)},
\end{displaymath}
for all $R \subset T^*M^n$.

Just as for the case of $\rho^{(N)}$, one can write $\rho^{(n)}$ as
\begin{displaymath}
\rho^{(n)} = f_n (x_1,\dots,x_n) \; \text{vol}^{(n)}
\end{displaymath}
where $f_n$ is commonly called the $n$-particle distribution function.

Starting with Eq.~\eqref{eqn:gle}, we proceed to derive and equation for the evolution of $\rho^{(n)}$.  Integrate Eq.~\eqref{eqn:gle} over $T^*M^{N-n}$ giving
\begin{equation}
\label{eqn:int_gle}
0 = \frac{1}{V^{N-n}} \int_{T^*M^{N-n}} {\left[\frac{\partial \rho^{(N)}}{\partial \tau} + \mathscr{L}_U \rho^{(N)}\right]},
\end{equation}
Commuting the $\partial / \partial \tau$ and the integration, then using Eq.~\eqref{eqn:gle}; rewrite Eq.~\eqref{eqn:int_gle} as
\begin{equation}
\label{eqn:time_int_gle}
\frac{\partial \rho^{(n)}}{\partial \tau} = - \frac{1}{V^{N-n}} \int_{T^*M^{N-n}} \mathscr{L}_U \rho^{(N)},
\end{equation}
We now wish to split the integral on the RHS into the sum of four terms, by splitting $U$ into parts and using the distributive properties of the Lie derivative.

Define the vector fields $u_i$, $u_{ij}$, $u^{(n)}$ and $u^{(n)}_\text{int}$ by
\begin{align}
\text{i}_{u_{i}}\omega^{(N)} &= -dH_1(x_i)\\
\text{i}_{u_{ij}}\omega^{(N)} &= -dH_2(x_i,x_j)\\
\text{i}_{u^{(n)}}\omega^{(N)} &= -dH^{(n)}(x_1,\dots,x_n)\\
\text{i}_{u_\text{int}^{(n)}}\omega^{(N)} &= -dH_\text{int}^{(n)}(x_1,\dots,x_{n+1}),
\end{align}
where
\begin{equation}
H_\text{int}^{(n)} \equiv \sum_{i=1}^{n}{H_2(x_i,x_{n+1})}
\end{equation}
One can now write Eq.~\eqref{eqn:time_int_gle} as
\begin{multline}
\label{eqn:time_int_gle_2}
- \frac{\partial \rho^{(n)}}{\partial \tau} = \frac{1}{V^{N-n}} \left[{ \int_{T^*M^{N-n}} {\mathscr{L}_{u^{(n)}} \rho^{(N)}}}\right. \\
+ \left.{(N-n) \int_{T^*M^{N-n}} {\mathscr{L}_{u_{n+1}} \rho^{(N)}}}\right. \\
+ \left.{(N-n) \int_{T^*M^{N-n}} {\mathscr{L}_{u^{(n)}_\text{int}} \rho^{(N)}}}\right. \\
+ \left.{\frac{(N-n)(N-n-1)}{2} \int_{T^*M^{N-n}} {\mathscr{L}_{u_{n+1,n+2}} \rho^{(N)}}}\right] .
\end{multline}
Here we have used the symmetry of $\rho^{(N)}$ under particle interchange to consolidate the last three terms in the square brackets.  For instance, 
\begin{equation}
\int_{T^*M^{N-n}} {\mathscr{L}_{u_{n+4}} \rho^{(N)}} = \int_{T^*M^{N-n}} {\mathscr{L}_{u_{n+1}} \rho^{(N)}}
\end{equation}
and
\begin{equation}
\int_{T^*M^{N-n}} {\mathscr{L}_{u_{n+6,n+20}} \rho^{(N)}} =  \int_{T^*M^{N-n}} {\mathscr{L}_{u_{n+1,n+2}} \rho^{(N)}}.
\end{equation}
We further reduce the RHS of Eq.~\eqref{eqn:time_int_gle_2} by commuting the parts of the integrals which do not involve the vectors in the Lie derivatives with the Lie derivatives, and by using Eq.~\eqref{eqn:def_rhon}.  This gives
\begin{multline}\label{eqn:time_int_gle_3}
- \frac{\partial \rho^{(n)}}{\partial \tau} =  \mathscr{L}_{u^{(n)}} \rho^{(n)} +  \frac{N-n}{V} \int_{T^*M} {\mathscr{L}_{u_{n+1}} \rho^{(n+1)}} \\
+ \frac{N-n}{V} \int_{T^*M} {\mathscr{L}_{u^{(n)}_\text{int}} \rho^{(n+1)}} \\
+ \frac{(N-n)(N-n-1)}{2V^2} \int_{T^*M^2} {\mathscr{L}_{u_{n+1,n+2}} \rho^{(n+2)}}.
\end{multline}
One can easily prove, using Cartan's formula for the Lie derivative that
\begin{equation}
\mathscr{L}_W \rho^{(n)} = d (\text{i}_W \rho^{(n)} )
\end{equation}
for any vector $W$.  Using this fact and Stoke's Theorem the second and fourth terms of Eq.~\eqref{eqn:time_int_gle_3} can be written as
\begin{multline}\label{eqn:time_int_gle_4}
- \frac{\partial \rho^{(n)}}{\partial \tau} =  \mathscr{L}_{u^{(n)}} \rho^{(n)} +  \frac{N-n}{V} \int_{\partial(T^*M)} {\text{i}_{u_{n+1}} \rho^{(n+1)}} \\
+ \frac{N-n}{V} \int_{T^*M} {\mathscr{L}_{u^{(n)}_\text{int}} \rho^{(n+1)}} \\
+ \frac{(N-n)(N-n-1)}{2V^2} \int_{\partial(T^*M^2)} {\text{i}_{u_{n+1,n+2}} \rho^{(n+2)}}.
\end{multline}
These two terms can be set equal to zero if
\begin{equation*}
\left. \text{i}_{u_{n+1}} \rho^{(n+1)} \right|_{\partial(T^*M)}  = \left. \text{i}_{u_{n+1,n+2}} \rho^{(n+2)} \right|_{\partial(T^*M^2)} = 0.
\end{equation*}
This condition is obviously met if either there are no particles on the boundary, or there is no flow across the boundary, or there is no boundary.

Neglecting these two terms we are left with the \underline{Generalized BBGKY Hierarchy}
\begin{equation}\label{eqn:bbgky}
\boxed{\frac{\partial \rho^{(n)}}{\partial \tau} +  \mathscr{L}_{u^{(n)}} \rho^{(n)} =  - n_0 \int_{T^*M} {\mathscr{L}_{u^{(n)}_\text{int}} \rho^{(n+1)}},}
\end{equation}
where $n_0 \equiv (N-n)/V$.  We now have a set of $n$ coupled equations for the $N$ different distributions forms, $\rho^{(n)}$ (for $n=1,\dots,N$), with the $n^\text{th}$ equation coupling $\rho^{(n)}$ to $\rho^{(n+1)}$.  This is what meant by a hierarchy of equations.

There is a straight forward geometric interpretation of Eq.~\eqref{eqn:bbgky}.  What appears on the LHS is the full time derivative of $\rho^{(n)}$ with respect to the $n$-particle Hamiltonian flow, $d \rho^{(n)} / d\tau$.  If the RHS were zero the $n$-particle distribution would evolve as if the other $N-n$ particles did not exist.  But, they do exist.  The RHS is not zero; it is a term which contains the Lie derivative with respect to $u^{(n)}_\text{int}$.  This vector can be rewritten as
\begin{equation*}
u^{(n)}_\text{int} = \sum_{i=1}^n {u_{i,n+1}}
\end{equation*}
which is the change in the flow of the first $n$ particles due to the $n+1$ particle.  This effect of this interaction is integrated over all possible locations of the $n+1$ particle in phase space weighted by the distribution (\textit{i.e.,} the probability it is at that point).  The geometric interpretation of the BBGKY Hierarchy is that one has $n$-particle Hamiltonian flow perturbed by the average interaction of the $n+1$ particle with the first $n$ particles.

The coordinate-free expression of the BBGKY Hierarchy given in Eq.~\eqref{eqn:bbgky} can be cast in more recognizable form by introducing generic canonical coordinates.  To keep things simple, we assume that $M$ is one dimensional.  At least locally we can find a coordinate $q$ for $M$ and the associated coordinates $(p,q)$ on $T^*M$.\citep{spivak2018calculus}  To have a more compact equation, we define $\widetilde{H}_i \equiv H_1(x_i)$ and $\widetilde{H}_{ij} \equiv H_2(x_i,x_j)$.  The first two equations in the hierarchy can now be pulled back and be written as
\begin{equation}\label{eqn:1d_bbgky_1}
\frac{\partial f_1}{\partial \tau} + \{ \widetilde{H}_1, f_1 \} = -n_0 \int{ dp_2 \; dq_2 \; \{\widetilde{H}_{12}, f_2 \}}
\end{equation}
and
\begin{equation}\label{eqn:1d_bbgky_2}
\begin{split}
\frac{\partial f_2}{\partial \tau} &+ \{ \widetilde{H}_1 +\widetilde{H}_2 + \widetilde{H}_{12}, f_2 \} \\
&= -n_0 \int{ dp_3 \; dq_3 \; \{ \widetilde{H}_{13} + \widetilde{H}_{23}, f_3 \}},
\end{split}
\end{equation}
where the Poisson bracket of the two functions is
\begin{equation*}
\{ g,h \} \equiv \sum_{i=1}^N { \frac{\partial g}{\partial p_i} \frac{\partial h}{\partial q_i} -  \frac{\partial g}{\partial q_i}  \frac{\partial g}{\partial p_i} } .
\end{equation*}

The remaining challenge is to find a systematic expansion in terms of a small parameter.  An expansion that will allow the hierarch to be truncated, giving just one or two equations for a like number of functions.  The expansion parameter could be $n_0b^3$ where $b$ is the range of short range binary interactions.  This allows one to use the ideas of Bogoliubov to reduce Eqns.~\eqref{eqn:1d_bbgky_1} and \eqref{eqn:1d_bbgky_2} to the Boltzmann Equation and the Master Equation.  The Master Equation can further be reduced, assuming that the transport step size is smaller than the scale on which the distribution function changes, to the Fokker-Plank Equation.  The details of these derivations will be shown in Sec.~\ref{sec:kinetic}.  If the interactions are not short range (\textit{i.e.,} $H_2 = |\vec{r}_1 - \vec{r}_2|$), one must then make the Mayer cluster expansion in terms of the weakness of particle correlations.  This allows one to manipulate Eq.~\eqref{eqn:1d_bbgky_1}  into the Vlasov Equation.  These expansions and the resulting equations form the practical framework around which kinetic theory can be developed.

\section{Generalized Variational Theory of Reaction Rates}
\label{sec:vrates}
The use of the BBGKY hierarchy that we develop in this section is not what is traditionally done in kinetic theory, but it is closely related to what is done in the Variational Theory of Reaction Rates (VTRR).  Inspired by the geometrical interpretation of Eq.~\eqref{eqn:bbgky}, we start by asking what is the flow out of a region $R \subset T^*M^n$ if we are given the $(n+1)$-particle distribution form, $\tilde{\rho}^{(n+1)}$.  This flow $d N^{(n)} / d \tau$ is
\begin{equation}
\begin{split}
\frac{d N^{(n)}}{d \tau} & = \int_{R} \frac{\partial \rho^{(n)}}{\partial \tau} +  \mathscr{L}_{u^{(n)}} \rho^{(n)} \\
& =  - n_0 \int_{R \times T^*M} {\mathscr{L}_{u^{(n)}_\text{int}} \tilde{\rho}^{(n+1)}} \\
& = - n_0 \int_{\partial{(R \times T^*M)}} {\text{i}_{u^{(n)}_\text{int}} \tilde{\rho}^{(n+1)}},
\end{split}
\end{equation}
where the region, $R$, is convected by the $n$-particle flow.  Making the easily satisfied assumption that
\begin{equation*}
\left. \text{i}_{u^{(n)}_\text{int}} \tilde{\rho}^{(n+1)} \right|_{R \times \partial (T^*M)} = 0,
\end{equation*}
we find that the flux is
\begin{equation}
\frac{d N^{(n)}}{d \tau} = - n_0 \int_{\partial{R} \times T^*M} {\text{i}_{u^{(n)}_\text{int}} \tilde{\rho}^{(n+1)}}.
\end{equation}

\begin{figure}
\noindent\includegraphics[width=17pc]{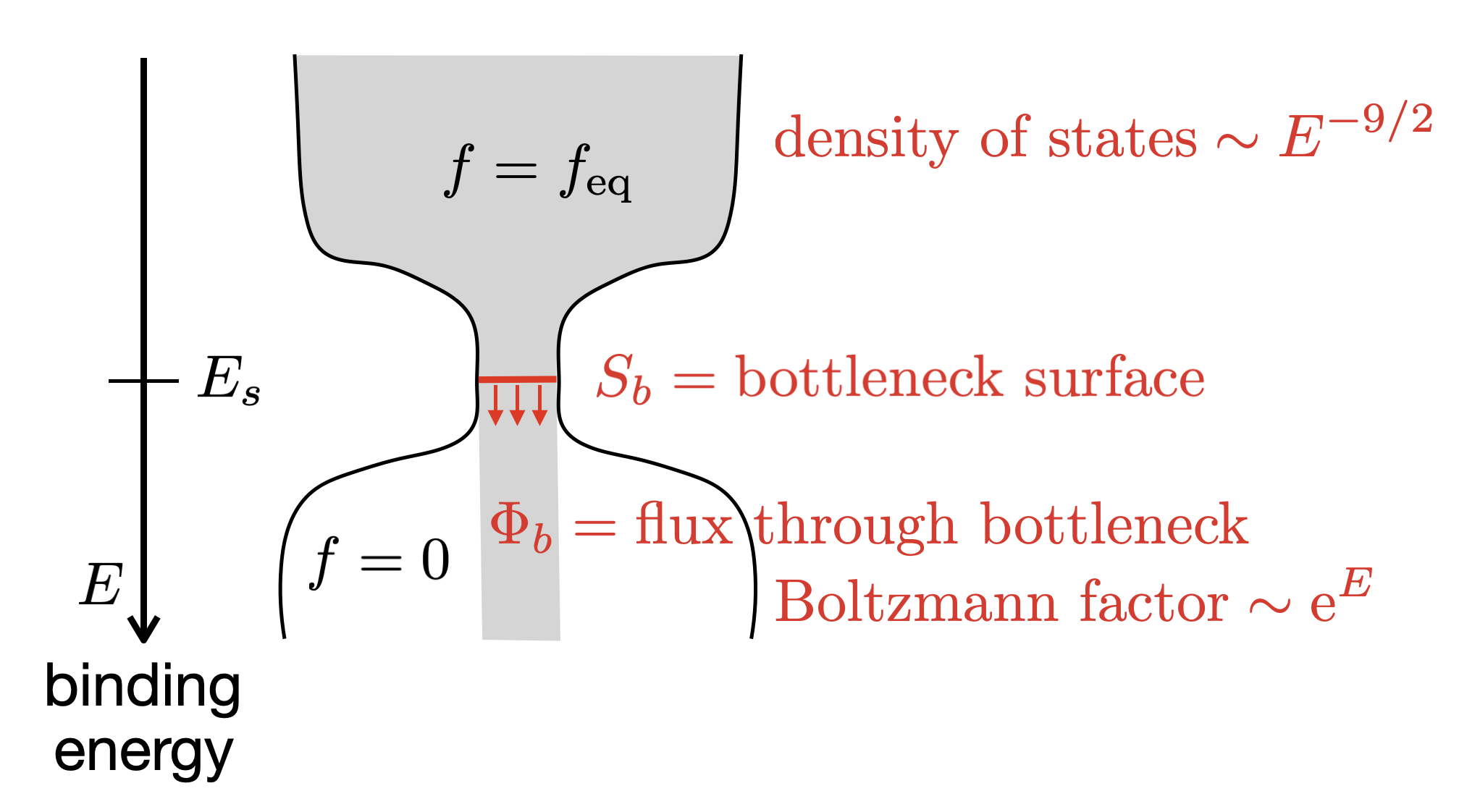}
\caption{\label{fig:bottleneck} Illustration of the geometry of the VTRR bound on the recombination flux, $\Phi_b$.  A bottleneck exists at energy surface, $S_b$, defined by bottleneck energy, $E_s$.  Above this surface the distribution is fully populated and dominated by the density of states that scale as $E_s^{-9/2}$.  The distribution is effectively zero below this surface where the equilibrium distribution is dominated by the Boltzmann factor of $\text{e}^{E_s}$.}
\end{figure}

We now assume that we can construct a continuous function $g$ on $T^*M^n$ advected by the flow $u^{(n)}$, that is $\mathscr{L}_{u^{(n)}} g=0$.  At least one such function exists since $\mathscr{L}_{u^{(n)}} H^{(n)}=0$.  One can foliate $T^*M^n$ by use of $g$ and the leaves of this foliation are invariant under the flow $u^{(n)}$.  We now consider a region, $R$, whose boundary is given by two sets of leaves $S_0$ and $S_1$.  We define an interior set of leaves, $S$, as a set of leaves which divide $R$ into two regions $R_0$ and $R_1$ such that $S+S_0 = \partial R_0$ and $S+S_1 = \partial R_1$.  We define the bottleneck set, $S_b$, as the interior set that gives the minimum value of the absolute flux, $\Phi_b$, in other words
\begin{equation}
\begin{split}
n_0 \int_{S \times T^*M} &{\left| \text{i}_{u^{(n)}_\text{int}} \tilde{\rho}^{(n+1)} \right|} \ge \Phi_b \\
&= n_0 \int_{S_b \times T^*M} {\left| \text{i}_{u^{(n)}_\text{int}} \tilde{\rho}^{(n+1)} \right|},
\end{split}
\end{equation}
for all $S$.  This geometry is shown in Fig.~\ref{fig:bottleneck}
%===============================%
%===============================%

If $\tilde{\rho}^{(n+1)}$ is an upper bound on $\rho^{(n+1)}$, that is
\begin{equation*}
\int_A \rho^{(n+1)} \le \int_A \tilde{\rho}^{(n+1)},
\end{equation*}
for all possible $\rho^{(n+1)}$ and $A \subset T^*M^{n+1}$;  then the steady state flux must be less than $\Phi_b$.  The proof of this statement starts by assuming that there is a steady state flux greater than $\Phi_b$.  Separating $R$ into two regions by $S_b$, one can readily see that $\rho^{(n+1)}$ will increase in one of these regions until it must exceed the upper bound $\tilde{\rho}^{(n+1)}$.

The flux $\Phi_b$ is the variational estimate of a rate;  the variation being over all interior sets of leaves $S$, and the rate being the steady state flux through $R$.

There are practical limitations to how good of a variational estimate one can obtain.  First, the larger number of $S$ one varies over;  the better the upper bound on the rate one will obtain.  Of course the larger the number of $S$, the more work one must do.  Second, the higher the order, $n$, of distribution form $\tilde{\rho}^{(n+1)}$ one uses;  the more one can take into account multiple particle correlations.  But the higher the order one uses;  the higher the dimension of the multiple integrals one must evaluate.  For most common applications one uses $\tilde{\rho}^{(2)}$, taking into account up to two-particle correlations.  The variation is taken over the one parameter family of $S_E$ such that $H^{(1)} = E$, where $E$ is a constant value for the one-particle energy.  The variational estimate for this case gives the formula for the new \underline{Generalized VTRR} (Generalized Variational Theory of Reaction Rates)
\begin{equation}
\label{eqn:vtrr.bound}
\boxed{\Phi_b = n_0 \int_{S_E \times T^*M} {\left| \text{i}_{u_{12}} \tilde{\rho}^{(2)} \right|}.}
\end{equation}
To cast this expression in more recognizable form, we use the same coordinates employed in Eq.~\eqref{eqn:1d_bbgky_1} and pullback Eq.~\eqref{eqn:vtrr.bound}.  This allows us to write
\begin{equation}
\label{eqn:vtrr.pullback}
\Phi_b = n_0 \int_{E=H_1} {dq_1 \; dp_2 \; dq_2 \; \tilde{f}_2 \left| \{H_{12},H_1\} \frac{\partial p_1}{\partial H_1} \right|}.
\end{equation}

\section{Reduction to common kinetic equations}
\label{sec:kinetic}
So far, the conservative Generalized Liouville Equation shown in Eq.~\eqref{eqn:gle}, which can be written as
\begin{equation}
    \frac{\text{D} \rho^{(N)}}{\text{D}\tau}=0,
\end{equation}
where
\begin{equation}
    \frac{\text{D}}{\text{D} \tau} \equiv \frac{\partial}{\partial \tau} + \mathscr{L}_u
\end{equation}
is the full time derivative that advects a quantity according to a vector field $u$ on the cotangent bundle, conserves phase space volume $\int{d\lambda}$.  This implies that the quantity $H^{(N)}$ is conserved and the flow follows a minimal trajectory, that is $\delta S=\delta \int{\lambda}=0$.  This equation was integrated to give an expression for $\rho^{(n)}$, the Mayer Cluster Expansion in the order of the correlation $n$ and the correlation parameter $\Gamma$ in Eq.~\eqref{eqn:bbgky}, which can be written as
\begin{equation}
\label{eqn:bbgky_D}
    \frac{\text{D}\rho^{(n)}}{\text{D}\tau} = \mathcal{C}[\rho^{(n+1)}] \thicksim \text{O}(\Gamma^3)
\end{equation}
where
\begin{equation}
    \mathcal{C}[\rho^{(n+1)}] \equiv - n_0 \int_{T^*M} {\mathscr{L}_{u^{(n)}_\text{int}} \rho^{(n+1)}}
\end{equation}
is the collision operator,
\begin{equation}
    \Gamma \equiv \frac{e^2/a}{k_\text{B}T} \thicksim g^{2/3} \thicksim \left( {n_0 b^3} \right)^{1/3}
\end{equation}
is the correlation parameter, $g \equiv 1/n_0 \lambda_\text{D}^3$ is the plasma parameter, $a \equiv (4 \pi n_0 / 3)^{-1/3}$ is the average distance between particles, $b \equiv e^2/k_\text{B}T$ is the distance of closest approach, and $\lambda_\text{D} \equiv (k_\text{B}T/4 \pi n_0 e^2)^{1/2}$ is the Debye length.  The Generalized BBGKY Hierarchy written in this form still is conservative.  Unfortunately, when we expand the distribution form in the weakness of correlation, $\Gamma \ll 1$, then truncate and close the hierarchy of equations, the system will no longer be conservative.  When the system is no longer conservative, the dynamics is no longer reversible, and there will be a dissipation (\textit{i.e.}, irreversible removal) of energy from the system.  The exception will be the Vlasov Equation.  As noted in Sec.~\ref{sec:intro}, this is not the case for the new generating functional approach of Glinsky.\citep{glinsky.24a}

\begin{figure}
\noindent\includegraphics[width=15pc]{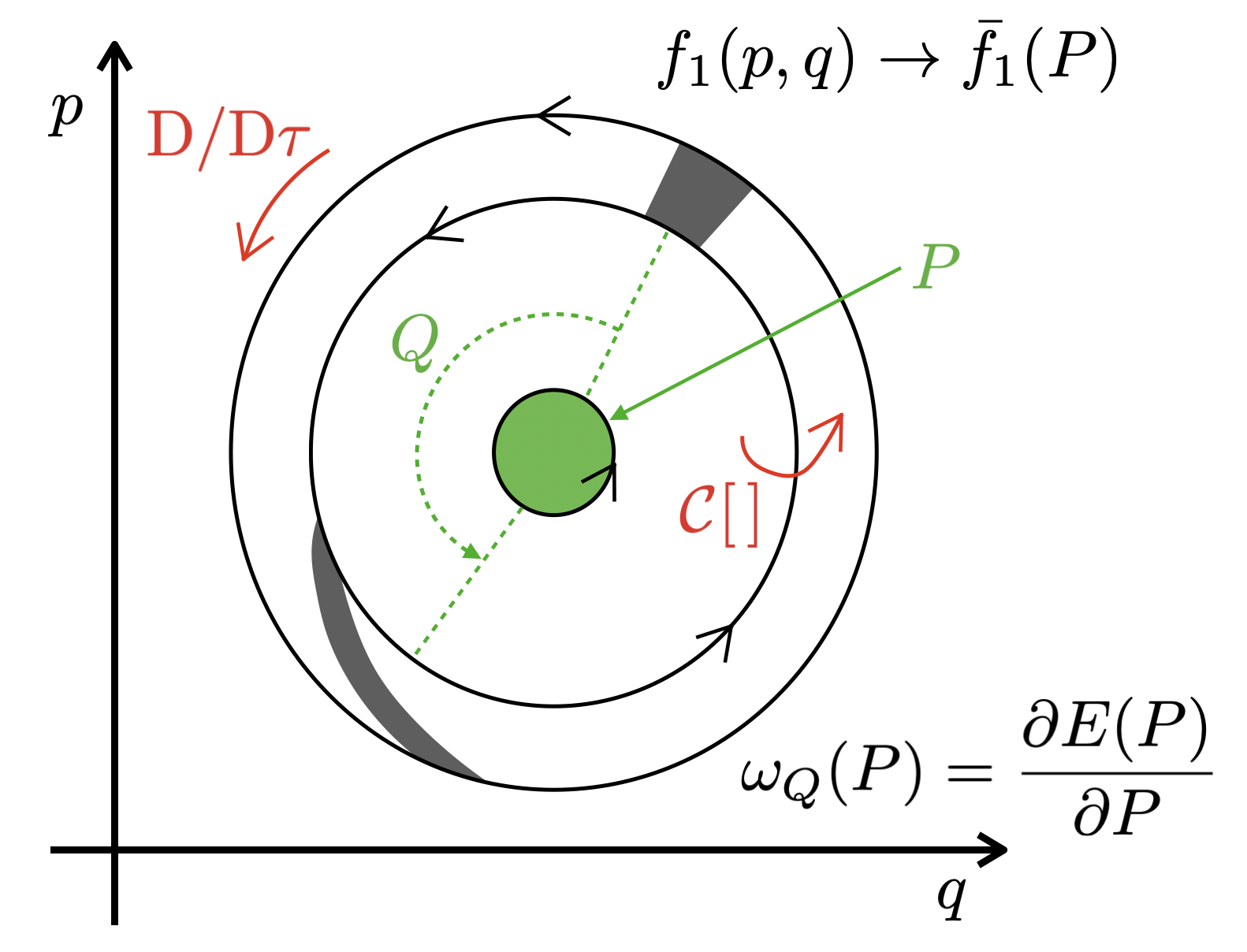}
\caption{\label{fig:b.mix} Diagram that shows how the one-particle distribution $f_1(p,q)$ evolves and relaxes to $\bar{f}_1(P)$.  Because of the $P$ dependence of $\omega_Q(P)$ there is a shear in a small volume that mixes the distribution uniformly in $Q$ so that the distribution relaxes to be a function of $P$ only.  The action, $P$, is the green area, and $Q$ is the green angle.  The motion in angle $Q$ is generated by the red canonical derivative $\text{D}/\text{D}\tau$, while the motion in action $P$ is generated by the red collision operator $\mathcal{C}[ \, ]$ that is an external thermal force that will establish a thermal equilibrium distribution $f_\text{eq}(P) \thicksim \exp{(-E(P)/k_\text{B}T)}$.  The distribution will relax on a dynamical time scale $\Omega \approx \omega_Q$ to the ``local'' equilibrium $\bar{f}_1(P)$, and on the collisional time scale $\nu_c$ to the thermal equilibrium $f_\text{eq}(P)$.  Note that the relaxation to a thermal equilibrium is not conservative, and does not happen in the new generating functional theory of Glinsky.\citep{glinsky.24a}}
\end{figure}

\begin{figure}
\noindent\includegraphics[width=12pc]{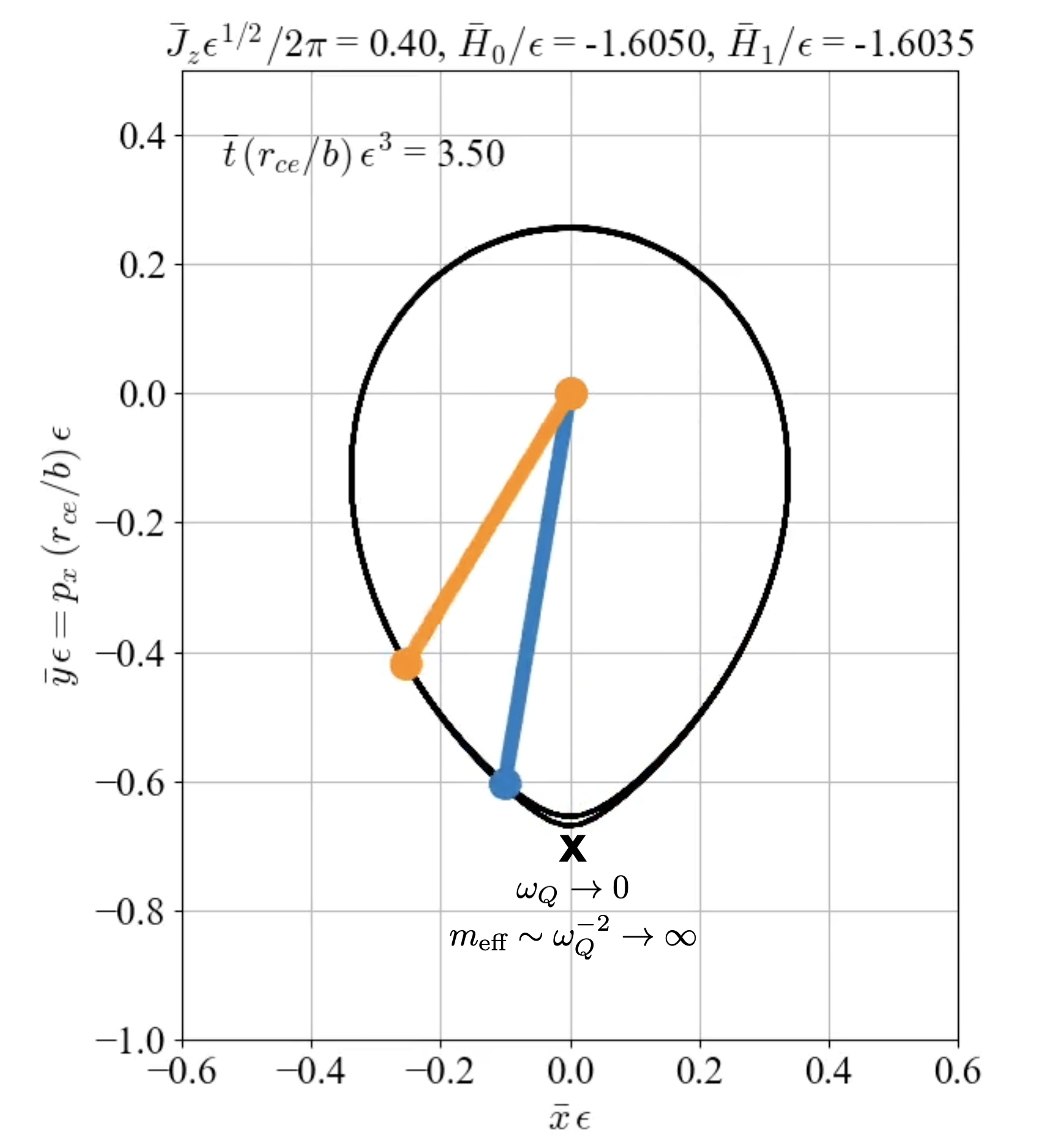}
\caption{\label{fig:gca.mix}  Phase space motion for two trajectories of the GCA, shown in Fig.~\ref{fig:gca.dp}:  (blue) $\bar{H}_0/\epsilon=-1.6050$ and $\omega_{Q0}=2.33$, (orange) $\bar{H}_1/\epsilon=-1.6035$ and $\omega_{Q1}=2.03$. The x-point is at $\bar{y}_s\epsilon=-0.7$ and $\bar{H}_s/\epsilon=-1.6022$.  Although the energies differ by 0.1\%, the frequencies differ by 13\%.  The large difference in frequencies causes a shear that rapidly mixes the distribution $f_1(J_D,\psi_D)$ to $\bar{f}_1(J_D)$, where $D$ references the drift motion of the GCA shown in this figure with frequency $\omega_Q$.  The frequency of the trajectory that goes to the x-point is zero, so that the effective mass is infinite, $m_\text{eff} \to \infty$, at the x-point.  The YouTube video of this figure can be found at this \href{https://youtu.be/YcYL6AWDAV8}{Link}.}
\end{figure}

We now discuss the Ideas of Bogoliubov on how the distribution forms relax to ``local'' equilibriums, so that the distributions will be uniform over the fundamental angles, $Q_i$, and only a function of the fundamental actions, $P_i$.  First note that the rate of evolution of the LHS of Eq.~\eqref{eqn:bbgky_D} for $n=1$, is $\Omega \equiv \bar{v}/b \thicksim \omega_Q(P)$ the dynamical rate, where $\bar{v} \equiv (k_\text{B}T/m)^{1/2}$ is the thermal velocity.  While the rate of evolution of the RHS is
\begin{equation}
    \nu_c \equiv \frac{1}{\Omega} \, \frac{d\Omega}{d\tau} = n_0 \bar{v} \, b^2
\end{equation}
the collision rate.  The rate of evolution of the LHS of Eq.~\eqref{eqn:bbgky_D} for $n=2$, is $\nu_c$ the collision rate.  While the rate of evolution of the RHS is
\begin{equation}
    \Omega_c \equiv \frac{1}{\Omega^2} \, \frac{d^2\Omega}{d\tau^2} = n_0^2 \bar{v} \, b^5
\end{equation}
the correlation rate.  Note that $\Omega \gg \nu_c \gg \Omega_c$.  The one-particle distribution, $f_1(P,Q)$, mixes along the dynamical orbit at the dynamical rate $\Omega$ to $\bar{f}_1(P)$ as shown in Fig.~\ref{fig:b.mix} and Fig.~\ref{fig:gca.mix}.  Then, $\bar{f}_1(P)$ evolves at the collision rate $\nu_c$.  Continuing, the two-particle distribution, $f_1(P_1,Q_1,P_2,Q_2)$, relaxes at the collision rate $\nu_c$ to $\bar{f}_2(P_1,P_2)$.  Then, $\bar{f}_2(P_1,P_2)$ evolves at the correlation rate $\Omega_c$, and so on.  Also, $f_2[\bar{f}_1]$, $f_2$ is a functional of $\bar{f}_1(\tau)$ with no other $\tau$ dependence.  This allows us to write
\begin{equation}
\label{eqn:f2.expansion}
\begin{split}
    f_2(p_1,q_1,p_2,q_2) &= f_1(p_1,q_1) \, f_1(p_2,q_2) + g(p_1,q_1,p_2,q_2) \\
    &= f_1(p_1,q_1) \, f_1(p_2,q_2) + \text{O}(\Gamma) \\
    &\approx f_1(p_1,q_1) \, f_1(p_2,q_2) \\
    &\to \bar{f}_1(P_1) \, \bar{f}_1(P_2).
\end{split}
\end{equation}
%===============================%
%===============================%
%===============================%
%===============================%

\subsection{Vlasov Equation}
\label{sec:Vlasov.equation}
To derive a general form of the Vlasov Equation, start with Eq.~\eqref{eqn:1d_bbgky_1} and substitute in the expression for $f_2$ from Eq.~\eqref{eqn:f2.expansion} to get
\begin{equation}
\begin{split}
    &\frac{\partial f_1(p_1,q_1)}{\partial \tau} + \{ \widetilde{H}_1(p_1,q_1), f_1(p_1,q_1) \}_1 \\
    &= -n_0 \int{ dp_2 \; dq_2 \; \{\widetilde{H}_{12}, f_1(p_1,q_1) \, f_1(p_2,q_2) \}_1} \\
    &= -n_0 \int{ dp_2 \; dq_2 \; f_1(p_2,q_2) \; \{\widetilde{H}_{12}, f_1(p_1,q_1) \}_1} \\
    &= - \left\{ n_0 \int{ dp_2 \; dq_2 \; f_1(p_2,q_2) \; \widetilde{H}_{12}}, f_1(p_1,q_1) \right\}_1 \\
    &= - \left\{ \left< \widetilde{H}_{12} \right>_2, f_1 \right\}_1.
\end{split}   
\end{equation}
Define
\begin{equation}
    \left< \widetilde{H}_{12} \right>_2 (p_1,q_1) \equiv - n_0 \int{ dp_2 \; dq_2 \; f_1(p_2,q_2) \; \widetilde{H}_{12}}
\end{equation}
to get the \underline{Generalized Vlasov Equation}
\begin{equation}
\label{eqn:vlasov}
    \boxed{\frac{\partial f_1}{\partial \tau} + \left\{ \widetilde{H}_1 + \left< \widetilde{H}_{12} \right>_2, f_1 \right\}_1 = 0.}
\end{equation}

Note that this equation is still conservative, since it can be written as 
\begin{equation}
\label{eqn:rho.1.vlasov}
    \frac{\text{D} \rho^{(1)}}{\text{D}\tau} = 0.
\end{equation}
For this reason, the Vlasov Equation is commonly called the Collisionless Boltzmann Equation, to be discussed in Sec.~\ref{sec:boltzmann} and Sec.~\ref{sec:VFP.equation}.  Unlike the following collisional equations of this section that truncate and close the conservative BBGKY Hierarchy in way that destroys the conservative nature and leads to dissipation, the Vlasov Equation remains conservative.  As such, the correlations of the plasma are maintained and the system is reversible.  This is why Landau damping \citep{landau.46} is not a dissipation, but a phase dispersion caused by the shear in phase space, shown in Fig.~\ref{fig:b.mix}.  This reversibility, that is conservative nature, is demonstrated by the existence of plasma wave echos,\citep{gould.67} and closed form solutions that have emergent behaviors \citep{oneil.65,oneil.71,morales.74} also known as BGK modes.\citep{bgk.57}

To write the Vlasov equation in more familiar form, assume that $\widetilde{H}_1=p_1^2$ and $\widetilde{H}_{12}=V(q_1-q_2)$.  The Vlasov Equation will be
\begin{equation}
    \frac{\partial f_1}{\partial \tau} + p_1 \frac{\partial f_1}{\partial q_1} - \frac{\partial \left< V \right>_2}{\partial q_1} \, \frac{\partial f_1}{\partial p_1} = 0 
\end{equation}
or in vector form using Cartesian non-relativistic coordinates
\begin{equation}
\label{eqn:free.vlasov}
    \frac{\partial f_1}{\partial t} + \mathbf{v} \cdot \nabla_\mathbf{x} f_1 + \left< \mathbf{a}_{12} \right> \cdot \nabla_\mathbf{v} f_1 = 0,
\end{equation}
where
\begin{equation}
\begin{split}
    \left< \mathbf{a}_{12} \right> &\equiv - \frac{\partial}{\partial q_1} \left( n_0 \int{ dp_2 \; dq_2 \; f_1(p_2,q_2) \; V(q_1-q_2)} \right)\\
    &= \int{d\mathbf{v}_2 \; d\mathbf{x}_2 \; f_1(\mathbf{v}_2,\mathbf{x}_2) \; \mathbf{a}_{12}} \\
    &= \frac{e}{m} \, \mathbf{E}.
\end{split}
\end{equation}
This is the common form of the Vlasov Equation for an electrostatic field when $V(q_1-q_2)=e/|\mathbf{x}_1-\mathbf{x}_2|$ and $m \, \mathbf{a}_{12}=e^2 \, (\mathbf{x}_1 - \mathbf{x}_2)/|\mathbf{x}_1 - \mathbf{x}_2|^3 = e \, \mathbf{E}$.  The advantage of the canonical exterior calculus viewpoint of Eq.~\eqref{eqn:vlasov} is the generality of the formalism, the simplicity and geometric interpretability of the result, and the ease of calculation.  For instance, it would be very easy to extend the Vlasov equation to include the full EM fields and to write it in relativistic helical coordinates.  It was also easy to recognize that it is still conservative by the form of Eq.~\eqref{eqn:rho.1.vlasov}.

\subsection{Boltzmann Equation}
\label{sec:boltzmann}
We again use the ideas of Bogoliubov to close the equations.  We start with the first two equations of the BBGKY Hierarchy, Eqns.~\eqref{eqn:1d_bbgky_1} and \eqref{eqn:1d_bbgky_2}, neglect the $f_3$ third order term on the RHS of the second equation and the $\partial_\tau f_2$ term on the LHS, to give
\begin{equation}
\begin{split}
    \frac{\partial f_1}{\partial \tau} + \{\widetilde{H}_1,f_1\} &= - n_0 \int{d \Gamma_2 \; \{\widetilde{H}_{12},f_2\}_1} \\
    \{\widetilde{H}_1+\widetilde{H}_2+\widetilde{H}_{12}, f_2\}_{12} &= 0,
\end{split}
\end{equation}
where $d \Gamma_2 \equiv d p_2 \; d q_2$.  Now substitute $\{ \widetilde{H}_{12},f_2\}_{12}$ from the second equation into the first equation, yielding
\begin{equation}
    \frac{\partial f_1}{\partial \tau} + \{\widetilde{H}_1,f_1\} = n_0 \int{d \Gamma_2 \; [\{\widetilde{H}_1,f_2\}_1 + \{\widetilde{H}_2,f_2\}_2]}.
\end{equation}
Transform, $(p_i,q_i) \to (J_i,\psi_i)$, into action-angle variables, where $\widetilde{H}_i=\widetilde{H}_i(J_i)$, so that $J_i = \text{constant}$ and $\psi_i=\omega_i \tau + \psi_{i0}$, since $\omega_i=\partial \widetilde{H}_i/\partial J_i=\text{constant}$ and $\partial \widetilde{H}_i/\partial \psi_i=0$.  Given that  $d \Gamma_i = d J_i \; d \psi_i$,
\begin{equation}
\begin{split}
    \frac{\partial f_1}{\partial \tau} &+ \omega_1 \frac{\partial f_1}{\partial \psi_1} = n_0 \int{d \Gamma_2 \; \left( \omega_1 \frac{\partial f_2}{\partial \psi_1} + \omega_2 \frac{\partial f_2}{\partial \psi_2} \right)} \\
    \frac{\partial f_1}{\partial \tau} &+ \frac{\partial}{\partial \psi_1} \left( \omega_1 f_1 \right) \\
    &= n_0 \int{d \Gamma_2 \; \left( \frac{\partial}{\partial \psi_1} \left( \omega_1 f_2 \right) + \frac{\partial}{\partial \psi_2} \left( \omega_2 f_2 \right) \right)}.
\end{split}
\end{equation}
Using the fact that $\Omega \gg \nu_c$, average over $\psi_1$ by $\int{d \psi_1/2\pi}$ and define
\begin{equation}
    \bar{f}_1(J_1) \equiv \int{\frac{d\psi_1}{2\pi}\, f_1(J_1,\psi_1)}
\end{equation}
to give
\begin{equation}
\begin{split}
    \frac{\partial \bar{f}_1(J_1)}{\partial \tau} &= n_0 \int{d J_2 \; d \psi_2 \; \int{ \frac{d \psi_1}{2\pi} \; \frac{\partial}{\partial \psi_2} \left( \omega_2 f_2 \right)}} \\
    &= n_0 \int{d J_2 \; \frac{d \psi_1}{2\pi} \int_{\partial V}{v_c(J_2) \, d \sigma_s \, (f_{2\text{f}}-f_{2\text{o}})}},
\end{split}
\end{equation}
where $f_{2\text{o}}= \bar{f}_1(J_1) \, \bar{f}_1(J_2)$ before the collision, $f_{2\text{f}}= \bar{f}_1(J_{1\text{f}}) \, \bar{f}_1(J_{2\text{f}})$ after the collision, $J_{1\text{f}}(J_1,\psi_1,J_2)$, and $J_{2\text{f}}(J_1,\psi_1,J_2)$.  So,
\begin{multline}
    \frac{\partial \bar{f}_1(J_1)}{\partial \tau} = n_0 \int{d J_2 \; \frac{d \psi_1}{2\pi} \int_{\partial V}{v_c(J_2) \, d \sigma_s}} \\
    [\bar{f}_1(J_{1\text{f}}(J_1,\psi_1,J_2)) \; \bar{f}_1(J_{1\text{f}}(J_1,\psi_1,J_2)) \\
    - \bar{f}_1(J_1) \; \bar{f}_1(J_2)].
\end{multline}
Now multiply by the identity
\begin{equation}
    I=\int{dJ'_1 \; dJ'_2 \; \delta(J'_1-J_{1\text{f}}) \; \delta(J'_2-J_{2\text{f}})},
\end{equation}
to give
\begin{multline}
    \frac{\partial \bar{f}_1(J_1)}{\partial \tau} = n_0 \int{d J_2 \; dJ'_2 \; dJ'_1 \; \frac{d \psi_1}{2\pi} \int_{\partial V}{v_c(J_2) \, d \sigma_s}} \\
    \delta(J'_1-J_{1\text{f}}) \; \delta(J'_2-J_{2\text{f}}) \\
    [\bar{f}_1(J'_1) \; \bar{f}_1(J'_2) - \bar{f}_1(J_1) \; \bar{f}_1(J_2)].
\end{multline}
By defining the differential cross section, $d \sigma/d\Omega$ and the scattering kernel $K(J_1,J_2|J'_1,J'_2)$
\begin{multline}
    |v| \frac{d \sigma}{d \Omega} \equiv K(J_1,J_2|J'_1,J'_2) \equiv |v| \frac{d\sigma}{dJ'_1 \; dJ'_2}  \\
    \equiv n_0 \int{ \frac{d \psi_1}{2\pi} \int_{\partial V}{v_c(J_2) \, d \sigma_s \; \delta(J'_1-J_{1\text{f}}) \; \delta(J'_2-J_{2\text{f}})}},
\end{multline}
where $d\Omega \equiv dJ'_1 \; dJ'_2$, we can derive two forms of the \underline{Generalized Boltzmann Equation}
\begin{equation}
\label{eqn:gen.boltzmann}
    \boxed{\frac{\partial \bar{f}_1(J_1)}{\partial \tau} = \int{dJ_2 \, d\Omega \, |v| \frac{d \sigma}{d \Omega} [\bar{f}_1(J'_1) \, \bar{f}_1(J'_2)-\bar{f}_1(J_1) \, \bar{f}_1(J_2)]}}
\end{equation}
or
\begin{multline}
    \frac{\partial \bar{f}_1(J_1)}{\partial \tau} = \int{dJ_2 \; dJ'_1 \; dJ'_2 \; \; K(J_1,J_2|J'_1,J'_2)} \\
   [\bar{f}_1(J'_1) \, \bar{f}_1(J'_2)-\bar{f}_1(J_1) \, \bar{f}_1(J_2)].
\end{multline}
This equation expresses the collision operator for the one-particle distribution function as the integral, over all final states of both particles and all initial states of the second particle, of either the differential scattering cross section $d \sigma/d\Omega$ or the scattering kernel $K(J_1,J_2|J'_1,J'_2)$, times the difference in $\bar{f}_2$.

\subsection{Master Equation}
\label{sec:master.equation}
We start by writing the Generalized Boltzmann Equation as
\begin{multline}
   \frac{\partial \bar{f}_1(J_1)}{\partial \tau} = \int{dJ_2 \; dJ'_1 \; dJ'_2 \; \; K(J_1,J_2|J'_1,J'_2)} \\
   [\bar{f}_1(J'_1) \, \bar{f}_1(J'_2)-\bar{f}_1(J_1) \, \bar{f}_1(J_2)] \\
   = \int{dJ_2 \, dJ'_1 \, dJ'_2 \, \bar{f}_\text{th}(J_1) \, \bar{f}_\text{th}(J_2) \,  K(J_1,J_2|J'_1,J'_2)} \\
   \left[ \frac{\bar{f}_1(J'_1) \, \bar{f}_1(J'_2)}{\bar{f}_\text{th}(J'_1) \; \bar{f}_\text{th}(J'_2)} - \frac{\bar{f}_1(J_1) \, \bar{f}_1(J_2)}{\bar{f}_\text{th}(J_1) \; \bar{f}_\text{th}(J_2)} \right]
\end{multline}
where we have used the fact that 
\begin{equation}
    \bar{f}_\text{th}(J'_1) \; \bar{f}_\text{th}(J'_2) = \bar{f}_\text{th}(J_1) \; \bar{f}_\text{th}(J_2)
\end{equation}
because of conservation of energy
\begin{equation}
    \widetilde{H}_1(J'_1) + \widetilde{H}_1(J'_2) = \widetilde{H}_1(J_1) + \widetilde{H}_1(J_2),
\end{equation}
where
\begin{equation}
     \bar{f}_\text{th}(J) \equiv \text{e}^{-\widetilde{H}_1(J) / k_\text{B}T} \text{   and   } \frac{\partial \bar{f}_\text{th}(J)}{\partial \tau}=0.
\end{equation}
Since the second particle is free, $\bar{f}_1(J_2)=\bar{f}_\text{th}(J_2)$ and $\bar{f}_1(J'_2)=\bar{f}_\text{th}(J'_2)$ so that
\begin{multline}
    \frac{\partial \bar{f}_1(J_1)}{\partial \tau} = \int{dJ'_1 \left[ \int{ dJ'_2 \, dJ_2 \, \bar{f}_\text{th}(J_2) \, K(J_1,J_2|J'_1,J'_2)} \right]} \\
   \bar{f}_\text{th}(J_1) \; \left[ \frac{\bar{f}_1(J'_1)}{\bar{f}_\text{th}(J'_1)} - \frac{\bar{f}_1(J_1)}{\bar{f}_\text{th}(J_1)} \right].
\end{multline}
Defining the equilibrium transition rate as
\begin{equation}
    k(J_1|J'_1) \equiv \int{ dJ'_2 \, dJ_2 \, \bar{f}_\text{th}(J_2) \, K(J_1,J_2|J'_1,J'_2)},
\end{equation}
then substituting it into the equation gives
\begin{equation}
     \frac{\partial \bar{f}_1(J_1)}{\partial \tau} = \int{dJ'_1 \; k(J_1|J'_1) \; \bar{f}_\text{th}(J_1) \left[ \frac{\bar{f}_1(J'_1)}{\bar{f}_\text{th}(J'_1)} - \frac{\bar{f}_1(J_1)}{\bar{f}_\text{th}(J_1)} \right]}.
\end{equation}
By TP symmetry (time reversal and spacial inversion),
\begin{equation*}
    K(J_1,J_2|J'_1, J'_2)=K(J'_1,J'_2|J_1, J_2),
\end{equation*}
and energy conservation, there is \underline{detailed balance} (see Lifshitz and Pitaevskii,\citep{ll.kinetics} Chapter I, Sections 2-3)
\begin{equation}
    \bar{f}_\text{th}(J) \; k(J|J') = \bar{f}_\text{th}(J') \; k(J'|J).
\end{equation}
Using detailed balance and substituting it into the equation, one gets the \underline{Generalized Master Equation}
\begin{equation}
\label{eqn:gen.master}
    \boxed{\frac{\partial \bar{f}_1(J)}{\partial \tau} = \int{dJ' \; \left[ k(J'|J) \; \bar{f}_1(J') - k(J|J') \; \bar{f}_1(J) \right]}.}
\end{equation}
This equation expresses the collision operator for the one-particle distribution function as the integral, over all final states of the first particle, of the incoming flux minus the outgoing flux.  It is assumed that the incoming states of the second particle are in thermal equilibrium, or equivalently $\bar{f}_1(J_2) = \bar{f}_\text{th}(J_2)$.

\subsection{Fokker-Planck Equation}
\label{sec:FP.equation}
\begin{figure}
\noindent\includegraphics[width=\columnwidth]{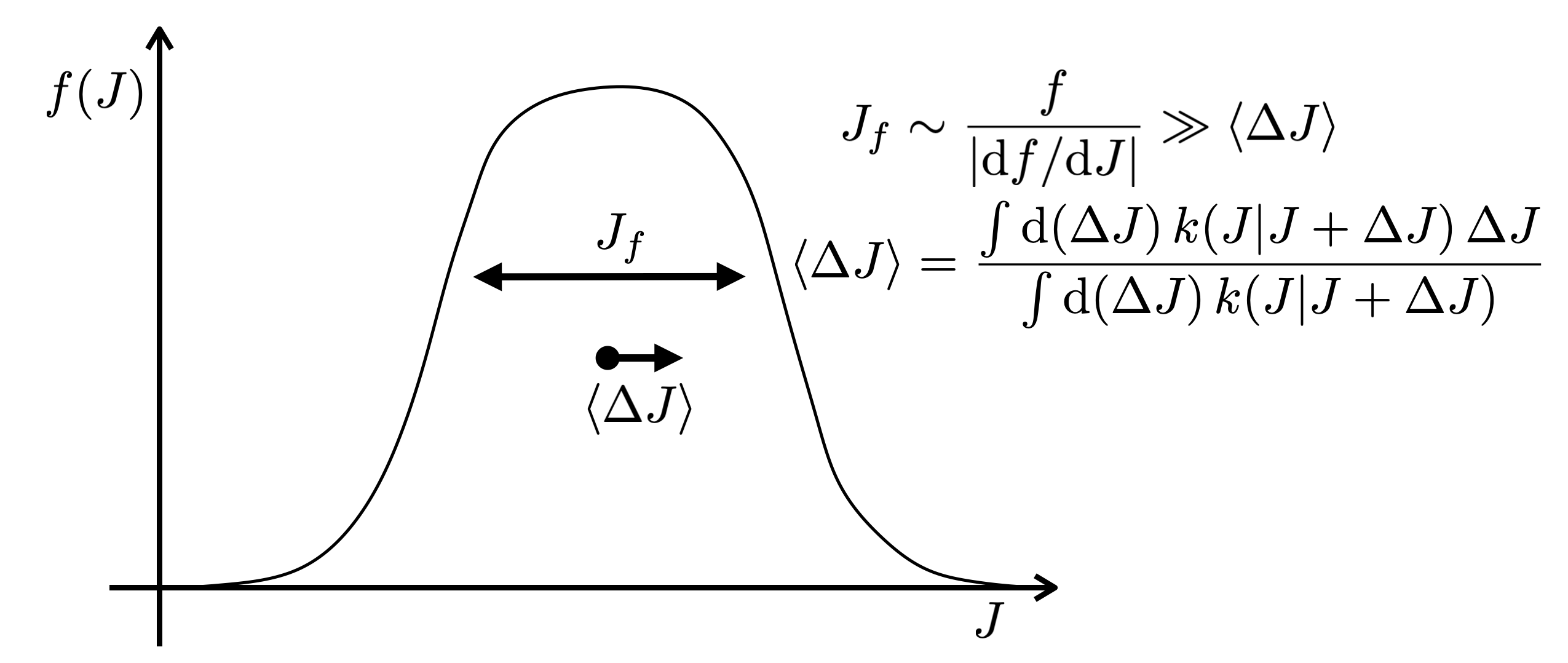}
\caption{\label{fig:fp.scale} One-particle distribution function, as a function of action, $f(J)$, for the small step size diffusive approximation of the Fokker-Planck equation.  The step size, $\left< \Delta J \right>$, is much smaller than the scale on which the distribution changes, $J_f$.}
\end{figure}

We start with the Generalized Master Equation, Eq.~\eqref{eqn:gen.master}, and make a moment expansion, in terms of the smallness of $\Delta J$ when compared to the scale on which the distribution function changes, as shown in Fig.~\ref{fig:fp.scale}.  This yields
\begin{equation}
\begin{split}
\label{eqn:fp.1}
    \frac{\partial \bar{f}_1(J)}{\partial \tau} &= \int{d(\Delta J) \; [ k((J+\Delta J|J) \, \bar{f}_1(J+\Delta J)} \\
    &- k((J|J+\Delta J) \, \bar{f}_1(J)] \\
    &= \int{d(\Delta J) \; [ k(J|J-\Delta J) \, \bar{f}_1(J)} \\
    &+ \Delta J \cdot \frac{\partial}{\partial J} ( k(J|J-\Delta J) \, \bar{f}_1(J)) \\
    &+ \frac{1}{2} \Delta J \, \Delta J : \frac{\partial^2}{\partial J \, \partial J} ( k(J|J-\Delta J) \, \bar{f}_1(J)) \\
    &- k(J|J-\Delta J) \, \bar{f}_1(J)].
\end{split}
\end{equation}
Define
\begin{equation}
\begin{split}
   \widetilde{A} &\equiv \int{d(\Delta J) \; k(J|J+\Delta J) \; \Delta J}, \\
   B &\equiv \frac{1}{2} \int{d(\Delta J) \; k(J|J+\Delta J) \; \Delta J \; \Delta J}, \\
   &\text{and} \\
   f(J) &\equiv \bar{f}_1(J),
\end{split}
\end{equation}
then substitute into Eq.~\eqref{eqn:fp.1} and integrate by parts to get
\begin{equation}
\begin{split}
    \frac{\partial f(J)}{\partial \tau} &= \frac{\partial}{\partial J} \cdot (f \, \widetilde{A}) + \frac{\partial^2}{\partial J \, \partial J} : (f \, B) \\
    &= \frac{\partial}{\partial J} \cdot \left[ f \, \widetilde{A} + \frac{\partial}{\partial J} (f \, B) \right] \\
    &= \frac{\partial}{\partial J} \cdot \left[ f \, \widetilde{A} + f \frac{\partial B}{\partial J} + B \cdot \frac{\partial f}{\partial J} \right] \\
    &= \frac{\partial}{\partial J} \cdot \left[ f \, A + B \cdot \frac{\partial f}{\partial J} \right] \\
    &= - \frac{\partial S}{\partial J},
\end{split}
\end{equation}
where
\begin{equation}
\begin{split}
    A &\equiv \widetilde{A} + f \frac{\partial B}{\partial J} \\
    &\text{and}\\
    S &\equiv -f \, A - B \cdot \frac{\partial f}{\partial J}.
\end{split}
\end{equation}
At equilibrium,
\begin{equation}
\begin{split}
    \frac{\partial f_\text{eq}}{\partial \tau} &= - \left. \frac{\partial S}{\partial J} \right|_\text{eq} = 0, \\
    S_\text{eq} &= 0, \\
    f_\text{eq} \, A + B \frac{\partial f_\text{eq}}{\partial J} &= 0, \\
    &\text{and} \\
    f_\text{eq} \, A &= - B \frac{\partial f_\text{eq}}{\partial J},
\end{split}
\end{equation}
so that
\begin{equation}
\begin{split}
    \frac{\partial f(J)}{\partial \tau} &= \frac{\partial}{\partial J} \cdot \left[ - \frac{f}{f_\text{eq}}B \cdot \frac{\partial f_\text{eq}}{\partial J} + B \cdot \frac{\partial f}{\partial J} \right] \\
    &= \frac{\partial}{\partial J} \cdot \left[ B \cdot \left( \frac{\partial f}{\partial J} - \frac{f}{f_\text{eq}} \frac{\partial f_\text{eq}}{\partial J} \right) \right].
\end{split}
\end{equation}
The leads to the \underline{Generalized Fokker-Planck Equation}
\begin{equation}
\label{eqn:gfp}
    \boxed{\frac{\partial \bar{f}_1(J)}{\partial \tau} = \frac{\partial}{\partial J} \cdot \left[ \left< \frac{\Delta J \, \Delta J}{\Delta \tau} \right> \cdot \left( \frac{\partial \bar{f}_1}{\partial J} - \frac{\bar{f}_1}{f_\text{eq}} \frac{\partial f_\text{eq}}{\partial J} \right) \right],}
\end{equation}
where
\begin{equation}
    \left< \frac{\Delta J \, \Delta J}{\Delta \tau} \right> \equiv B
\end{equation}
is the diffusion tensor.
%===============================%
%===============================%

This generalized equation can be reduced to the more familiar drift diffusion form by assuming that $f_\text{eq}= \text{e}^{-V(J)}$ so that
\begin{equation}
    \frac{1}{f_\text{eq}} \, \frac{\partial f_\text{eq}}{\partial J} = - \frac{d V}{d J} = \text{restoring force},
\end{equation}
yielding the familiar drift diffusion equation
\begin{equation}
    \frac{\partial f(J)}{\partial \tau} = \frac{\partial}{\partial J} \cdot \left[ B \cdot \left( \frac{\partial f}{\partial J} + \frac{d V}{d J} \, f \right) \right].
\end{equation}

\subsection{Vlasov-Fokker-Planck Equation}
\label{sec:VFP.equation}
Now consider correlated modes where $\omega_1 \ll \omega_2 = \nu_c = n_0 \bar{v} \, b^2$.  The fastest modes of the collisionless Vlasov equation are of order $\omega_p \thicksim \text{O}(\Gamma^{3/2}\bar{v}/b) \thicksim \text{O}(\Gamma^{-3/2} n_0 \bar{v} \, b^2)$, so that for these modes one can neglect the effects of collisions.  For low frequency modes, one can include the Fokker-Planck collision operator with the LHS of the Vlasov equation giving the \underline{Generalized VFP Equation} or Generalized Vlasov-Fokker-Planck Equation
\begin{equation}
\label{eqn:gvfp}
    \boxed{\frac{\partial f_1}{\partial \tau} + \left\{ \widetilde{H}_1 + \left< \widetilde{H}_{12} \right>, f_1 \right\} = \frac{\partial}{\partial J} \cdot \left[ B \cdot \left( \frac{\partial \bar{f}_1}{\partial J} + \frac{d V}{d J} \bar{f}_1 \right) \right],}
\end{equation}
where the averaging in $\bar{f}_1$ are taken over the collision frequency $n_0 \bar{v} \, b^2$.

Many times the gradients in the distribution function $f_1$ are too large for the VFP Equation to be used.  In that case one needs to drop back to the form of the collision operator given by the Master Equation
\begin{multline}
    \frac{\partial f_1}{\partial \tau} + \left\{ \widetilde{H}_1 + \left< \widetilde{H}_{12} \right>_2, f_1 \right\}_1 \\
    = \int{dJ' \; \left[ k(J'|J) \; \bar{f}_1(J') - k(J|J') \; \bar{f}_1(J) \right]},
\end{multline}
where the frequency of 
\begin{equation*}
    \left< \widetilde{H}_{12} \right>_2 \thicksim \text{O}(\omega_p) \thicksim \text{O}(\Gamma^{3/2}),
\end{equation*}
the frequency of the collision operator on the RHS is $\text{O}(\nu_c) \thicksim \text{O}(\Gamma^3)$, and $\omega_p=(4\pi n_0 e^2/m)^{1/2}$ is the plasma frequency.  Here instead of the operator being summarized by the diffusion in the canonical momentum (for instance, the action, $J$, or the velocity, $\mathbf{v}$) determined by the diffusion tensor, $B=\left< \Delta J \, \Delta J / \Delta \tau \right>$, it is characterized by the scattering kernel, $k(J|J')$.  One can also use the Boltzmann operator if $\bar{f}_1(J_2) \ne f_\text{th}(J_2)$, using $K(J_1,J_2|J'_1,J'_2)$.

\subsection{Fluid Equations}
\label{sec:fluid.equations}
To derive the Generalized Fluid Equations, start with the Generalized Vlasov Equation, Eq.~\eqref{eqn:vlasov}, for $f_1(P,Q,t)$
\begin{equation}
    \frac{\partial f_1}{\partial t} + \left\{ \widetilde{H}_1 + \left< \widetilde{H}_{12} \right>_2, f_1 \right\}_1 = 0,
\end{equation}
where $\widetilde{H}_1(P)$, so that $\dot{P}=-\partial \widetilde{H}_1/\partial Q=0$ and $\dot{Q}=\partial \widetilde{H}_1/\partial P=\omega_Q(P)=\text{constant}$.  Integrate the Vlasov equation, $\int{dP}$, using the definitions
\begin{equation*}
    n \equiv \int{dP \, f_1}
\end{equation*}
and
\begin{equation*}
    \left< \omega_Q \right>_P \equiv \bar{\omega}_Q \equiv \frac{1}{n} \int{dP \, \omega_Q(P) \, f_1},
\end{equation*}
so that
\begin{equation}
\begin{split}
    0 = \frac{\partial n}{\partial t} &+ \int{dP \left( \frac{\partial \widetilde{H}_1}{\partial P} \frac{\partial f_1}{\partial Q} - \frac{\partial \widetilde{H}_1}{\partial Q} \frac{\partial f_1}{\partial P}  \right)} \\
    &+ \int{dP \left\{ \left< \widetilde{H}_{12} \right>,f_1 \right\} } \\
    = \frac{\partial n}{\partial t} &+ \int{dP \, \omega_Q(P) \, \frac{\partial f_1}{\partial Q}} \\
    = \frac{\partial n}{\partial t} &+ \frac{\partial}{\partial Q} \left[ \int{dP \, \omega_Q(P) \, f_1(P,Q,t)} \right],
\end{split}
\end{equation}
where integration by parts has been used to get the final expression.  This yields the \underline{Generalized Continuity Equation}
\begin{equation}
\label{eqn:gen.continuity}
    \boxed{\frac{\partial n}{\partial t} + \frac{\partial}{\partial Q} \left(n \, \bar{\omega}_Q \right)=0.}
\end{equation}
Operate on the Vlasov equation with $\int{\omega_Q \, dP}$, using the definition
\begin{equation*}
    \left< g \right> \equiv \bar{g} \equiv \frac{1}{n} \int{dP \, g \, f_1},
\end{equation*}
so that
\begin{equation}
\begin{split}
    0 = \frac{\partial}{\partial t} \left( n \, \bar{\omega}_Q \right) &+ \int{dP \, \omega_Q(P) \, \omega_Q(P) \frac{\partial f_1}{\partial Q}} \\
    &+ \int{dP \, \omega_Q \, \left\{ \left< \widetilde{H}_{12} \right>,f_1 \right\}} \\
    = \frac{\partial}{\partial t} \left( n \, \left< \omega_Q \right>_P \right) &+ \frac{\partial}{\partial Q} \left( n \, \left< \omega_Q  \, \omega_Q \right>_P \right) \\
    &+ \int{dP \, \omega_Q \, \left\{ \left< \widetilde{H}_{12} \right>_2,f_1 \right\}},
\end{split}
\end{equation}
where integration by parts has been used to get the final expression.  This yields the \underline{Generalized Momentum (Force) Equation}
\begin{equation}
\label{eqn:gen.force}
    \boxed{\frac{\partial}{\partial t} \left( n \, \bar{\omega}_Q \right) + \frac{\partial}{\partial Q} \left( n \, \overline{\omega_Q \, \omega_Q} \right) = - \int{dP \, \omega_Q \left\{ \left< \widetilde{H}_{12} \right>,f_1 \right\}}}
\end{equation}
Note that the Fluid Equations are no longer conservative, unlike the Vlasov Equation.

To reduce the generalized equations to the traditional expressions, assume
\begin{equation*}
    \widetilde{H}_1 = \frac{P_1^2}{2} \text{  and  } \widetilde{H}_{12} = V(Q_1 - Q_2),
\end{equation*}
so that
\begin{equation*}
\begin{split}
   \omega_Q = P_1 &= \mathbf{v} \\
    \frac{\partial}{\partial Q} &= \nabla \cdot \\
    &\text{and} \\
    - \int{dP \, \omega_Q \left\{ \left< \widetilde{H}_{12} \right>,f_1 \right\}} &= -\frac{\partial V}{\partial Q} n = n \, \left< \mathbf{a}_{12} \right>.
\end{split}
\end{equation*}
This leads to the continuity equation
\begin{equation}
\label{eqn:free.fluid.1}
    \frac{\partial n}{\partial t} + \nabla \cdot \left( n \, \mathbf{v} \right) = 0
\end{equation}
and the momentum equation
\begin{equation}
\label{eqn:free.fluid.2}
    \frac{\partial \left( n \, \mathbf{v} \right)}{\partial t} + \nabla \cdot \left( n \, \overline{\mathbf{v} \, \mathbf{v}} \right) = n \, \left< \mathbf{a}_{12} \right>
\end{equation}
or for a magnetized plasma with isotropic pressure
\begin{equation}
    \frac{\partial \left( n \, \mathbf{v} \right)}{\partial t} + \frac{1}{m} \nabla p = \frac{q}{m} \, n \, \left(\mathbf{E} + \frac{\mathbf{v}}{c} \times \mathbf{B} \right),
\end{equation}
where $\overline{\mathbf{v} \, \mathbf{v}} = \bar{v}^2 \, \overleftrightarrow{I}$, $T \equiv m \, \bar{v}^2$, and $p \equiv \text{pressure} = m \, n \, \bar{v}^2$.  The momentum equation can be extended to account for a drifting Maxwellian
\begin{equation}
    m \, n \frac{\partial \mathbf{v}}{\partial t} + m \, n \left( \mathbf{v} \cdot \nabla \right) \mathbf{v}  = - \nabla p + q \, n \left(\mathbf{E} + \frac{\mathbf{v}}{c} \times \mathbf{B} \right),
\end{equation}
where $\overline{\mathbf{v} \, \mathbf{v}} = \mathbf{v} \, \mathbf{v} + \bar{v}^2 \, \overleftrightarrow{I}$.  To complete the closure an expression, that is Equation of State (EoS), for the pressure, $p$, needs to be found.  There are two common limiting cases.  If the evolution of the plasma is slower than the temperature equilibration, then the evolution is isothermal, so that
\begin{equation}
    \nabla p = \nabla \left( n \, T \right) =T \, \nabla n.
\end{equation}
If the evolution of the plasma is faster than the temperature equilibration, then the evolution is adiabatic, so that
\begin{equation}
    \nabla p = \nabla \left( n \, T \right) = n \nabla T + T \, \nabla n = \gamma \, T \, \nabla n,
\end{equation}
where $\gamma= (2+D)/D$ is the ratio of the specific heats, and $D$ is the number of dimensions which share a temperature.  Usually $D=3$, so that $\gamma=5/3$.  The system of equations are completed by Maxwell's equations, where the charge density is $\rho_c \equiv \sum{q_s \, n_s}$, the total current density is $\mathbf{J} \equiv \sum{q_s \, n_s \, \mathbf{v}_s}$, and the sum is over the species in the plasma $s$.

\subsection{MHD Equations}
\label{sec:MHD.equations}
What we wish to do is find a form of the ``fluid'' equations which would be valid if $\omega_Q \ll n_0 \bar{v} b^2$, that is for low frequency.  In this case we need to start with the \underline{Collisional Vlasov Equation} based on the VFP Equation~\eqref{eqn:gvfp},
\begin{equation}
\label{eqn:coll.vlasov}
    \frac{\partial f_1}{\partial t} + \left\{ \widetilde{H}_1 + \left< \widetilde{H}_{12} \right>, f_1 \right\} = \mathcal{C}[\bar{f}_1],
\end{equation}
where $\mathcal{C}[\bar{f}_1]$ can be a Boltzmann, Master, or Fokker-Planck collision operator, as shown in Eqns.~\eqref{eqn:gen.boltzmann}, \eqref{eqn:gen.master}, and \eqref{eqn:gvfp}.  We will assume the most general form, the Boltzmann form, for the collision operator.

We start by integrating the Collisional Vlasov equation over $P$, $\int{dP}$, to get the continuity equation as before,
\begin{equation}
    \frac{\partial n}{\partial t} + \frac{\partial}{\partial Q} \left(n \, \bar{\omega}_Q \right)=0,
\end{equation}
since $\int{dP \, \mathcal{C}[\bar{f}_1]}=0$.  Next operate on the Collisional Vlasov Equation with $\int{\omega_Q \, dP}$ to get,
\begin{equation}
\begin{split}
    \frac{\partial}{\partial t} \left( n \, \bar{\omega}_Q \right) = - \frac{\partial}{\partial Q} \left( n \, \overline{\omega_Q \, \omega_Q} \right) &- \int{dP \, \omega_Q \left\{ \left< \widetilde{H}_{12} \right>,f_1 \right\}} \\
    &+ \int{dP \, \omega_Q \, \mathcal{C}[\bar{f}_1]},
\end{split}
\end{equation}
the momentum equation, where the first term on the RHS is the thermal pressure force, the second term on the RHS is the two-particle interaction force, and the third term on the RHS is the collisional force.  The pressure tensor can be defined as $p=n \, \overline{\omega_Q \, \omega_Q}$.

We now assume that we have a two-fluid model, where the fluids are of the species $s=\{+,-\} = \{i,e\}$.  Now proceed to the one-fluid model, also known as MagnetoHydroDynamics (MHD).  Define,
\begin{equation*}
\begin{split}
    \text{mass density} &\equiv \rho_M \equiv m_+ n_+ + m^- n^- \approx m_+ n_+ \\
    \text{charge density} &\equiv \rho_c \equiv q_+ n_+ + q_- n_- \\
    \text{center of mass velocity} &\equiv \Omega_M \equiv \frac{1}{\rho_M} \left( m_+ n_+ \omega_+ + m_- n_- \omega_- \right) \\
    &\text{and} \\
    \text{current density} &\equiv J_c \equiv \left( q_+ n_+ \omega_+ + q_- n_- \omega_- \right).
\end{split}
\end{equation*}
Multiply the ion continuity equation by $m_+$, and multiply the electron continuity equation by $m_-$, then add together to get,
\begin{equation}
\label{eqn:gen.mass}
    \boxed{\frac{\partial \rho_M}{\partial \tau} + \frac{\partial}{\partial Q} \left( \rho_M \, \Omega_M \right) = 0,}
\end{equation}
the \underline{Generalized Mass Conservation Equation}.  Next multiply the ion continuity equation by $q_+$, and multiply the electron continuity equation by $q_-$, then add together to get
\begin{equation}
\label{eqn:gen.charge}
    \boxed{\frac{\partial \rho_c}{\partial \tau} + \frac{\partial}{\partial Q} \left( J_c \right) = 0,}
\end{equation}
the \underline{Generalized Charge Conservation Equation}.  Now, multiply the ion momentum equation by $m_+$, and multiply the electron continuity equation by $m_-$, then add together to get,
\begin{equation}
\label{eqn:gen.mhd.force}
    \boxed{\frac{\partial}{\partial \tau} \left( \rho_M \Omega_M \right) + \frac{\partial}{\partial Q} \left( p_\text{t} \right) = F_\text{ti},}
\end{equation}
the \underline{Generalized Force Equation}, where
\begin{equation*}
\begin{split}
    p_\text{t} &\equiv m_+ n_+ \overline{\omega_+ \omega_+} + m_- n_- \overline{\omega_- \omega_-} \\
    &= \text{total pressure tensor} = p_+ + p_-, \\
    F_\text{tp} &\equiv - \frac{\partial}{\partial Q} \left( p_\text{t} \right) = \text{total thermal pressure force}, \\
   F_\text{ti} &\equiv - \int{dP \, m_+ \omega_+ \left\{ \left< H_{12} \right>,f_{1+} \right\} + m_- \omega_- \left\{ \left< H_{12} \right>,f_{1-} \right\}} \\
    &= \text{total two-particle interaction force}, \\
    &\text{and} \\
     K_+ &\equiv \int{dP \, m_+ \omega_+ \mathcal{C}[f_{1+}] } = - \int{dP \, m_- \omega_- \mathcal{C}[f_{1-}] } \\
    &\equiv - K_- .
\end{split}
\end{equation*}
Finally, multiply the ion momentum equation by $q_+$, and the electron momentum equation by $q_-$, then add together to get,
\begin{equation}
\label{eqn:gen.mhd.ohms}
    \boxed{\frac{\partial J_c}{\partial \tau} + \frac{\partial}{\partial Q} \left( p_\text{qt} \right) = F_\text{qti} + F_\text{qtc},}
\end{equation}
the \underline{Generalized Ohm's Law}, where
\begin{equation*}
\begin{split}
    p_\text{qt} &\equiv q_+ n_+ \overline{\omega_+ \omega_+} + q_- n_- \overline{\omega_- \omega_-} \\
    &= \frac{q_+}{m_+} p_+ + \frac{q_-}{m_-} p_- = \text{total charge pressure tensor}, \\
    F_\text{qtp} &\equiv - \frac{\partial}{\partial Q} \left( p_\text{qt} \right) \\
    &= \text{total charge thermal pressure force}, \\
    F_\text{qti} &\equiv - \int{dP \, q_+ \omega_+ \left\{ \left< H_{12} \right>,f_{1+} \right\} + q_- \omega_- \left\{ \left< H_{12} \right>,f_{1-} \right\}} \\
    &= \text{total two-particle charge interaction force}, \\
    &\text{and} \\
    F_\text{qtc} &\equiv \int{dP \, q_+ \omega_+ \mathcal{C}[f_{1+}] + q_- \omega_- \mathcal{C}[f_{1-}] } \\
    &= \frac{q_+ m_- - q_- m_+}{m_+ m_-} K_+ \\
    &= \text{total charge collisional force}.
\end{split}
\end{equation*}
This makes four equations and four unknowns, $\rho_M$, $\rho_c$, $\Omega_M$, and $J_c$.

As was done for the fluid equations, let's reduce the generalized expressions to the traditional expressions.  remember that
\begin{equation*}
\begin{split}
    \frac{\partial}{\partial Q} &= \nabla \cdot, \\
    \Omega_M &= \mathbf{V}, \\
    J_c &= \mathbf{J}, \\
    \tau &= t, \\
    m_+ &= m_i, \, m_- = m_e, \\
    &\text{and} \\
    q_+ &= e, \, q_- = -e,
\end{split}
\end{equation*}
so that the mass conservation equation is
\begin{equation}
\label{eqn:free.MHD.1}
    \frac{\partial \rho_M}{\partial t} + \nabla \cdot \left( \rho_M \mathbf{V} \right) = 0,
\end{equation}
the charge conservation equation is
\begin{equation}
\label{eqn:free.MHD.2}
    \frac{\partial \rho_c}{\partial t} + \nabla \cdot \mathbf{J} = 0,
\end{equation}
the force equation is
\begin{equation}
\label{eqn:free.MHD.3}
    \rho_M \frac{\partial \mathbf{V}}{\partial t} = - \nabla p + \rho_c \mathbf{E} + \frac{1}{c} \, \mathbf{J} \times \mathbf{B},
\end{equation}
and Ohm's Law is
\begin{multline}
\label{eqn:free.MHD.4}
    \frac{m_e m_i}{\rho_M e^2} \, \frac{\partial \mathbf{J}}{\partial t} = \frac{m_i}{2 \rho_M e} \nabla p + \mathbf{E} + \frac{1}{c} \mathbf{V} \times \mathbf{B} \\
    - \frac{m_i}{\rho_M ec} \mathbf{J} \times \mathbf{B} - \frac{\mathbf{J}}{\sigma},
\end{multline}
where
\begin{equation*}
\begin{split}
    p_\text{t} &= p = p_i + p_e, \\
    p_\text{qt} &= \frac{e}{2m} p, \\
    F_\text{ti} &= \rho_c \mathbf{E} + \frac{1}{c} \mathbf{J} \times \mathbf{B}, \\
    F_\text{qti} &= \frac{\rho_M e^2}{m_e m_i} \left( \mathbf{E} + \frac{1}{c} \mathbf{V} \times \mathbf{B} \right) - \frac{e}{m_e} \mathbf{J} \times \mathbf{B}, \\
    &\text{and} \\
    F_\text{qtc} &= - \frac{\rho_M e^2}{m_e m_i \sigma} \mathbf{J}.
\end{split}
\end{equation*}
This makes four equations and six unknowns, $\rho_M$, $\rho_c$, $\mathbf{V}$, $\mathbf{J}$, $\mathbf{E}$, and $\mathbf{B}$.  The system of equations is completed by Maxwell's equations,
\begin{equation*}
\begin{split}
    \nabla \times \mathbf{E} &= - \frac{1}{c} \, \frac{\partial \mathbf{B}}{\partial t}, \\
    \nabla \times \mathbf{B} &= \frac{4 \pi}{c} \, \mathbf{J} + \frac{1}{c} \, \frac{\partial \mathbf{E}}{\partial t}.
\end{split}
\end{equation*}

For very low frequencies the plasma becomes quasi-neutral, so that $\rho_c=0$, leading to the \underline{Ideal Resistive MHD Equations},
\begin{equation}
\begin{split}
    \frac{\partial \rho_M}{\partial t} + \nabla \cdot \left( \rho_M \mathbf{V} \right) &= 0, \\
    \nabla \cdot \mathbf{J} &= 0, \\
    \rho_M \frac{\partial \mathbf{V}}{\partial t} &=  \frac{\mathbf{J} \times \mathbf{B}}{c} - \nabla p, \\
    \mathbf{E} + \frac{\mathbf{V} \times \mathbf{B}}{c} &= \frac{\mathbf{J}}{\sigma}.
\end{split}
\end{equation}

This coordinate-free approach has been used in the implementation of the transport of both magnetic fields and thermal energy in Finite Element (FE) Arbitrary Lagrangian Eulerian (ALE) MHD computer simulations.\citep{castillo.04,koning.20}  It allows for conservative FE implementations, such as divergence-free magnetic fields, on the ALE mesh.

\section{Specific examples}
\label{sec:examples}
The first specific example will be that of a free particle with an electrostatic interaction.  The one-particle Hamiltonian is
\begin{equation*}
    H_1 = \frac{\mathbf{v}_1^2}{2},
\end{equation*}
and the two-particle interaction Hamiltonian is
\begin{equation*}
    H_{12} = \frac{1}{\left| \mathbf{x}_2 - \mathbf{x}_2 \right|}.
\end{equation*}
For this example, the expression for the Vlasov Equation was given in Sec.~\ref{sec:Vlasov.equation} as Eq.~\eqref{eqn:free.vlasov}, the expression for the Fluid Equations was given in Sec.~\ref{sec:fluid.equations} as Eqns.~\eqref{eqn:free.fluid.1} and \eqref{eqn:free.fluid.2}, and the expression for the MHD equations was given in Sec.~\ref{sec:MHD.equations} as Eqns.~\eqref{eqn:free.MHD.1}, \eqref{eqn:free.MHD.2}, \eqref{eqn:free.MHD.3} and \eqref{eqn:free.MHD.4}.  To obtain the expression for the Boltzmann Equation, remember that
\begin{equation*}
\begin{split}
    \bar{F}_1(J_1) &\to f(\mathbf{v}_1,t), \\
    dJ'_1 \, dJ'_2&\to d^3 \mathbf{v}'_1 \, d^3 \mathbf{v}'_2, \\
    &\text{and} \\
    dJ_2 &\to d^3 \mathbf{v}_2,
\end{split}
\end{equation*}
so that the Boltzmann Equation for this case can be written as
\begin{multline}
    \frac{\partial f(\mathbf{v}_1, t)}{\partial t} \\
    = \int{\left| \mathbf{v}_2 - \mathbf{v}_2 \right| \, \left[ f(\mathbf{v}'_1) \, f(\mathbf{v}'_2) - f(\mathbf{v}_1) \, f(\mathbf{v}_2) \right] \, d \sigma \, d^3 \mathbf{v}_2}.
\end{multline}
Now for the Master Equation, identify
\begin{equation*}
\begin{split}
    J &\to \mathbf{v}, \\
    \bar{f}_1(J) &\to f(\mathbf{v},t), \\
    d J' &\to d^3 \mathbf{v}', \\
    d J_2 \, d J'_2 &\to d^3 \mathbf{v}_2 \, d^3 \mathbf{v}'_2, \\
    f_\text{th}(J_2) &= f_\text{th}(\mathbf{v}_2) = \text{e}^{- v^2_2 / 2}, \\
    &\text{and} \\
    K(\mathbf{v}_1,\mathbf{v}_2 | \mathbf{v}'_1,\mathbf{v}'_2) &= \left| \mathbf{v}_2 - \mathbf{v}_1 \right| \, d \sigma,
\end{split}
\end{equation*}
so that the Master Equation can be written as
\begin{equation}
    \frac{\partial f(\mathbf{v}, t)}{\partial t} = \int{d^3 \mathbf{v}' \, \left[ k(\mathbf{v}' | \mathbf{v}) \, f(\mathbf{v}') - k(\mathbf{v} | \mathbf{v}') \, f(\mathbf{v}) \right]},
\end{equation}
where
\begin{multline}
    k(\mathbf{v} | \mathbf{v}') = \int{\left| \mathbf{v}_2 - \mathbf{v}_1 \right| \, f_\text{th}(\mathbf{v}_2) } \\
    \delta(\mathbf{v}'-\mathbf{v}_f(\mathbf{v},\sigma_s,\psi_1)) \, \frac{d \psi_1}{2\pi} \, d \sigma_s \, d^3 \mathbf{v}_2.
\end{multline}
Finally for the Fokker-Planck Equation, identify
\begin{equation*}
\begin{split}
    f_\text{eq}(J) &= f_\text{eq}(\mathbf{v}) = \text{e}^{- v^2 / 2}, \\
    &\text{and} \\
    \frac{1}{f_\text{eq}} \, \frac{\partial f_\text{eq}}{\partial \mathbf{v}} &= - \mathbf{v},
\end{split}
\end{equation*}
so that the Fokker-Planck Equation is
\begin{equation}
    \frac{\partial f(\mathbf{v}, t)}{\partial t} = \frac{\partial}{\partial \mathbf{v}} \cdot \left[ \left< \frac{\Delta \mathbf{v} \, \Delta \mathbf{v}}{\Delta t} \right> \cdot \left( \frac{\partial f}{\partial \mathbf{v}} + \mathbf{v} \, f \right) \right],
\end{equation}
where
\begin{equation}
    \left< \frac{\Delta \mathbf{v} \, \Delta \mathbf{v}}{\Delta t} \right> = \frac{1}{2} \int{d(\Delta \mathbf{v}) \, k(\mathbf{v}|\mathbf{v}+\Delta \mathbf{v}) \, \Delta \mathbf{v} \, \Delta \mathbf{v}}.
\end{equation}

As will be discussed in detail in Sec.~\ref{sec:app_vtrr}, the Hamiltonian for a guiding center electron and ion in a strong magnetic field, can be written as $H(J_z,J_\text{D})$, where $J_z$ is the $z$-bounce action and $J_\text{D}$ is the $E \times B$ drift action perpendicular to the strong magnetic field.  Identify
\begin{equation*}
\begin{split}
    \widetilde{H}_1 &= H(J_z,J_\text{D}), \\
    \widetilde{H}_{12} &= \frac{1}{\mathbf{x}_2 - \mathbf{x}_1}, \\
    \tau &\to t, \\
    \bar{f}_1(J_1) &\to f(J_z,J_\text{D},t), \\
    d J_2 &\to d p_{2z}, \\
    &\text{and} \\
    d J'_1 \, d J'_2 &\to d J'_z \, d J'_\text{D} \, d p'_{2z},
\end{split}
\end{equation*}
so that
\begin{equation*}
    \left| v_{2z} \right| \, d \sigma = K(J_z,J_\text{D},p_{2z}|J'_z,J'_\text{D},p'_{2z}) \, d J'_z \, d J'_\text{D} \, d p'_{2z},
\end{equation*}
yielding the Boltzmann Equation,
\begin{multline}
    \frac{\partial f(J_z,J_\text{D},t)}{\partial t} = \int{ \left| v_{2z} \right| \, \left[ f(J'_z,J'_\text{D}) \, f(p'_{2z}) \right.} \\
    \left. - f(J_z,J_\text{D}) \, f(p_{2z}) \right] d \sigma \, dp_{2z},
\end{multline}
and the Master Equation,
\begin{multline}
\label{eqn:gca.master}
    \frac{\partial f(J_z,J_\text{D},t)}{\partial t} = \int{dJ'_z \, dJ'_\text{D} \, \left[ k(J'_z,J'_\text{D}|J_z,J_\text{D}) \ f(J'_z,J'_\text{D}) \right.} \\
    \left. - k(J_z,J_\text{D}|J'_z,J'_\text{D}) \ f(J_z,J_\text{D}) \right],
\end{multline}
where
\begin{multline*}
    k(J_z,J_\text{D}|J'_z,J'_\text{D}) = \int{\left| v_{2z} \right| \, f_\text{th}(p_{2z}) \,} \\
    \delta(J'_z-J_{z\text{f}}(J_z,J_\text{D},p_{2z},\psi_z,\psi_\text{D},\sigma_s)) \\
    \delta(J'_\text{D}-J_\text{Df}(J_z,J_\text{D},p_{2z},\psi_z,\psi_\text{D},\sigma_s)) \, \\
    \frac{d \psi_z}{2\pi} \frac{d \psi_\text{D}}{2\pi} d \sigma_s \, dp_{2z},
\end{multline*}
and
\begin{equation*}
    f_\text{th}(p_{2z}) = \text{e}^{-p_{2z}^2 / 2}.
\end{equation*}
Furthermore, defining
\begin{equation*}
\begin{split}
    \mathbf{J} &= (J_z,J_\text{D}), \\
    \vec{\omega} &= (\omega_z,\omega_\text{D}), \\
    f_\text{eq} &= \text{e}^{-H(\mathbf{J})}, \\
    &\text{and} \\
    \frac{1}{f_\text{eq}} \frac{\partial f_\text{eq}}{\partial \mathbf{J}} &= - \frac{\partial H(\mathbf{J})}{\partial \mathbf{J}} = - \vec{\omega},
\end{split}
\end{equation*}
yields the Fokker-Planck Equation
\begin{equation}
    \frac{\partial f(\mathbf{J}, t)}{\partial t} = \frac{\partial}{\partial \mathbf{J}} \cdot \left[ \left< \frac{\Delta \mathbf{J} \, \Delta \mathbf{J}}{\Delta t} \right> \cdot \left( \frac{\partial f}{\partial \mathbf{J}} + \vec{\omega} \, f \right) \right],
\end{equation}
where
\begin{equation}
    \left< \frac{\Delta \mathbf{J} \, \Delta \mathbf{J}}{\Delta t} \right> = \frac{1}{2} \int{d(\Delta \mathbf{J}) \, k(\mathbf{J}|\mathbf{J}+\Delta \mathbf{J}) \, \Delta \mathbf{J} \, \Delta \mathbf{J}}.
\end{equation}
\citet{glinsky1991guiding} evaluated the evolution of $\bar{f}_1(J_z,J_\text{D})$ according to the Master Equation using a Monte Carlo technique and estimated the three-body recombination rate for GCAs to be $0.070(10) \, n_e^2 \bar{v}_e b^5$.  They also derived an expression for the three-body recombination for GCAs based on the Fokker-Planck equation.  Unfortunately, the diffusion approximation leading to the Fokker-Planck Equation is violated because both $\Delta J_z$ and $\Delta J_\text{D}$ are large.  The diffusion coefficient is very sensitive to the impact parameter for the small impact parameters that contribute the most to the kinetics, and the diffusion coefficient goes to infinity logarithmically as the impact parameter goes to infinity.  They also estimated the three-body recombination rate for GCAs using the VTRR.  All this work was done using vector calculus in Cartesian coordinates.  Based on the coordinate-free approach of this paper, the Generalized VTRR given by Eq.~\eqref{eqn:vtrr.pullback} is used, in Sec.~\ref{sec:app_vtrr}, to estimate the three-body recombination rate for GCAs, including the $E \times B$ drift flux, and for the general case with an ion of finite mass and velocity.

\section{Application of Variational Theory of Reaction Rates}
\label{sec:app_vtrr}
\begin{figure}
\noindent\includegraphics[width=\columnwidth]{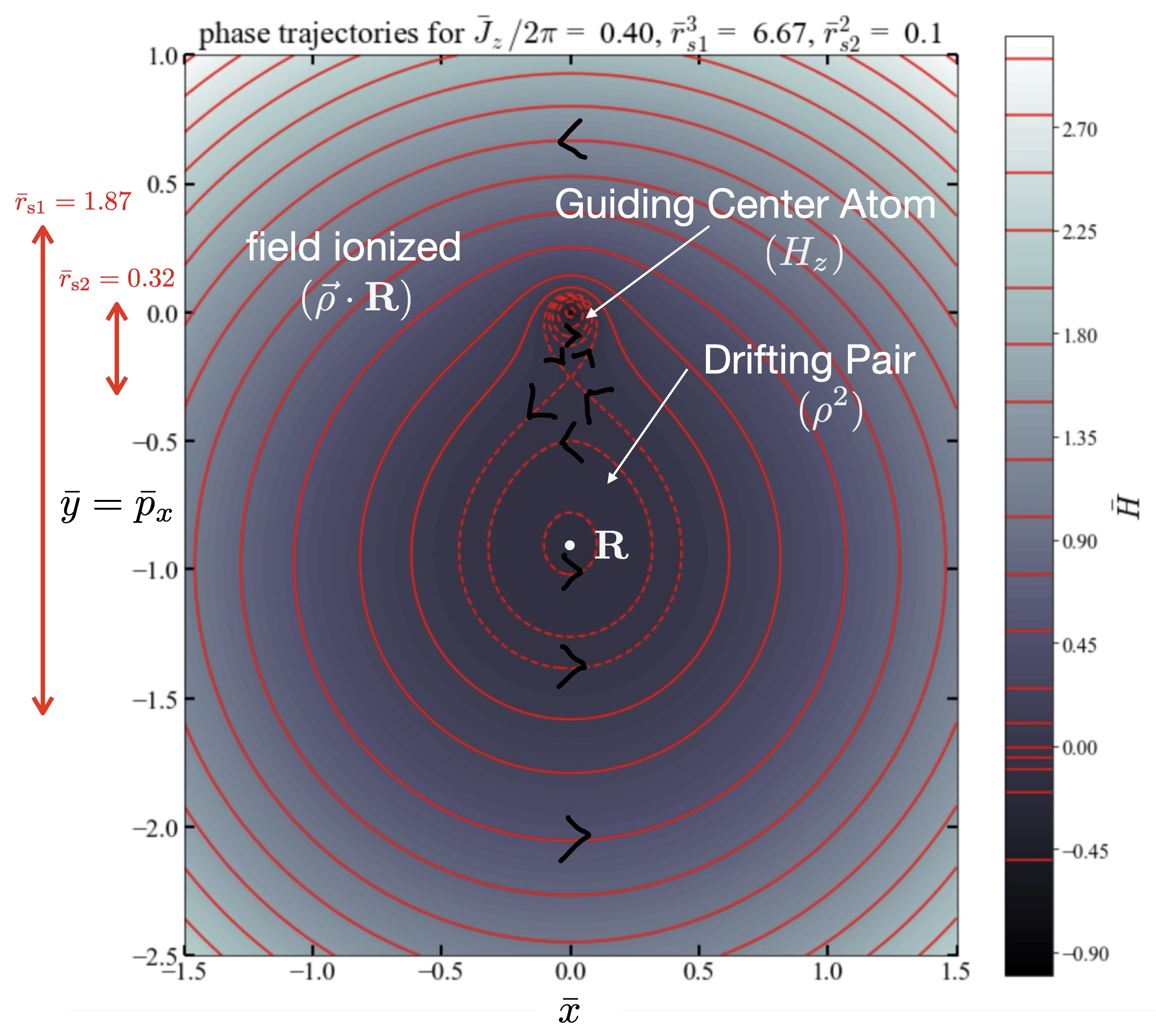}
\caption{\label{fig:full.phase.space} Phase space trajectories for motion of an ion and a guiding center electron in a strong magnetic field.  Shown is the $E \times B$ drift motion in the plane perpendicular to the magnetic field with coordinates given by $x$ and $y \sim p_x$.  The trajectories are shown by the solid red lines for $H_1>0$ and by the dotted red lines for $H_1<0$, for $J_z/2\pi = 0.40$, the size of the first DP separatrix $\bar{r}^3_\text{s1}=6.67$, and the size of the second GCA separatrix $\bar{r}^2_\text{s2}=0.1$.  The values of $a=0.15$, $\mathbf{R}=- \hat{y}$, $1/a=6.67$, and $R/a=10$.}
\end{figure}

We consider the three-body recombination of an electron and an ion in a magnetic field strong enough that the electron undergoes guiding center dynamics.  Let us start by considering the ``one-particle'' Hamiltonian of the electron and the ion.  The Hamiltonian can be reduced to two degrees of freedom ($z$ and $E \times B$ dynamics),
\begin{multline}
    H_1(R;\psi,p_\psi,z,p_z) = \\
    \Omega_i \left[ m_e \Omega_e \frac{R^2}{2} - R \sqrt{2 m_e \Omega_e p_\psi} \cos{\psi} + p_\psi \right] \\
    + \frac{p_z^2}{2 \mu} - \frac{e^2}{\sqrt{\frac{2 p_\psi}{m_e \Omega_e} + z^2}},
\end{multline}
where $m_e$ is the electron mass, $m_i$ is the ion mass, $\Omega_e=e B / m_e c$ is the electron cyclotron frequency, $\mu=m_e m_i / (m_e + m_i)$ is the reduced electron mass, and $R$ is the distance between the electron and ion guiding centers.  The motion can be further ``averaged'' over $z$ to give the Hamiltonian with one degree of freedom,
\begin{multline}
    H_1(R,J_z;\bar{\psi},\bar{p}_\psi) = \\
    \Omega_i \left[ m_e \Omega_e \frac{R^2}{2} - R \sqrt{2 m_e \Omega_e \bar{p}_\psi} \cos{\bar{\psi}} + \bar{p}_\psi \right] \\
    + H_z(J_z,\bar{p}_\psi),
\end{multline}
where $J_z$ is the $z$-bounce action.  Defining the radius as $\rho^2 \equiv 2 \bar{p}_\psi / (m_e \Omega_e)$ with angle $\theta=\bar{\psi}$, then normalizing $\vec{\rho}$ and $\mathbf{R}$ by $b$, $H$ by $k_\text{B} T_e$, and $J$ by $m_e \bar{v}_e b$, so that
\begin{equation*}
\begin{split}
    \vec{\rho} &\equiv \rho \, (\cos{\theta},\sin{\theta},0), \\
    \frac{1}{a} &\equiv \left( \frac{b}{r_\text{ce}} \right)^2 \, \frac{m_e}{m_i} = \frac{\Omega_i}{\omega_{E \times B}} = \bar{r}_\text{s1}^{-3}, \\
    &\text{and} \\
    \frac{R}{a} &\equiv \left( \frac{b}{r_\text{ce}} \right) \, \frac{\bar{v}_{i \perp}}{\bar{v}_e} = \bar{r}_\text{s2}^{-2} = \varepsilon^2,
\end{split}
\end{equation*}
gives the dimensionless Hamiltonian,
\begin{equation}
    H_1(J_z,\mathbf{R},a;\vec{\rho}) = \frac{1}{2} \, \left( \vec{\rho} - \mathbf{R} \right)^2 + a \, H_z(J_z,\rho)
\end{equation}
or
\begin{multline}
\label{eqn:ei.full.hamiltonian}
    H_1\left( J_z,\frac{R}{a},\frac{1}{a};\rho,\theta \right) \\
    = \left( \frac{1}{a} \right) \, \frac{\rho^2}{2} + \left( \frac{R}{a} \right) \, \rho \, \sin{\theta} + H_z(J_z,\rho).
\end{multline}
If $1/a \ll 1$ so that $m_i \to \infty$ and $R/a \ll 1$ so that $\bar{v}_{i \perp} \to 0$, then
\begin{equation}
\label{eqn:GCA.Hamiltonian}
    H_1 = H_z(J_z,\rho),
\end{equation}
where $\rho$ is cyclic so that $\rho=\text{constant}$ along with $J_z$ and the trajectories are circles.  If neither of these conditions are met, then one gets trajectories in phase space as shown in Figs.~\ref{fig:full.phase.space} and \ref{fig:gca.dp.plot}.  The circular GCA trajectories, governed by Eq.~\eqref{eqn:GCA.Hamiltonian}, are within the separatrix of size $\bar{r}_\text{s2}= (R/a)^{-1/2}$, about the origin.  Electrons outside this separatrix are not bound and are carried to $\rho$ values of order the ion cyclotron radius.
%===============================%
%===============================%

\begin{figure}
\noindent\includegraphics[width=\columnwidth]{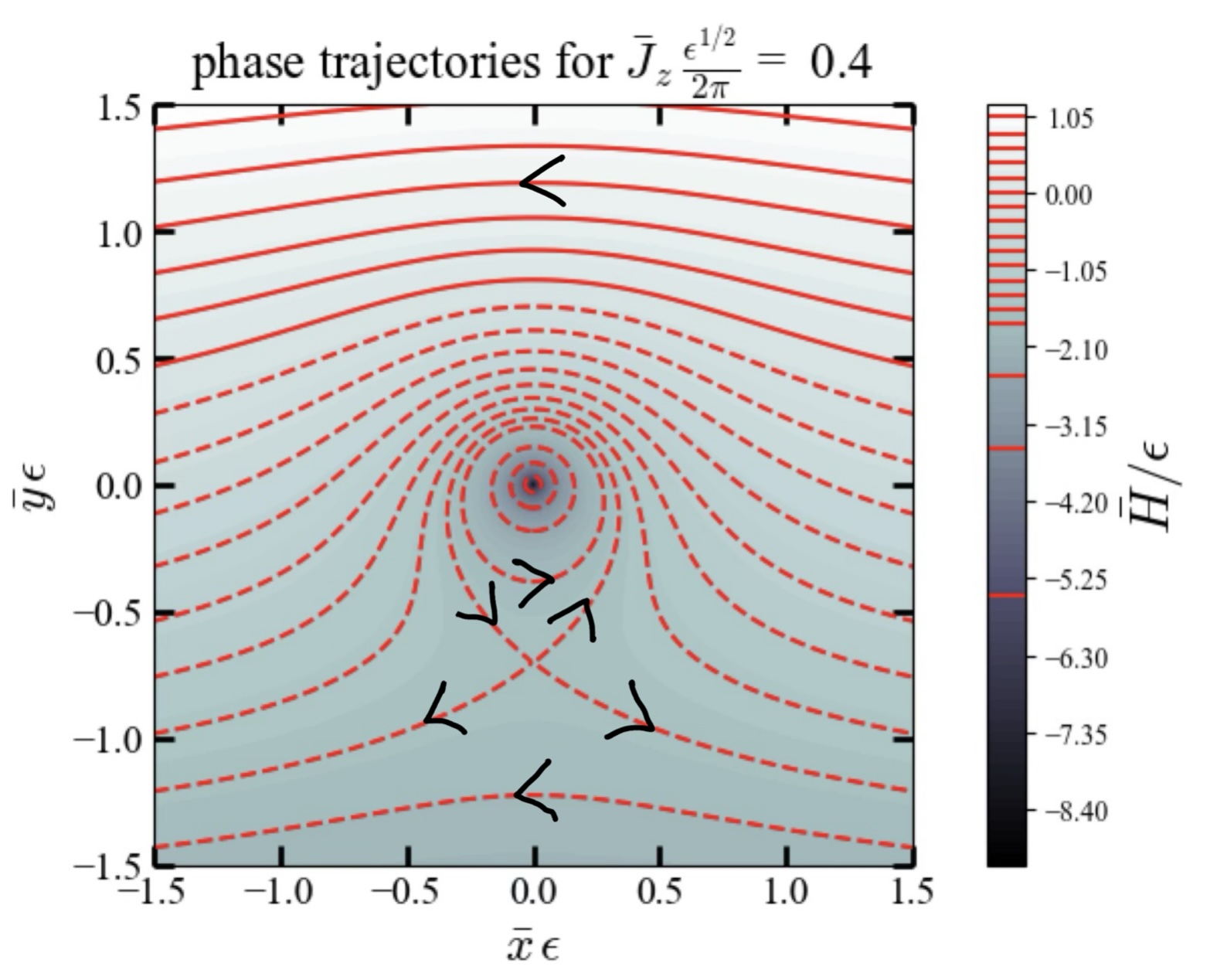}
\caption{\label{fig:gca.phase.space} Phase space trajectories for a GCA in an electric field.  Shown is the $E \times B$ drift motion in the plane perpendicular to the magnetic field with coordinates given by $x$ and $y \sim p_x$.  The trajectories are shown by the solid red lines for $H_1>0$ and by the dotted red lines for $H_1<0$, for $J_z/2\pi = 0.40$.}
\end{figure}

\begin{figure}
\noindent\includegraphics[width=15pc]{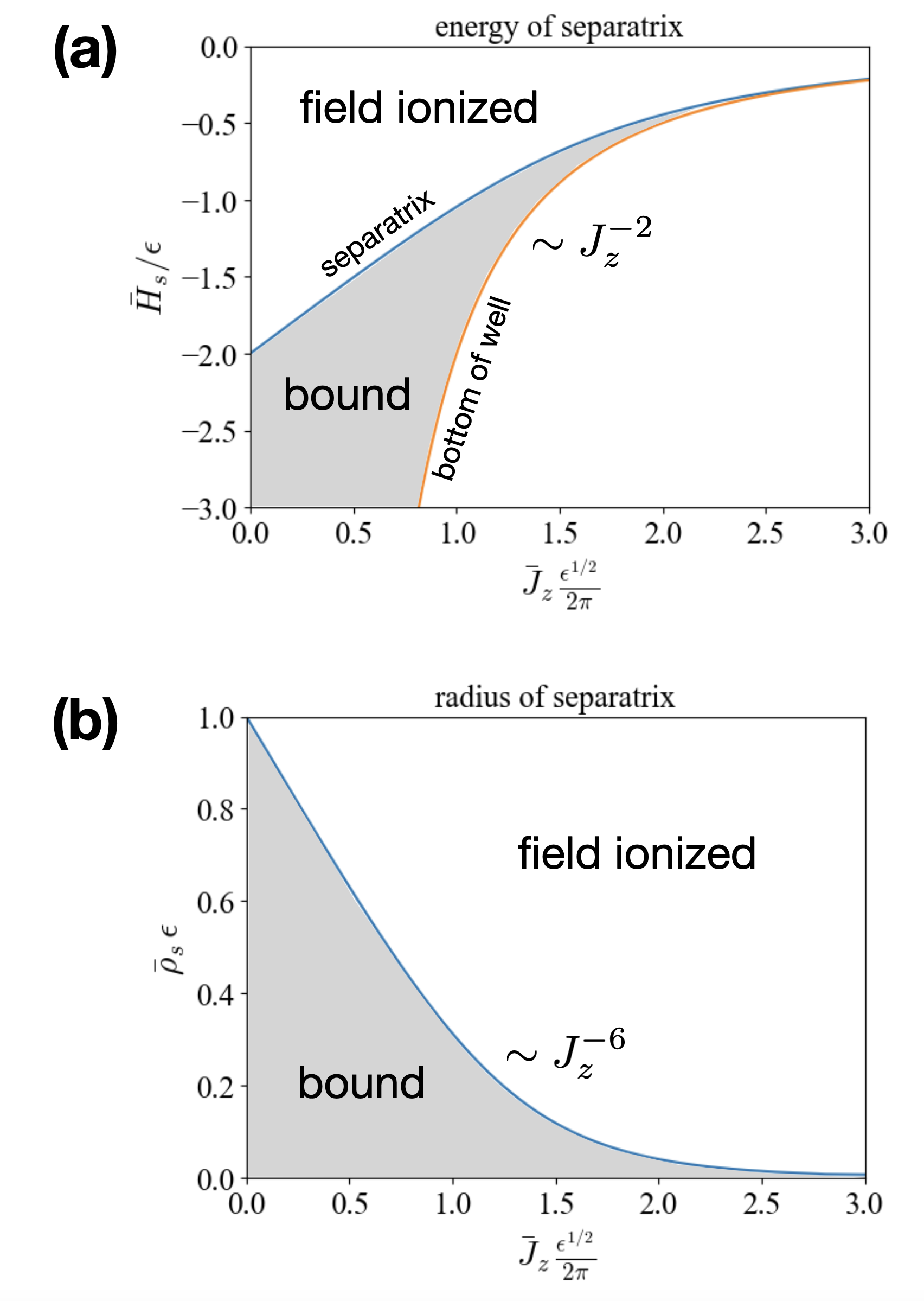}
\caption{\label{fig:gca.field.ionize} The field ionization effect of a constant electric field on the GCA.  (a) The energy of the separatrix and the bottom of the GCA potential well as functions of the $z$-bounce action $J_z/2\pi$.  This divides the space into an area of bound GCA, and an area of field ionized electrons.  (b) The radius of the separatrix, $\rho_\text{s}=\bar{r}_\text{s2}$, as a function of $J_z/2\pi$.}
\end{figure}

The character of the separatrix can be examined in more detail by assuming that $m_i \to \infty$, that is $1/a \to 0$, while allowing $R \to \infty$, so that $R/a$ and $v_{i \perp}$ are finite.  this effectively takes the ion cyclotron radius to infinity.  The parameter $R/a=\varepsilon^2$ can be scaled out of the problem, giving the Hamiltonian
\begin{equation}
    H_1(J_z;\rho,\theta) = \rho \, \sin{\theta} + H_z(J_z,\rho),
\end{equation}
where we have scaled
\begin{equation*}
\begin{split}
    &H \text{ by } \varepsilon \;\;\; (k_\text{B} T_e \text{ units}), \\
    &\rho \text{ by } \varepsilon^{-1} \;\;\; (b \text{ units}), \\
    &J_z \text{ by } \varepsilon^{-1/2} \;\;\; (m_e \bar{v}_e b \text{ units}), \\
    &p_z \text{ by } \varepsilon^{1/2} \;\;\; (m_e \bar{v}_e \text{ units}), \\
    &p_x \text{ by } \left( \frac{b}{r_\text{ce}} \right) \varepsilon^{-1} \;\;\; (m_e \bar{v}_e \text{ units}), \\
    &\omega_z \text{ by } \varepsilon^{3/2} \;\;\; (\bar{v}_e / b \text{ units}), \\
    &\omega_\text{D} \text{ by } \left( \frac{r_\text{ce}}{b} \right) \varepsilon^3 \;\;\; (\bar{v}_e / b \text{ units}),  \Rightarrow \frac{\omega_\text{D}}{\omega_z} \sim \frac{r_\text{ce}}{b} \varepsilon^{3/2}, \\
    &\text{and} \\
    &J_\text{D} \text{ by } \left( \frac{b}{r_\text{ce}} \right) \varepsilon^{-2} \;\;\; (m_e \bar{v}_e b \text{ units}).
\end{split}
\end{equation*}
Examples of the phase space trajectories are shown in Figs.~\ref{fig:gca.phase.space} and \ref{fig:gca.mix}.  The effect of finite ion thermal velocity is to electric field, $E_x=e \bar{v}_{i \perp} B / c$, ionize the atoms that are bound with $\rho \gtrsim 1/ \varepsilon$ and $|H| \lesssim \varepsilon$, where $\varepsilon \ll 1$, as shown in Fig.~\ref{fig:gca.field.ionize}.
%===============================%
%===============================%
%===============================%
%===============================%

\subsection{Stationary ion}
\label{sec:vtrr.fixed.ion}
The effect of this field ionization and subsequent transport to radii of order the ion cyclotron radius will be examined in Sec.~\ref{sec:vtrr.full}, after we calculate the rate for the $m_i \to \infty$ and $\bar{v}_{i \perp} \to 0$ limits (that is, the $1/a \to 0$ and $R/a \to 0$ limits) in this section.  In this limit we have the following forms for the Hamiltonians:
\begin{equation}
\begin{split}
    H_1(z,p_z,x,p_x) &= \frac{p_z^2}{2m} - \frac{e^2}{\sqrt{x^2 + \left( \frac{p_x}{m \Omega} \right)^2 + z^2}} \\
    &= \frac{p_z^2}{2m} - \frac{e^2}{r_1}
\end{split}
\end{equation}
and
\begin{equation}
\begin{split}
    H_{12} &= \frac{e^2}{\sqrt{\left( x_1 - x_2 \right)^2 + \left( \frac{p_{x1} - p_{x2}}{m \Omega} \right)^2 + \left( z_1 - z_2 \right)^2}} \\
    &= \frac{e^2}{r_{12}},
\end{split}
\end{equation}
where $\mathbf{B}=(mc \, \Omega/e) \, \hat{z}$, and we have dropped the subscript $e$ for the electron in this subsection.  We set $\widetilde{f}_2$ equal to 0 for $H_1<E$ and equal to a thermal distribution for $H_1>E$, that is
\begin{equation}
    \widetilde{f}_2 \sim \frac{N}{2\pi (m\bar{v})^2 \, (m\Omega)^2} \, \text{e}^{-(H_1+H_2+H_{12})}.
\end{equation}
The constraints are:
\begin{equation*}
\begin{split}
    H_1 &= - E_\text{s} \quad \text{(at bottleneck surface or separatrix)} \\
    H &\ge -E_2 \quad \text{(particle \#2 is unbound)} \\
    \text{i}_v \text{vol} &\ge 0 \quad \text{(} \widetilde{f}_2=0 \text{ below bottleneck)} \\
    \bar{\rho}_{12} &\le \rho_0 \quad \text{(adiabatic cutoff, limits multiple crossings)}.
\end{split}
\end{equation*}
Now calculate
\begin{equation*}
    \frac{\partial p_{z1}}{\partial H_1} = \frac{m}{p_{z1}}
\end{equation*}
and
\begin{equation*}
\begin{split}
    \{ H_{12}, H_1 \} &= \frac{e^2}{m} \frac{p_{1z} (z_1-z_2)}{r_{12}^3} + \left( \frac{e^2}{m\Omega} \right)^2 \left( \frac{x_2 p_{1x} - x_1 p_{2x}}{r_1^3 \, r_{12}^3} \right) \\
    &= (\mathbf{p}_1 \cdot \mathbf{F}_{12})_{\parallel \text{ to } \mathbf{B}} +  (\mathbf{p}_1 \cdot \mathbf{F}_{12})_{\perp \text{ to } \mathbf{B}, \text{ or } \mathbf{E} \times \mathbf{B}} \\
    &= \frac{d K_\parallel}{d t} + \frac{d K_\perp}{d t}.
\end{split}
\end{equation*}
Now, this leads to the following expression for the flux, using Eq.~\eqref{eqn:vtrr.pullback},
\begin{multline}
    \Phi_b = \frac{n}{V} \int_{E_\text{s}= -H_1}{\frac{4N}{2\pi(m \bar{v})^2(m\Omega)^2} \, \text{e}^{-(H_1+H_2+H_{12})}} \\
    \left[ \frac{e^2(z_1-z_2)}{r_{12}^3} + \frac{m}{p_{1z}} \left( \frac{e^2}{m\Omega} \right)^2 \left( \frac{x_2 p_{1x} - x_1 p_{2x}}{r_1^3 \, r_{12}^3} \right) \right] \\
    dx_1 \, dz_1 \, dp_{1x} \, dx_2 \, dz_2 \, dp_{2x} \, dp_{2z}.
\end{multline}
Make the change to dimensionless variables
\begin{align*}
    \varepsilon_\text{s} &\equiv - \frac{E_\text{s}}{m \bar{v}^2} & \rho_1 &\equiv r_1/b & \bar{x} &\equiv x/b \\
    \varepsilon_{2z} &\equiv \frac{p_{2z}}{2m\bar{v}^2} & \rho_2 &\equiv r_2/b & \bar{y} &\equiv p_x/m\Omega b \\
    \mu_2 &\equiv \frac{1}{\rho_2} - \frac{1}{\rho_{12}} & \rho_{12} &\equiv r_{12}/b & \bar{z} &\equiv z/b \\
    r_\text{ce}^2 &= \left( \frac{\bar{v}}{\Omega} \right)^2 & b &= \frac{e^2}{m \bar{v}^2} & \mathbf{r}_{21} &= \mathbf{x}_1 - \mathbf{x}_2,
\end{align*}
so that the flux is
\begin{multline}
    \Phi_b = n^2\bar{v} \, b^5 \frac{4}{2\pi \sqrt{2}} \, \text{e}^{\varepsilon_\text{s}} \int_{S^7}{\text{e}^{-\varepsilon_{2z} + \mu_2}} \\
    \left[ \frac{\bar{z}_1 - \bar{z}_2}{\rho_{12}^3} + \left( \frac{r_\text{ce}}{\sqrt{2}b} \right) \frac{\bar{x}_2 \bar{y}_1 - \bar{x}_1 \bar{y}_2}{\rho_1^3 \, \rho_{12}^3 \sqrt{1/\rho_1 - \varepsilon_\text{s}}} \right] \\
    \varepsilon_{2z}^{-1/2} \, d\varepsilon_{2z} \, d\bar{x}_1 \, d\bar{y}_1 \, d\bar{z}_1 \, d\bar{x}_2 \, d\bar{y}_2 \, d\bar{z}_2.
\end{multline}
Doing four of the integrals leaves us with
\begin{multline*}
    \Phi_b = R_0 \, (2\pi)^{3/2} \frac{\text{e}^{\varepsilon_\text{s}}}{\varepsilon_\text{s}^4} \int_{\bar{\rho}_1=0}^{1}{\bar{\rho}_1^2 \, d\bar{\rho}_1} \\
    \int_{\cos\zeta=-1}^{1}{d(\cos\zeta) \, (1+a)} \int_{\bar{\rho}_{12}=0}^{\rho_0}{d\bar{\rho}_{12} \, F(\mu_2)}
\end{multline*}
where
\begin{equation*}
\begin{split}
    R_0 &\equiv n^2 \bar{v} \, b^5, \\
    a &\equiv \varepsilon_\text{s}^{3/2} \left( \frac{r_\text{ce}}{\sqrt{2}b} \right) \frac{\sin\zeta}{\bar{\rho}_1 \sqrt{\bar{\rho}_1 (1-\bar{\rho}_1)}}, \\
    \mu_2 &\equiv \varepsilon_\text{s} \, \left( \frac{1}{\bar{\rho}_2} - \frac{1}{\bar{\rho}_{12}} \right) =  \varepsilon_\text{s} \, \left( \frac{\bar{\rho}_{12} - \bar{\rho}_2}{\bar{\rho}_2 \, \bar{\rho}_{12}} \right), \\
    \bar{\rho}_2^2 &= \bar{\rho}_1^2 + \bar{\rho}_{12}^2 - 2 \bar{\rho}_1 \bar{\rho}_{12} \cos\zeta, \\
    F(\mu_2) &\equiv \text{e}^{\mu_2} \, \text{erfc}(\sqrt{\text{max}(0,\mu_2)}), \\
    &\text{and} \\
    x &\equiv \cos\zeta.
\end{split}
\end{equation*}
We are left with the simple expressions
\begin{align*}
    \Phi_b  &= R_0 \, \frac{\text{e}^{\varepsilon_\text{s}}}{\varepsilon_\text{s}^4} \, f(r_\text{ce}/\sqrt{2}b,\varepsilon_\text{s},\rho_0) \\
    &= R_0 \, \frac{\text{e}^{\varepsilon_\text{s}}}{\varepsilon_\text{s}^4} \, \left[ g(\varepsilon_\text{s},\rho_0) + \left( \frac{r_\text{ce}}{\sqrt{2}b} \right) \varepsilon_\text{s}^{3/2} \, h(\varepsilon_\text{s},\rho_0) \right],
\end{align*}
where
\begin{multline*}
    g(\varepsilon_\text{s},\rho_0) \equiv (2\pi)^{3/2} \int_{\bar{\rho}_1=0}^{1}{\bar{\rho}_1^2 \, d\bar{\rho}_1} \int_{\bar{\rho}_{12}}^{\rho_0}{d\bar{\rho}_{12}} \\
    \int_{x=-1}^{1}{dx \, F(\mu_2)},
\end{multline*}
and
\begin{multline*}
    h(\varepsilon_\text{s},\rho_0) \equiv (2\pi)^{3/2} \int_{\bar{\rho}_1=0}^{1}{\frac{\bar{\rho}_1 \, d\bar{\rho}_1}{\sqrt{\bar{\rho}_1 \, (1-\bar{\rho}_1)}}} \int_{\bar{\rho}_{12}}^{\rho_0}{d\bar{\rho}_{12}} \\
    \int_{x=-1}^{1}{dx \, \sqrt{1-x^2} \, F(\mu_2)}.
\end{multline*}
Finally, we are left with the following expression for the flux to numerically evaluate
\begin{equation}
\label{eqn:3body.num.rate}
    \Phi_b = R_0 \, \rho_0 \, \frac{\text{e}^{\varepsilon_\text{s}}}{\varepsilon_\text{s}^4} \, 2(2\pi)^{3/2} \, \left[ \bar{g}(\varepsilon_1,\varepsilon_2) + \varepsilon_0 \, \bar{h}(\varepsilon_1,\varepsilon_2) \right]
\end{equation}
where
\begin{align*}
    \varepsilon_0 &\equiv \frac{\varepsilon_\text{s}^{3/2}}{\sqrt{2}} \, \left( \frac{r_\text{ce}}{b} \right),  & \varepsilon_1 &\equiv 1/\rho_0, & \varepsilon_2 &\equiv \varepsilon_\text{s}^{1/2} / \rho_0,
\end{align*}
\begin{multline}
\label{eqn:g.def}
    \bar{g}(\varepsilon_1,\varepsilon_2) \equiv \int_{\xi=0}^{\pi/2}{d\xi \, \sin\xi \cos^5\xi} \int_{y=-1}^{1}{dy} \\
    \int_{z=0}^{1}{dz \, F\left( \varepsilon_2^2 \frac{(z-x)/\varepsilon_1}{xz} \right)},
\end{multline}
\begin{multline}
\label{eqn:h.def}
     \bar{h}(\varepsilon_1,\varepsilon_2) \equiv \int_{\xi=0}^{\pi/2}{d\xi \, \cos^2\xi} \int_{y=-1}^{1}{dy \, \sqrt{1-y^2}} \\
    \int_{z=0}^{1}{dz \, F\left( \varepsilon_2^2 \frac{(z-x)/\varepsilon_1}{xz} \right)},
\end{multline}
and
\begin{equation*}
    x^2 = \varepsilon_1^2 \cos^4\xi + z^2 -2 \varepsilon_1 y z \cos^2\xi.
\end{equation*}
%===============================%
\begin{figure}
\noindent\includegraphics[width=\columnwidth]{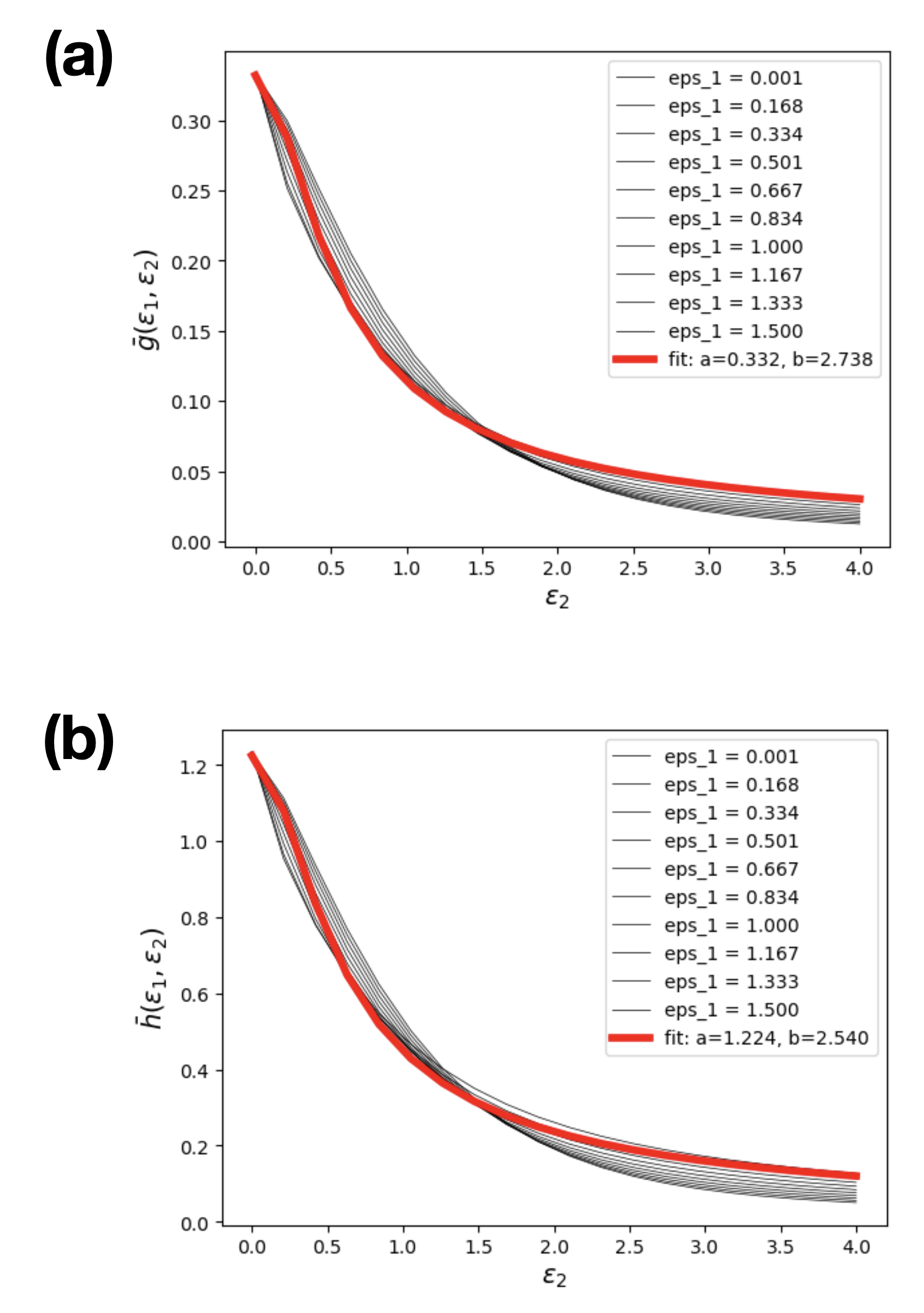}
\caption{\label{fig:g.h.functions} Two dimensionless functions that give the scaling of the flux, $\Phi_b$, with the energy of the ``bottleneck'' surface, $\varepsilon_\text{s}$, and the radius of the impact parameter cutoff, $\rho_0$.  The definitions are given by Eq.~\eqref{eqn:g.def} for $\bar{g}$ and by Eq.~\eqref{eqn:h.def} for $\bar{h}$, and their use in the calculation of the flux, $\Phi_b$, is shown in Eq.~\eqref{eqn:3body.num.rate}.  They are plotted as a function of $\varepsilon_2$ for a range of $\varepsilon_1$ as thin black lines.  The fits, according to Eq.~\eqref{eqn:g.fit} and Eq.~\eqref{eqn:h.fit}, are shown as the thick red lines.  (a) The function $\bar{g}(\varepsilon_1,\varepsilon_2)$, where $\varepsilon_1 \equiv 1/\rho_0$ and $\varepsilon_2 \equiv \sqrt{\varepsilon_\text{s}} / \rho_0$.  (b) The function $\bar{h}(\varepsilon_1,\varepsilon_2)$.}
\end{figure}
%===============================%
%===============================%
\begin{figure}
\noindent\includegraphics[width=17pc]{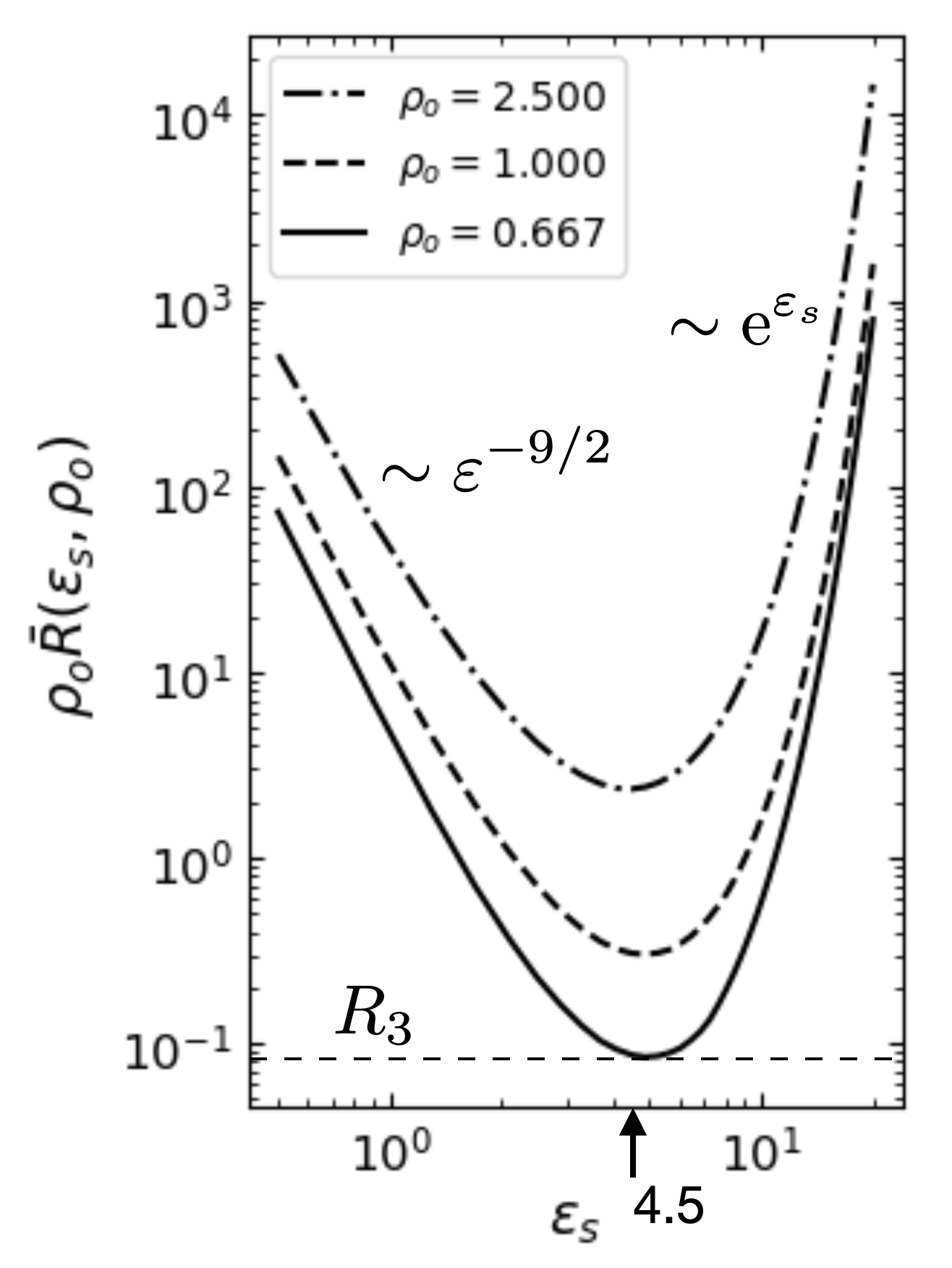}
\caption{\label{fig:3body.R.e.rho} The dimensionless $z$-bounce contribution to the flux, $\rho_0 \bar{R}(\varepsilon_\text{s},\rho_0)$.  The estimate of the flux, $\Phi_z$, is plotted as a function of the energy of the ``bottleneck'' surface, $\varepsilon_\text{s}$, for various values of the radius of the impact parameter cutoff, $\rho_0$.  For small values of $\varepsilon_\text{s}$, the rate scales as the density of states, $\varepsilon_\text{s}^{-4.5}$.  For large values of $\varepsilon_\text{s}$, the rate scales as the Boltzmann factor, $\exp{(\varepsilon_\text{s}})$.  This leads to a minimum or bottleneck at $\varepsilon_\text{sm}(\rho_0) \thickapprox 4.5$.  Also shown as the horizontal black dashed line is the three-body recombination rate, $R_3$, calculated by a Monte Carlo simulation of the Master Equation, Eq.~\eqref{eqn:gca.master}.  This rate is matched by the VTRR flux estimate at the ``bottleneck'' for an impact parameter cutoff of $\rho_0=2/3$.}
\end{figure}
%===============================%
%===============================%
\begin{figure}
\noindent\includegraphics[width=15pc]{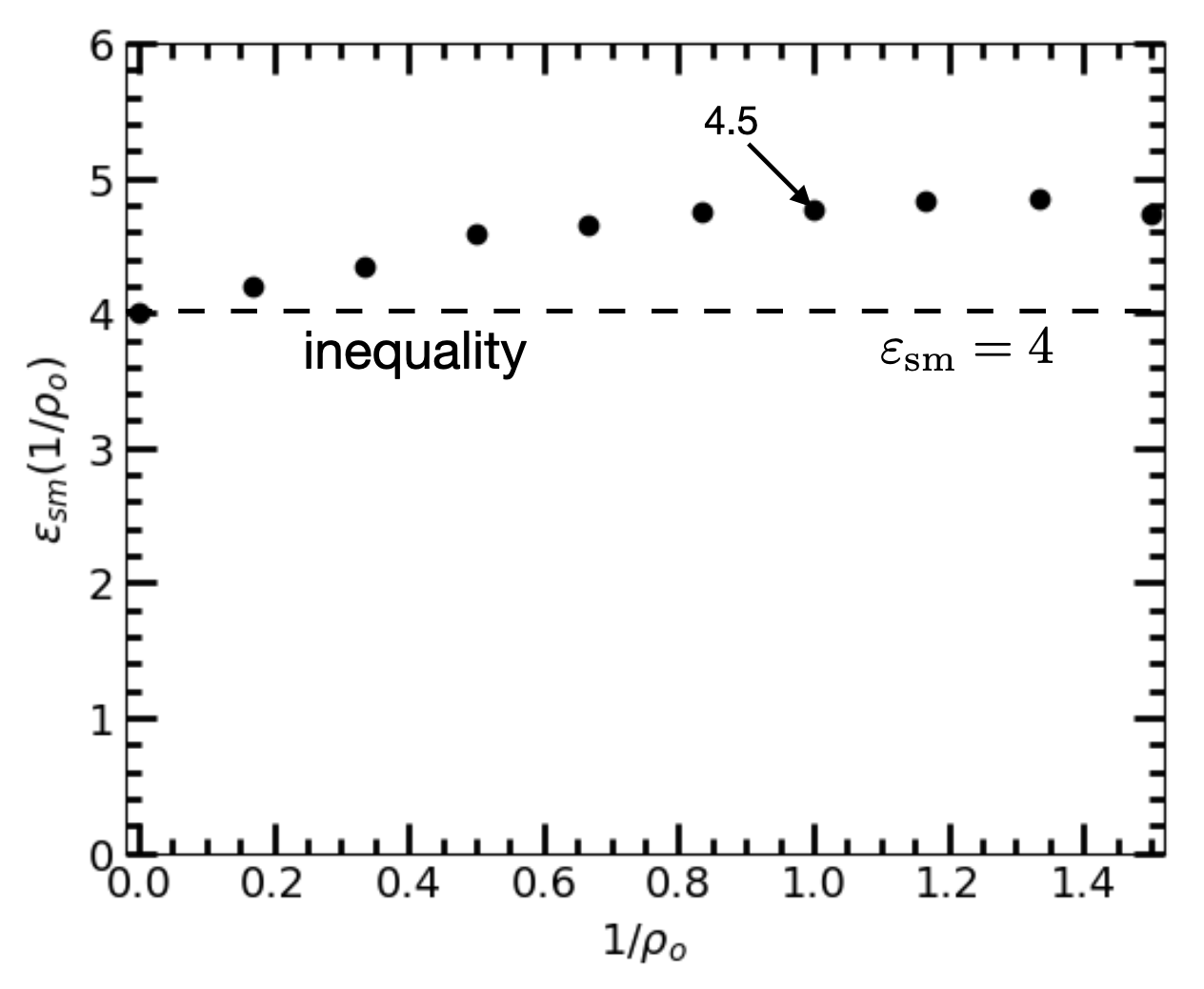}
\caption{\label{fig:3body.esm} The energy of the ``bottleneck'' surface, $\varepsilon_\text{sm}$, plotted as a function of the impact parameter cutoff radius, $\rho_0$.  Shown is the upper bound as the dashed black line labeled ``inequality'' at the value of $\rho_0 \to \infty$.  The asymptotic value of $4.5$, as $\rho_0 \to 0$, is labeled.}
\end{figure}
%===============================%
%===============================%
\begin{figure}
\noindent\includegraphics[width=15pc]{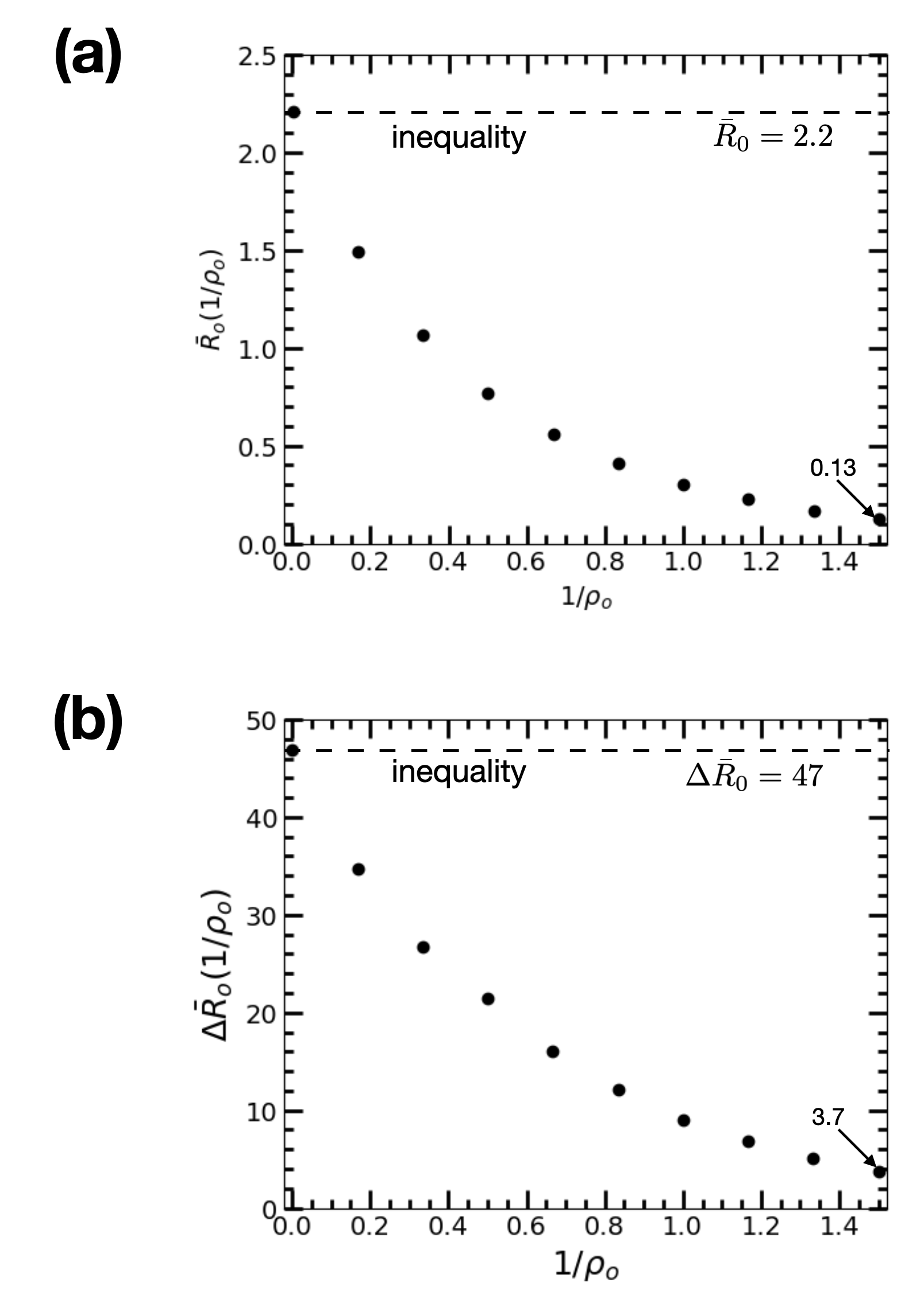}
\caption{\label{fig:3body.rates} The two contributions ($z$-bounce, $\bar{R}_0$, and $E \times B$ drift, $\Delta\bar{R}_0$) to the VTRR estimate of the three-body recombination rate, given in Eq.~\eqref{eqn:3body.flux}, are plotted as a function of the impact parameter cutoff, $\rho_0$.  Shown is the upper bound as the dashed black line labeled ``inequality'' at the value of $\rho_0 \to \infty$.  The two contributions at the value of the impact parameter cutoff that matches $R_3$, that is $1/\rho_0=1.5$, are labeled ($\bar{R}_0=0.13$ and $\Delta\bar{R}_0=3.7$).  (a) The $z$-bounce flux, $\Phi_z$, estimate given by $\bar{R}_0$.  (b) The $E \times B$ drift flux, $\Phi_\text{D}$, estimate given by $\Delta\bar{R}_0$.}
\end{figure}
%===============================%

The functions $\bar{g}(\varepsilon_1,\varepsilon_2)$ and $\bar{h}(\varepsilon_1,\varepsilon_2)$ can both be evaluated numerically and table interpolated.  It was found that both are weak functions of $\varepsilon_1$, as shown in Fig.~\ref{fig:g.h.functions}, and can be approximated as
\begin{equation}
\label{eqn:g.fit}
    \bar{g}(\varepsilon_1,\varepsilon_2) \thickapprox \frac{1/3}{\sqrt{1+A^2 \frac{\varepsilon_\text{s}}{\rho_0^2}}}
\end{equation}
and
\begin{equation}
\label{eqn:h.fit}
    \bar{h}(\varepsilon_1,\varepsilon_2) \thickapprox \frac{\pi^2 / 8}{\sqrt{1+B^2 \frac{\varepsilon_\text{s}}{\rho_0^2}}},
\end{equation}
where $A \thickapprox 2.7$ and $B \thickapprox 2.5$.  The flux can be separated into the $z$-bounce flux $\Phi_z$ and the $E \times B$ drift flux $\Phi_\text{D}$, with the following limits
\begin{equation*}
    \Phi_b = \Phi_z + \Phi_\text{D},
\end{equation*}
\begin{equation}
\label{eqn:z.flux.approx}
\begin{split}
    \Phi_z &\thickapprox R_0 \, \rho_0 \, \frac{\text{e}^{\varepsilon_\text{s}}}{\varepsilon_\text{s}^4} \, \frac{2(2\pi)^2}{3} \, (1+A^2 \varepsilon_\text{s} / \rho_0^2 )^{-1/2} \\
    &\xrightarrow[\rho_0 \to \infty]{} R_0 \, \rho_0 \, \frac{\text{e}^{\varepsilon_\text{s}}}{\varepsilon_\text{s}^4} \, \frac{2(2\pi)^2}{3} \quad \text{at } \varepsilon_\text{sm}=4 \\
    &\xrightarrow[\rho_0 \to 0]{} R_0 \, \rho_0 \, \frac{\text{e}^{\varepsilon_\text{s}}}{\varepsilon_\text{s}^{4.5}} \, \frac{2(2\pi)^2}{3A} \quad \text{at } \varepsilon_\text{sm}=4.5,
\end{split}
\end{equation}
and
\begin{equation*}
\begin{split}
    \Phi_\text{D} &\thickapprox R_0 \, \rho_0 \, \frac{\text{e}^{\varepsilon_\text{s}}}{\varepsilon_\text{s}^{5/2}} \, \frac{(2\pi)^{3/2}}{4 \sqrt{2}} \pi^2 \left( \frac{r_\text{ce}}{b} \right) \, (1+B^2\varepsilon_\text{s} / \rho_0^2)^{-1/2} \\
    &\xrightarrow[\rho_0 \to \infty]{} R_0 \, \rho_0 \, \frac{\text{e}^{\varepsilon_\text{s}}}{\varepsilon_\text{s}^{5/2}} \, \frac{(2\pi)^{3/2}}{4 \sqrt{2}} \pi^2 \left( \frac{r_\text{ce}}{b} \right) \quad \text{at } \varepsilon_\text{sm}=4 \\
    &\xrightarrow[\rho_0 \to 0]{} R_0 \, \rho_0 \, \frac{\text{e}^{\varepsilon_\text{s}}}{\varepsilon_\text{s}^3} \, \frac{(2\pi)^{3/2}}{4 \sqrt{2}} \frac{\pi^2}{B} \left( \frac{r_\text{ce}}{b} \right) \quad \text{at } \varepsilon_\text{sm}=4.5,
\end{split}
\end{equation*}
where $\Phi_\text{D}/\Phi_z \sim \omega_\text{D}/\omega_z \sim (r_\text{ce}/b) \, \varepsilon_\text{s}^{3/2}$ and $\varepsilon_\text{sm}=(\varepsilon_\text{s})_\text{min}$.  This leaves us with the following expression for the bound on the flux
\begin{equation}
\label{eqn:3body.flux}
\begin{split}
    \Phi_b &= R_0 \, \rho_0  \left[ \bar{R}_0(1/\rho_0) + \left( \frac{r_\text{rc}}{b} \right) \, \Delta \bar{R}_0(1/\rho_0) \right] \\
    &\le  R_0 \, \rho_0  \left[ \bar{R}(\varepsilon_\text{s},\rho_0) + \left( \frac{r_\text{rc}}{b} \right) \, \Delta \bar{R}(\varepsilon_\text{s},\rho_0) \right],
\end{split}
\end{equation}
where $\bar{R}_0(1/\rho_0)=\bar{R}(\varepsilon_\text{sm},\rho_0)$ and $\Delta\bar{R}_0(1/\rho_0)=\Delta\bar{R}(\varepsilon_\text{sm},\rho_0)$.  Figure~\ref{fig:3body.R.e.rho} shows how $\bar{R}(\varepsilon_\text{s},\rho_0)$ has a minimum in $\varepsilon_\text{s}$ at $\varepsilon_\text{sm}(1/\rho)$.  The three-body recombination rate calculated by the Monte Carlo simulation of Eq.~\eqref{eqn:gca.master}, the Master Equation, corresponds to $\bar{R}_0(1/\rho_0)$ at $\rho=2/3$.  The value of $\varepsilon_\text{sm}$ as a function of $1/\rho_0$ is shown in Fig.~\ref{fig:3body.esm}.  Note that $\varepsilon_\text{sm}=4$ for $1/\rho_0=0$, and $\varepsilon_\text{sm}=4.5$ for $\rho_0=0$.  The lower bounds on the flux, $\bar{R}_0$ and $\Delta\bar{R}_0$, as functions of $1/\rho_0$ are shown in Fig.~\ref{fig:3body.rates}.  Note that $\bar{R}_0 \le 2.2$ and $\Delta\bar{R}_0 \le 47$ with equality at $1/\rho_0=0$, and that $\bar{R}_0 = 0.13$ and $\Delta\bar{R}_0 \le 3.7$ at $1/\rho_0=1.5$.  This leads to the numerical bound on the flux of
\begin{equation}
    \Phi_b \le R_0 \, \rho_0 \, \left[ (2.2) + (r_\text{ce} / b)(47) \right]
\end{equation}
with equality as $\rho_0 \to \infty$.  The best numerical Variational Theory of Reaction Rates estimate of the three-body recombination rate, for $\rho_0=2/3$, is
\begin{equation}
    \boxed{\Phi_b = n^2 \, \bar{v} \, b^5 \, \left[ (0.085) + (r_\text{ce} / b)(2.5) \right].}
\end{equation}

\subsection{Finite ion mass and velocity}
\label{sec:vtrr.full}
Let us now turn our attention to the case where there is finite ion velocity, $\bar{v}_{i \perp}$, and finite ion mass, $m_i$.  For this case, both $R/a$ and $1/a$ are both not much less than 1, so that the full Hamiltonian shown in Eq.~\eqref{eqn:ei.full.hamiltonian} and Fig.~\ref{fig:full.phase.space} must be taken into account.  Three-body recombination into GCA only considers the $H_z(J_z,\rho)$ term and circular orbits with a bottleneck at $\varepsilon_\text{s}=4.5$.  For the case of finite ion mass and velocity, the GCA will be field ionized for radii larger than the separatrix at $\bar{r}_\text{s2}=(R/a)^{-1/2}$, that is for energies less than $1/\bar{r}_\text{s2}$.  These field ionized atoms will continue to go to larger radius until they reach the ion cyclotron radius, $\bar{r}_\text{ci}$.  Therefore, $\text{e}^{\varepsilon_\text{s}} \to \text{e}^{1/\bar{r}_\text{ci}}$ and $\varepsilon_\text{s}^{-9/2} \to \bar{r}_\text{s2}^{9/2}$ in the formula for the three-body recombination flux, $\Phi_b$, given in Eq.~\eqref{eqn:z.flux.approx}.  Remembering that $\varepsilon_\text{sm}$ asymptotes to $4.5$, and recognizing that $\Phi_b$ can not be faster than $R_0$, the formula for the three-body recombination flux can be modified to be
\begin{equation}
\label{eqn:3body.full.rate}
\begin{split}
    \Phi_b(\bar{r}_\text{ci},\bar{r}_\text{s2}) &= R_0 \, \min(1, \text{e}^{1/\bar{r}_\text{ci}} \, \bar{r}_\text{s2}^{9/2}) \\
    &= R_3(\bar{r}_\text{s1},\bar{r}_\text{s2}) = R_3(\eta,\xi).
\end{split}
\end{equation}
Here, we have recognized that the three-body recombination rate, $R_3$, is a function of two dimensionless variables, either the two dimensionless separatrix radii,
\begin{equation*}
    \bar{r}_\text{s1} = \left( \frac{1}{a} \right)^{-1/3} = \eta = \left[ \frac{m_i}{m_e} \left( \frac{r_\text{ce}}{b} \right)^2 \right]^{1/3} = \left( \frac{\omega_{E \times B}}{\Omega_i} \right)^{1/3}
\end{equation*}
for the DP and
\begin{equation*}
    \bar{r}_\text{s2} = \left( \frac{R}{a} \right)^{-1/2} = \eta^{3/4} \xi^{-1/4} = \left[ \frac{\bar{v}_e}{\bar{v}_{i \perp}} \left( \frac{r_\text{ce}}{b} \right) \right]^{1/2}
\end{equation*}
for the GCA, or the dimensionless variables that we will use to plot $R_3$,
\begin{equation*}
    \eta = \left( \frac{1}{a} \right)^{-1/3} = \bar{r}_\text{s1} = \left[ \frac{m_i}{m_e} \left( \frac{r_\text{ce}}{b} \right)^2 \right]^{1/3}
\end{equation*}
and
\begin{equation*}
    \xi = \left( \frac{1}{a} \right)^{-1} \left( \frac{R}{a} \right)^2 = \frac{\bar{r}_\text{s1}^3}{\bar{r}_\text{s2}^4} = \frac{T_i}{T_e}.
\end{equation*}
In terms of these variables the dimensionless ion cyclotron radius is
\begin{equation*}
    \bar{r}_\text{ci} = \frac{\bar{r}_\text{s1}^3}{\bar{r}_\text{s2}^2} = \eta^{3/2} \xi^{1/2} = \frac{m_i \bar{v}_{i \perp}}{m_e \bar{v}_e} \left( \frac{r_\text{ce}}{b} \right).
\end{equation*}

\begin{figure}
\noindent\includegraphics[width=17pc]{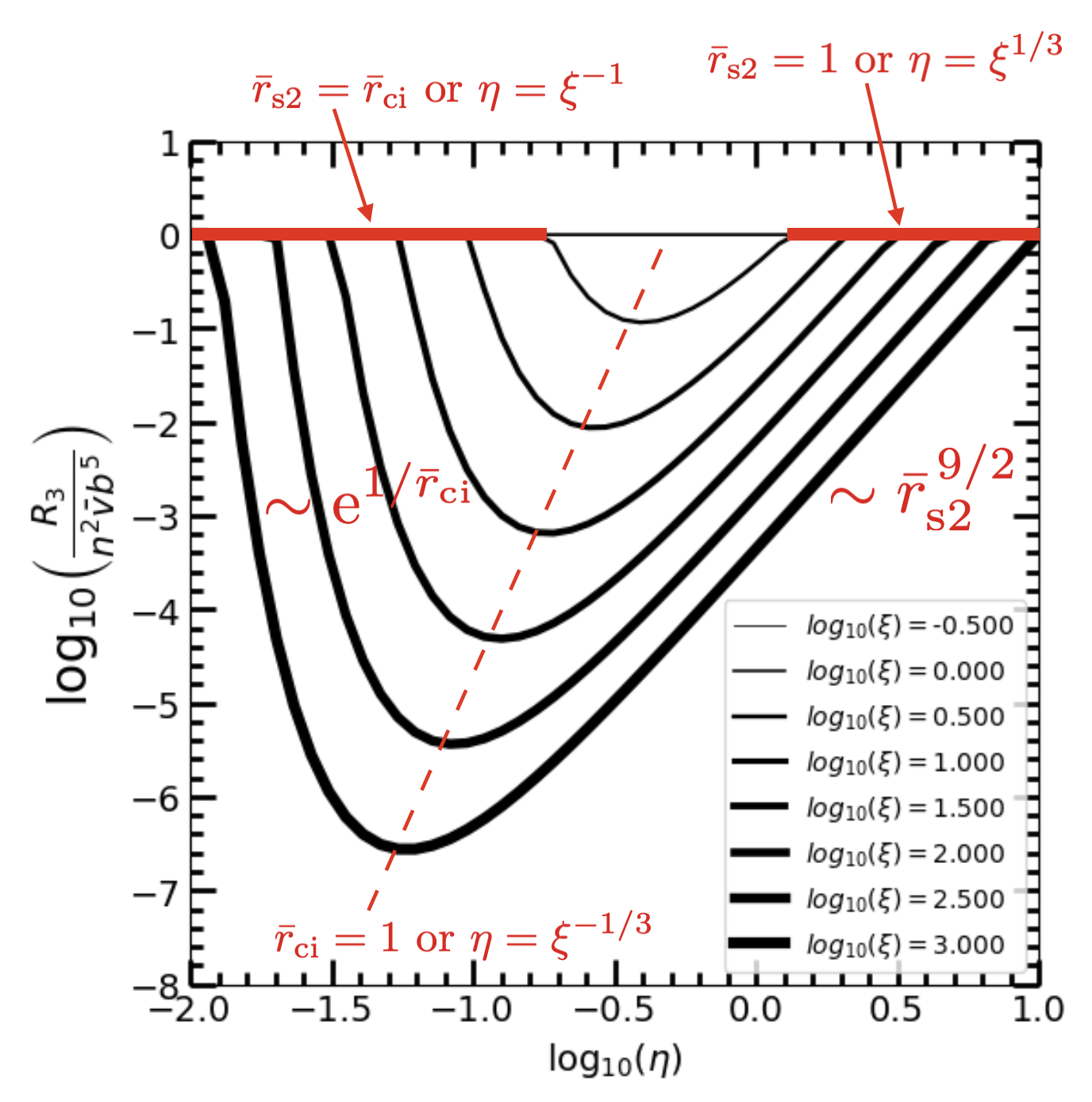}
\caption{\label{fig:3body.full.R3.v.eta} The three-body recombination rate $R_3$ as a function of $\eta$, for various values of $\xi$.  The rate scales as $R_3 \sim \bar{r}_\text{s2}^{9/2}$ for $\eta > \xi^{-1/3}$ or the field ionization of GCA with a separatrix, and as $R_3 \sim  \exp{(1/\bar{r}_\text{ci})}$ for $\eta < \xi^{-1/3}$ or the two bottleneck model.  The rate is reduced from $R_0$ when $\xi^{-1}<\eta<\xi^{1/3}$.  The red dashed line, $\eta \sim \xi^{-1/3}$, shows the transition from field ionization to the two bottleneck model.  It is also near the minimum $R_3$.}
\end{figure}

\begin{figure}
\noindent\includegraphics[width=\columnwidth]{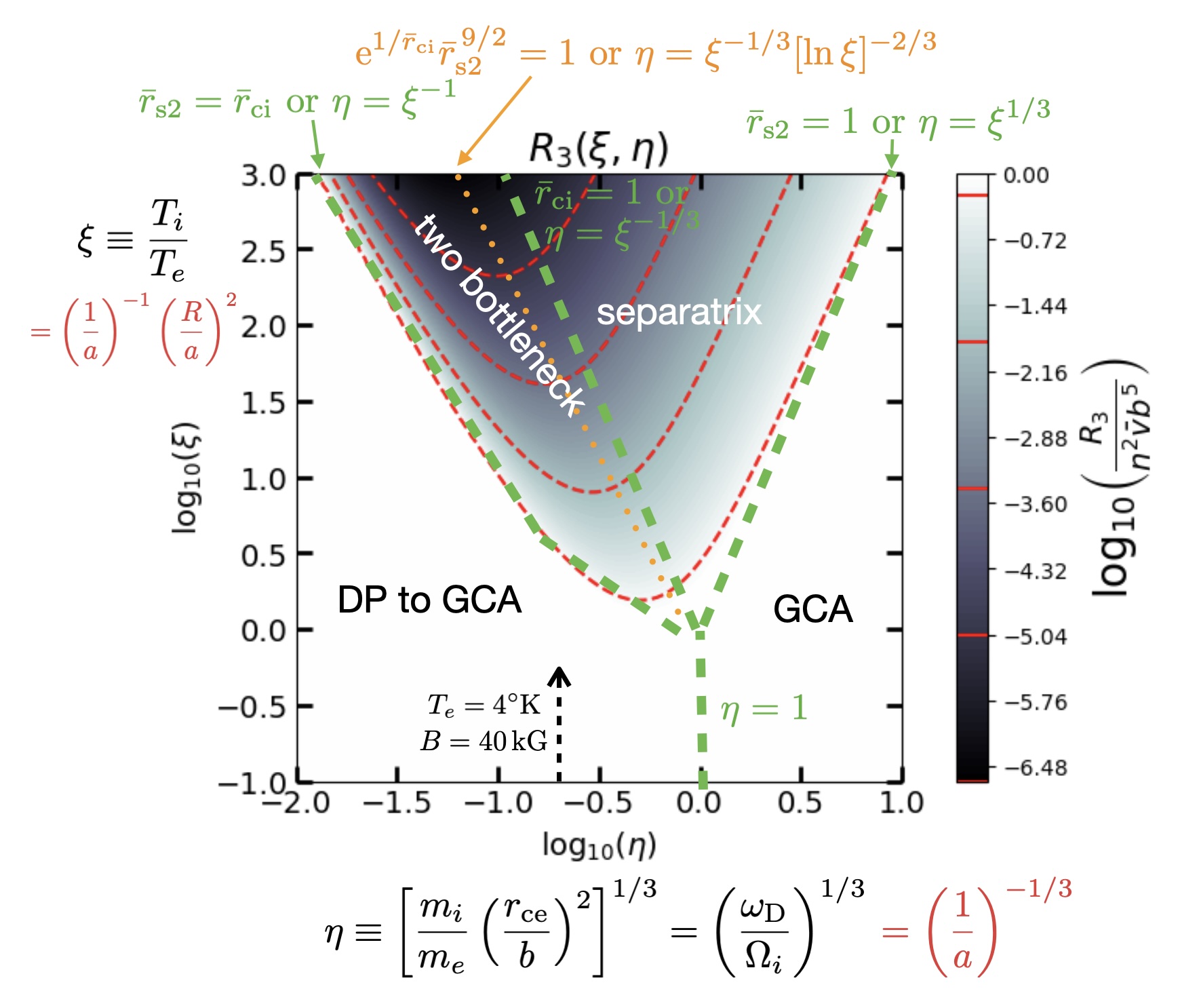}
\caption{\label{fig:3body.full.xi.v.eta} Contour plot of the three-body recombination rate $R_3$ as a function of $\eta$ and $\xi$, showing the domains of different recombination physics.  The thick green dashed lines divide the plane into the four different domains:  (1) GCA recombination with a single bottleneck, (2) DP to GCA recombination with a single bottleneck, (3) field ionization of the GCA with a separatrix, and (4) two bottleneck recombination.  The minimum recombination rate in $\eta$ is shown as the orange dotted line.  The condition determining the scaling and the $\xi$ scaling of the boundaries are shown.  The $\eta$ value for the CERN anti-hydrogen experiments is shown as the black dashed arrow.}
\end{figure}

The formula for the three-body recombination rate as a function of $\eta$ and $\xi$, $R_3(\eta,\xi)$, given in Eq.~\eqref{eqn:3body.full.rate}, is plotted in Fig.~\ref{fig:3body.full.R3.v.eta} and Fig.~\ref{fig:3body.full.xi.v.eta}.  Note that the rate is reduced when $\xi>1$ and $\xi^{-1}<\eta<\xi^{1/3}$.  The minimum rate occurs at $\eta=\xi^{-1/3} (\ln\xi)^{-2/3}$.  The condition determining this minimum is
\begin{equation*}
    \text{e}^{1/\bar{r}_\text{ci}} \, \bar{r}_\text{s2}^{9/2} = 1.
\end{equation*}
If $\eta>1$, then the dynamics is always that of a GCA or a field ionized GCA.  For $\eta<1$, the dynamics is that of DP at large radius, then transitions to GCA for smaller radii.
%===============================%
%===============================%
%===============================%
%===============================%

\begin{figure}
\noindent\includegraphics[width=\columnwidth]{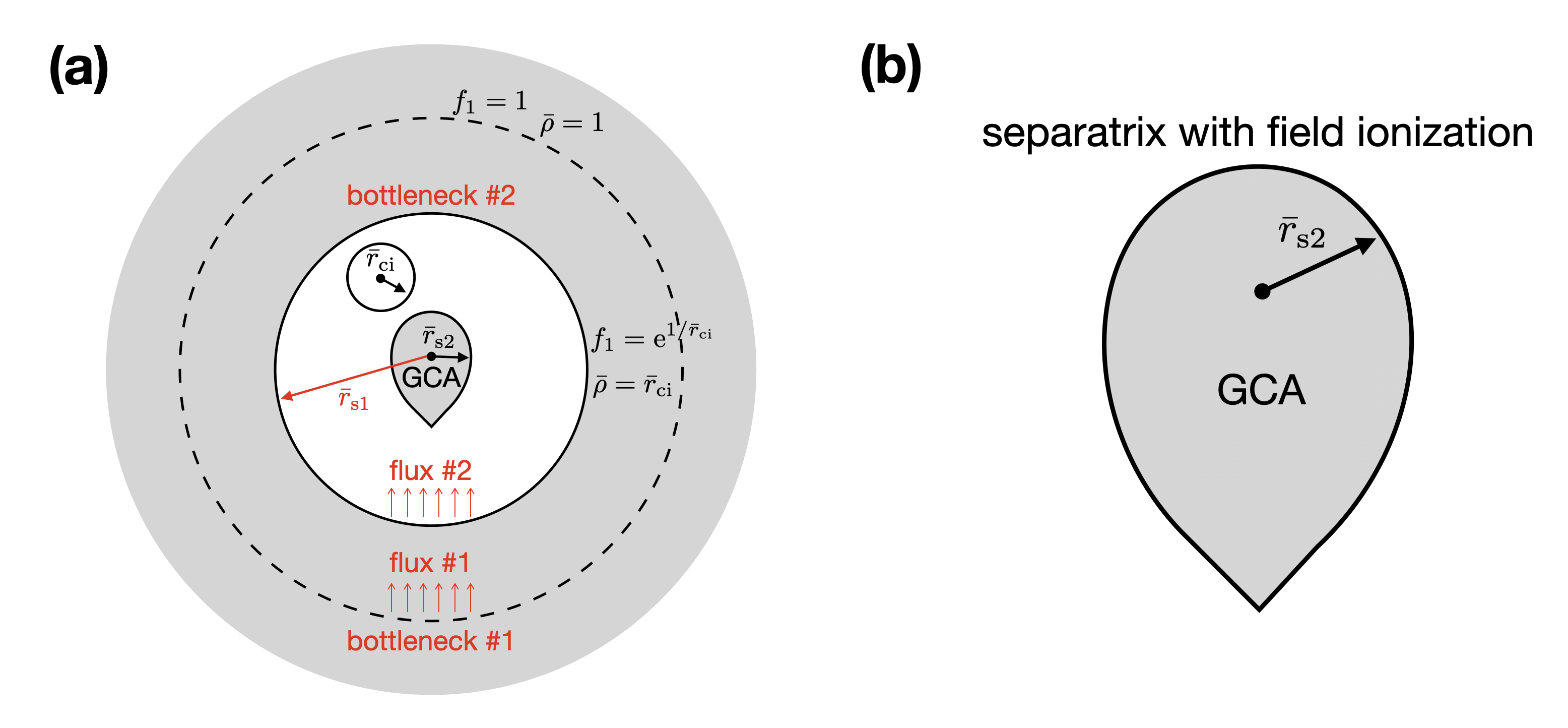}
\caption{\label{fig:3body.physics} Illustration of the physics of three-body recombination of an electron and an ion in a strong magnetic field.  (a) The 2-step or bottleneck model where there is a second bottle neck at $\bar{r}_\text{s1}$ down to which there is a thermal equilibrium.  The atoms undergo an ion cyclotron motion that carries them close to the domain of bound GCAs, where they can be bound as a GCA via three-body interaction.  (b) Field ionization model, where the GCA is field ionized for radii larger than $\bar{r}_\text{s2}$.}
\end{figure}

For $\xi^{-1/3}<\eta<\xi^{1/3}$, the rate is being reduced by electric field ionization of the GCA, illustrated in Fig.\ref{fig:3body.physics}b.  The rate is scaling as $R_3 \sim \bar{r}_\text{s2}^{9/2}$, the phase space volume inside the separatrix, that is the ionization radius, as shown in  Fig.\ref{fig:3body.physics}b.  The upper limit on the $\eta$ range is determined by the condition $\bar{r}_\text{s2}=1$, and the lower limit by the condition $\bar{r}_\text{ci}=1$.  
%===============================%
%===============================%

For $\xi^{-1}<\eta<\xi^{-1/3}$, the rate is being reduced by the a 2-step or two bottleneck model, illustrated in Fig.\ref{fig:3body.physics}a.  The rate is scaling as $R_3 \sim  \exp{(1/\bar{r}_\text{ci})}$, the Boltzmann factor at the ion cyclotron radius.  The upper limit on the $\eta$ range is determined by the condition $\bar{r}_\text{ci}=1$, and the lower limit by the condition $\bar{r}_\text{s2} = \bar{r}_\text{ci}$.  The three-body recombination rate is being determined by a second kinetic bottleneck at the ion cyclotron radius $\bar{\rho}=\bar{r}_\text{ci}$, as shown in the figure.  The distribution will be filled to thermal equilibrium down to the ion cyclotron radius, then the atom will undergo a cyclotron orbit until it is close the field ionization radius where it will be transported by a three-body collision into a bound GCA.

\begin{figure}
\noindent\includegraphics[width=17pc]{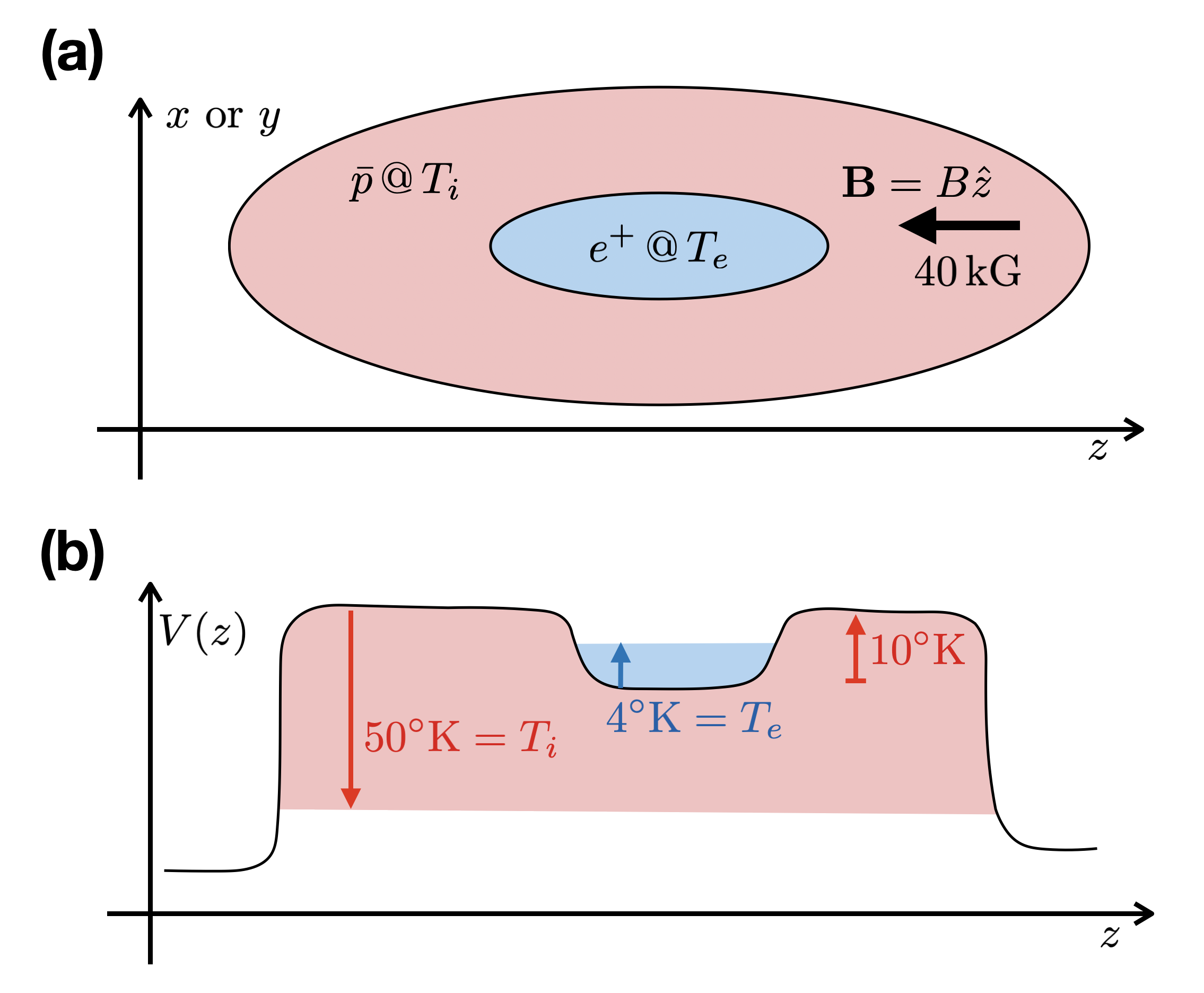}
\caption{\label{fig:3body.penning} The geometry of the nested Penning traps for the formation of anti-hydrogen at CERN.  (a) Spacially in $x$, $y$ and $z$.  (b) Potential energy structure in $z$, $V(z)$.}
\end{figure}

\begin{figure}
\noindent\includegraphics[width=17pc]{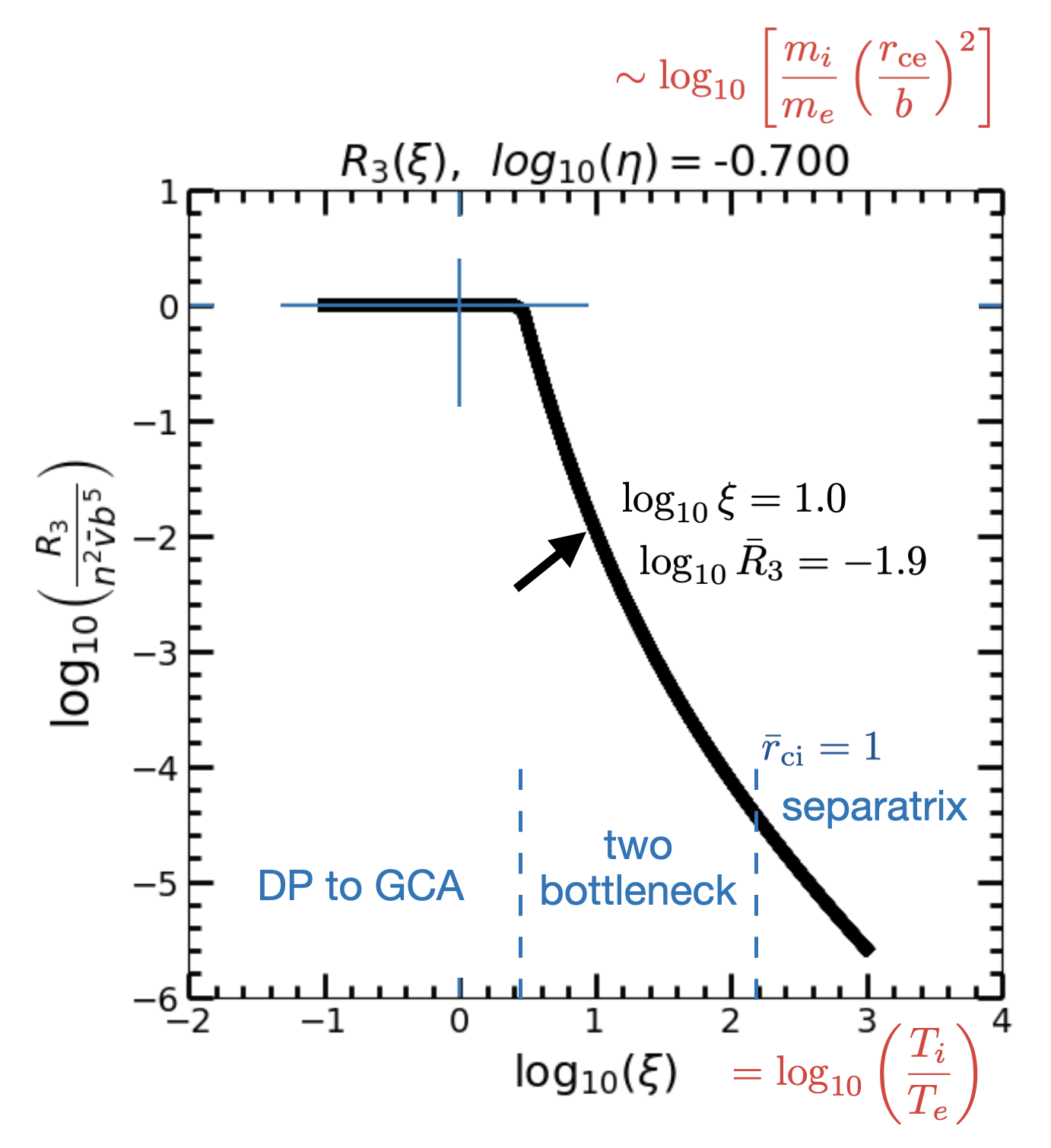}
\caption{\label{fig:3body.full.R3.xi} The three-body recombination rate $R_3$ as a function of $\xi$ for the $\eta=-0.7$ value corresponding to the conditions of Fig.~\ref{fig:3body.penning}.  A black arrow shows the value of $\xi=1.0$, corresponding to the conditions of Fig.~\ref{fig:3body.penning}, where the rate $R_3$ is reduced by a factor of $10^{1.9}$.  The three regimes of the physics are shown by the dashed blue lines.}
\end{figure}

The application that was the inspiration for this research, are the experiments that are forming anti-hydrogen at CERN.  These experiments are forming anti-hydrogen via three-body recombination in a nested Penning trap, as shown in Fig.~\ref{fig:3body.penning}.  These experiments have a magnetic field of $B=40 \, \text{kG}$, an electron temperature of $T_e = 4^\circ \text{K}$, and an ion temperature of $T_i=50^\circ \text{K}$.  This leads to $\log{\eta}=-0.7$ and $\log{\xi}=1.0$.  The three-body recombination rate as a function of $\xi$, for this case, is plotted in Fig.~\ref{fig:3body.full.R3.xi}.  For $\xi>1$, the three-body recombination scales as $R_3 \sim \xi^{-2}$, so that the three-body recombination rate is reduced by a factor of 100 from $R_0=n_e^2 \bar{v}_e b^5$ for these experiments.
%===============================%
%===============================%
%===============================%
%===============================%

\section{Conclusions}
\label{sec:conclusions}
This paper has recast plasma theory, that is the theory of collective systems, in a coordinate-free manner using exterior calculus and a sympletic geometry.  Using this approach, the physics is not obscured by the details of the coordinate dependent notation.  The derivations can be done in a general way, independent of the coordinate system or the form of the Hamiltonian, once and for all coordinate systems and Hamiltonians.  It started by assuming a canonical, that is symplectic or conservative, flow on the $N$-particle phase space that conserves phase space volume.  This was expressed as the Generalized Liouville Equation, Eq.~\eqref{eqn:gle}.  This equation was then integrated over $(N-n)$-particles to give the equations for the $n$-particle distribution forms, $\rho^{(n)}$.  These equations are also known as the Generalized BBGKY Hierarchy, Eq.~\eqref{eqn:bbgky}.  It should be noted that this hierarchy is an expansion in the weakness of correlation, $\Gamma \ll 1$.  This series is only asymptotically convergent, and the expansion parameter is of order one or greater $\Gamma \gtrsim 1$ for many situations.  In fact, these are the most interesting situations.  Recently, there has been developments using a generating functional approach,\citep{glinsky.24a} that is the Heisenberg Scattering Transformation (HST), that does not make the BBGKY expansion, so that the HST based theory can address the behavior of collective systems with more fidelity, simplicity, and correlation.   Despite this being said, there has been, and will continue to be, much that can be understood using the BBGKY expansion for weakly correlated plasmas.

The BBGKY Hierarchy was pulled-back to canonical coordinates and truncated at first order, leading to the Generalized Vlasov Equation, Eq.~\eqref{eqn:vlasov}.  This is an equation for evolution of the one-particle distribution function, $f_1(P,Q)$, due to interaction, that is correlation, with a second particle.  The interaction is modeled as an effective conservative force due to an average interaction Hamiltonian, $\left< H_{12} \right>$.  Therefore, the evolution is still conservative, but does not take into account three-body or greater interaction.  In order to account for three-body interaction, the BBGKY Hierarchy was truncated and averaged over the cyclic canonical coordinate, $Q$, to give three expressions for the temporal evolution of the one-particle distribution function averaged over $Q$, $\bar{f}_1(P)$.  This is expressed in terms of a collision functional, $\mathcal{C}[\bar{f}_1(P)]$, as
\begin{equation}
    \frac{\partial \bar{f}_1(P)}{\partial \tau} = \mathcal{C}[\bar{f}_1(P)].
\end{equation}
There is the most general form of the collision functional, the Boltzmann operator,
\begin{multline}
    \mathcal{C}_\text{b}[\bar{f}_1(P)] = \int{dP_2 \; dP'_1 \; dP'_2 \; \; K(P,P_2|P'_1,P'_2)} \\
   [\bar{f}_1(P'_1) \, \bar{f}_1(P'_2)-\bar{f}_1(P) \, \bar{f}_1(P_2)],
\end{multline}
where the distribution for the colliding particle is an arbitrary function of $P_2$, $\bar{f}_1(P_2)$.  There is the Master Equation collision operator,
\begin{equation}
    \mathcal{C}_\text{m}[\bar{f}_1(P)]=\int{dP' \; \left[ k(P'|P) \; \bar{f}_1(P') - k(P|P') \; \bar{f}_1(P) \right]},
\end{equation}
where the distribution for the colliding particle is a thermal distribution in $P_2$,
\begin{equation*}
    \bar{f}_\text{th}(P_2) \sim \text{e}^{-E(P_2)}.
\end{equation*}
Finally, there is the Fokker-Planck collision operator, when the distribution for the colliding particle is a thermal distribution and the change in $P$ is small compared to scale on which the distribution in $P$, $\bar{f}_1(P)$, changes,
\begin{equation}
    \mathcal{C}_\text{fp}[\bar{f}_1(P)]=\frac{\partial}{\partial P} \cdot \left[ \left< \frac{\Delta P \, \Delta P}{\Delta \tau} \right> \cdot \left( \frac{\partial \bar{f}_1}{\partial P} - \frac{\bar{f}_1}{f_\text{eq}} \frac{\partial f_\text{eq}}{\partial P} \right) \right].
\end{equation}
For all three cases, the collision operators are not conservative, but act as external forces, changing only the action, $P$, or energy, $E(P)$, of the collective system.  For low frequency modes of the plasma where $\omega_Q \ll \nu_c$, the Vlasov Equation must be extended to include collisions, to give the Collisional Vlasov Equation, Eq.~\eqref{eqn:coll.vlasov}.  When $\mathcal{C}[\bar{f}_1(P)]=\mathcal{C}_\text{fp}[\bar{f}_1(P)]$, this is the Vlasov-Fokker-Planck Equation, Eq.~\eqref{eqn:gvfp}.  Given the geometric interpretation of the Lie derivative, $\text{D}/\text{D} \tau$, a variational estimate was made of the collisional flux, $\Phi_\text{b}$.  The coordinate-free expression for this flux is given in Eq.~\eqref{eqn:vtrr.bound}, the Variational Theory of Reaction Rates (VTRR).

For plasma behavior that does not come from high order structure in $P$, a $P$-moment expansion can be made in the distribution function, $f_1(P,Q)$.  This yields the two fluid equations:  the Generalized Continuity Equation given by Eq.~\eqref{eqn:gen.continuity}, and the Generalized Momentum (Force) Equation given by Eq.~\eqref{eqn:gen.force}.  The Continuity Equation is the zeroth-order moment of the Vlasov Equation, Eq.~\eqref{eqn:vlasov}, and the Momentum Equation is first-order moment of the Vlasov Equation.  These equations are partial differential equations for:  (1) the zeroth-order moment of the one-particle distribution function, $n=\int{dP \, f_1(P,Q)}$, also known as the density, (2) the first-order moment, $n \, \bar{\omega}_Q=\int{dP \, \omega_Q(P) \, f_1(P,Q)}$, also known as the fluid momentum, and (3) the second-order moment, $p=n \, \overline{\omega_Q \, \omega_Q}= \int{dP \, \omega_Q(P) \, \omega_Q(P) \, f_1(P,Q)}$, also known as the pressure tensor.  The third equation is given by an Equation of State (EoS).  The Momentum Equation can be written as
\begin{equation}
    \frac{\partial}{\partial \tau} \left( n \, \bar{\omega}_Q \right) = F_\text{p} + F_\text{i},
\end{equation}
where the partial time derivative of the fluid momentum is equal to the sum of the thermal pressure force $F_\text{p} \equiv -\partial p / \partial Q$, and the two-particle interaction force, 
\begin{equation*}
    F_\text{i} \equiv - \int{dP \, \omega_Q(P) \left\{ \left< \widetilde{H}_{12} \right>,f_1(P) \right\}}.
\end{equation*}
The fluid equations are valid just so that collisions can be neglected, that is $\omega_Q \gg \nu_c$.  

This is the case for most situations, but there are plasma modes for which $\omega_Q \lesssim \nu_c$, so that collisions must be included and the Collisional Vlasov Equation, Eq.~\eqref{eqn:coll.vlasov}, used.  When the moments of the Collisional Vlasov Equation are taken, the continuity equation remains the same, and the momentum equation can be written as
\begin{equation}
\begin{split}
    \frac{\partial}{\partial \tau} \left( n \, \bar{\omega}_Q \right) &= F_\text{p} + F_\text{i} + F_\text{c} \\
    &= F^{(1)} + F^{(2)} + F^{(3)},
\end{split}
\end{equation}
where
\begin{equation*}
    F_\text{c} \equiv \int{dP \, \omega_Q(P) \, \mathcal{C}[\bar{f}_1(P)]}
\end{equation*}
is the collisional force, $F^{(1)}=F_\text{p}$ is the one-body interaction force,  $F^{(2)}=F_\text{i}$ is the two-body interaction force, and $F^{(3)}=F_\text{c}$ is three-body interaction force.  

The MagnetoHydroDynamic (MHD) equations are derived by transforming to center of mass, and center of charge coordinates.  This results in four equations: (1) the mass conservation equation, Eq.~\eqref{eqn:gen.mass}, (2) the charge conservation equation, Eq.~\eqref{eqn:gen.charge}, (3) the mass momentum (force) equation, Eq.~\eqref{eqn:gen.mhd.force}, and (4) the charge momentum (Ohm's Law) equation, Eq.~\eqref{eqn:gen.mhd.ohms}.  There are four unknowns: (1) the mass density, $\rho_M$, (2) the charge density, $\rho_c$, (3) the fluid momentum, $\rho_M \, \Omega_M$, and (4) the charge current density, $J_c$.

Finally, the VTRR was developed and applied to the physical system of an electron and an ion in a strong magnetic field.  The magnetic field is strong enough that the electron undergoes guiding center motion.  First, it was assumed that the ion was stationary.  In this limit it was found that the following bound can be put on the three-body recombination rate
\begin{equation}
    R_3 \lesssim \Phi_\text{b} = n_e^2 \, \bar{v}_e \, b^5 \, \left[ (0.085) + (2.5)(r_\text{ce} / b) \right],
\end{equation}
where the first term is due to $z$-bounce dynamics and the second term is due to drift dynamics.  For an ion with finite mass, $m_i$, and temperature, $T_i$, a two bottleneck model was developed.  For ion temperatures greater than the electron temperature, $T_e$, the three-body recombination is reduced by
\begin{equation}
    \frac{R_3}{n_e^2 \, \bar{v}_e \, b^5} \sim \left( \frac{T_i}{T_e} \right)^{-2}.
\end{equation}
This theory has been useful in the efforts to form anti-hydrogen at CERN.

\begin{acknowledgments}
First of all the author would like to thank Prof. Thomas O'Neil for all the mentorship in the traditional theoretical approach to plasma theory.  Further education given by Prof. Ted Frankel on the coordinate-free methods of exterior calculus and how they can be applied to physics, was an invaluable foundation for this research.  Finally, the friendship of fellow graduate student Poul Hjorth.  The many discussions we had on the foundations of plasma theory and exterior calculus were invaluable.  Above all, the author would like to thank Prof. Poul Hjorth for the invitation for a stay at DTH in Lyngby Denmark, during which parts of the first draft of this paper were written.  The computer software used to generate some of the figures is available on GitHub \citep{github.bbgky_dg}.  Lecture notes for a plasma physics course based on this work can be found on the internet \citep{glinsky.notes.22}.  These lecture notes give more detail and context for the content of this paper.
\end{acknowledgments}

\bibliography{BBGKY_DG_References.bib}

\end{document}